\documentclass[floatfix,prl,aps,twocolumn,superscriptaddress,10pt]{revtex4-2}

\usepackage{silence}
\usepackage[utf8]{inputenc}
\usepackage[T1]{fontenc}
\usepackage{color}
\usepackage{amsmath}
\usepackage{amsfonts}
\usepackage{hyperref}
\usepackage{amsthm}
\usepackage{bm}
\usepackage{float}
\usepackage{textcomp}
\usepackage{upgreek}
\usepackage{textgreek}
\usepackage{setspace}
\usepackage{amssymb,graphicx}
\usepackage[percent]{overpic}
\hypersetup{
  colorlinks = true,
  allcolors= blue,
}
\usepackage{tikz,pgfplots}
\usepackage{blkarray}
\usepackage{array}
\usepackage{multirow}
\usepackage[normalem]{ulem} 

\usepackage{ctable}

\usetikzlibrary{patterns}
\usetikzlibrary{fadings}
\usetikzlibrary{decorations.pathmorphing,patterns}
\tikzset{snake it/.style={decorate, decoration=snake}}
\usetikzlibrary{arrows, decorations.markings}
\pgfplotsset{compat=1.14}
\tikzset{
vecArrow/.style={
  thick,
  decoration={markings,mark=at position
   1 with {\arrow[scale=2,thin]{open triangle 60}}},
  double distance=1.4pt, shorten >= 10.5pt,
  preaction = {decorate},
  postaction = {draw,line width=1.4pt, white,shorten >= 4.5pt}
  },
innerWhite/.style={
  semithick,
  white,
  line width=1.4pt,
  shorten >= 4.5pt
  }
}

\DeclareMathAlphabet{\mathsfit}{\encodingdefault}{\sfdefault}{m}{sl}
\SetMathAlphabet{\mathsfit}{bold}{\encodingdefault}{\sfdefault}{bx}{sl}
\newcommand{\tens}[1]{\bm{\mathsfit{#1}}}
\newcommand{\tenscomp}[1]{\mathsfit{#1}}

\makeatletter
\usepackage{comment}
\let\wfs@comment@comment\comment
\let\comment\@undefined

\usepackage{changes}
\let\wfs@changes@comment\comment
\let\comment\@undefined

\newcommand\comment{%
    \ifthenelse{\equal{\@currenvir}{comment}}
    {\wfs@comment@comment}
    {\wfs@changes@comment}%
}
\makeatother
\begin{document}
\setcitestyle{super}

\title{Thermal Conductivity Predictions with Foundation Atomistic Models}
\author{Balázs Póta}
\affiliation{Theory of Condensed Matter Group of the Cavendish Laboratory, University of Cambridge, Cambridge, UK}
\author{Paramvir Ahlawat}
\affiliation{Theory of Condensed Matter Group of the Cavendish Laboratory, University of Cambridge, Cambridge, UK}
\author{Gábor Csányi}
\affiliation{Engineering Laboratory, University of Cambridge, Cambridge, UK} 
\author{Michele Simoncelli}
\email{michele.simoncelli@columbia.edu}
\affiliation{Theory of Condensed Matter Group of the Cavendish Laboratory, University of Cambridge, Cambridge, UK}
\affiliation{Department of Applied Physics and Applied Mathematics, Columbia University, New York (USA)} 

\begin{abstract}
Advances in machine learning have led to the development of foundation models for atomistic materials chemistry, enabling quantum-accurate descriptions of interatomic forces across chemically diverse compounds at reduced computational cost. 
Hitherto, the accuracy and utility of these models have been assessed relying on descriptors based on formation energies or idealized harmonic atomic vibrations. Yet, the rigorous and physically interpretable quantification of their capability to describe both realistic anharmonic atomic dynamics and technologically relevant observables remains a pressing problem.
Here, we address this problem, leveraging the Wigner formulation of heat transport and the Grüneisen approach to thermal expansion to connect the atomic-physics awareness of foundation models to their utility in predicting experimentally observable thermomechanical properties, presenting standards and fine-tuning protocols needed to achieve first-principles accuracy.
We apply our framework to a database of 103 solids with diverse compositions and structures, 
demonstrating that it overcomes the major bottlenecks of current methods for designing heat-management materials --- high cost, limited transferability, or lack of physics awareness --- and its potential to discover materials for next-gen technologies ranging from thermal insulation to neuromorphic computing.
\end{abstract}

\maketitle
Over the past decades, intense research efforts have been dedicated to tackle the challenging task of fitting the Born-Oppenheimer potential energy surface (PES) as a function of atomic coordinates\cite{blank_neural_1995,behler_generalized_2007,bartok_gaussian_2010,rupp_fast_2012,zhang_deep_2018,drautz_atomic_2019,seko_group-theoretical_2019,deringer_gaussian_2021,batzner_e3-equivariant_2022,kocer_neural_2022}. These resulted in the development of machine-learning potentials (MLPs), which allow us to reproduce atoms' interaction energies and forces with nearly quantum accuracy and computational cost orders-of-magnitude lower than first-principles methods.
These advances have enabled scientists to make progress in understanding how atomistic structure and composition influence macroscopic observables --- an example is the thermal conductivity, a property determined by atomic structure and forces, and relevant for technologies ranging from energy harvesting\cite{qian_phonon-engineered_2021} to efficient neuromorphic computing\cite{nataf_using_2024}.
Major drawbacks of methods based on MLPs are the significant work required to generate a first-principles database needed to train and validate MLPs, as well as the applicability limited to specific materials' compositions or structural phases.
Past works attempted to bypass these limitations by employing end-to-end machine-learning methods, which predict microscopic \cite{okabe_virtual_2024,Guo2023-pm} or macroscopic \cite{ojih2024graph,rodriguez2023million,srivastava2023end} materials' properties very efficiently, but without necessarily accounting for the fundamental physics that determines them.
Fittingly, recent work \cite{rose_three_2023} has formally demonstrated the possibility to obtain a physics-aware and mathematically complete description of atomic environments (and thus of forces) by employing Message Passing Networks \cite{gilmer_neural_2017} with many-body messages \cite{batatia_mace_2023} (with the MACE architecture \cite{batatia_mace_2023} this can be achieved, e.g., using 4-body terms and one message pass).
This breakthrough has enabled the development of foundation machine-learning potentials (fMLPs) \cite{chen_universal_2022,deng_chgnet_2023,batatia_foundation_2024,park_scalable_2024,yang2024mattersimdeeplearningatomistic,merchant_scaling_2023, ALIGNN, MEGNET, Voronoi_RF, BOWSR, CGCNN, CGCNN+P, Wrenformer}, which are trained across nearly all chemical elements and can be combined to describe almost arbitrary compounds.  
Recent works have assessed the accuracy of fMLPs in predicting, e.g., atoms’ interaction energies (thermodynamic stability\cite{riebesell_matbench_2024}) or harmonic vibrations\cite{yu_systematic_2024,deng_overcoming_nodate,lee_accelerating_2024}.
However, the rigorous and physically interpretable quantification of the fMLP's capability to describe atomic-level features of the PES (i.e., "their physics awareness") and their overall utility to determine macroscopic observables are pressing questions that are largely unexplored, due to the complexity of the relation between many-body interatomic forces and observables\cite{simoncelli_unified_2019,simoncelli_wigner_2022}.

Here we introduce a framework to rigorously address such a question, quantifying how the accuracy of fMLPs in describing interatomic forces and PES derivatives connects to their accuracy in predicting experimentally observable thermomechanical properties. We elucidate how fundamental insights into the physics awareness, accuracy, and overall utility of fMLPs can be derived from microscopic analysis of the linear-response solution of the Wigner Transport Equation (WTE) for heat conduction in crystals and glasses\cite{simoncelli_unified_2019,simoncelli_wigner_2022}, and the quasi-harmonic Grüneisen approach to describe thermal expansion \cite{ritz_thermal_2019,barron_thermal_1980,born_dynamical_1998,wallace_thermodynamics_1972}.
We leverage these insights to define physically interpretable, robust, and stringent benchmark metrics that expose subtle artifacts or inaccuracies in the smoothness (second and third derivatives) of the PES encoded by fMLPs.
As these derivatives expose even subtle discontinuities in the PES, compared to established energy-based metrics, our metrics 
provide information that is physically richer, more robust, and stricter for measuring both the utility of fMLPs and the physical accuracy (smoothness) of the encoded PES \cite{matbench_discovery_url}.
We showcase the capabilities of this combined fMLP-WTE-Grüneisen framework in state-of-the-art (SOTA) non-proprietary fMLPs trained on the Materials Project database \cite{materialsproject} (hereafter collectively referred to as `mp-fMLPs') --- M3GNet \cite{chen_universal_2022}, CHGNet \cite{deng_chgnet_2023}, MACE-MP-0 \cite{batatia_foundation_2024}, SevenNET \cite{park_scalable_2024}, and ORB-v1-MPtraj\cite{orb_github} --- assessing their performance over a database of first-principles reference data for 103 chemically and structurally diverse compounds~\cite{phono3py,seko_prediction_2015}.
We discuss how these metrics can be used to assess and improve the expressivity, training, and fine-tuning of fMLPs, enabling first-principles-level predictions of anharmonic vibrational properties and heat conductivity in solids with arbitrary compositions and structures. We conclude by showing how these developments enable the theory-driven design of heat-management materials with microscopic heat-transport physics that violates the semiclassical Boltzmann regime, which are critical for energy and information-management technologies.\\

\noindent
\textbf{Atomic vibrations \& thermomechanical observables}\\
The recently developed WTE \cite{simoncelli_unified_2019, simoncelli_wigner_2022} offers a comprehensive approach to predict from atomistic vibrational properties the thermal conductivity of anharmonic crystals\cite{PhysRevX.10.011019}, disordered glasses \cite{simoncelli_thermal_2023}, as well as the intermediate case of anharmonic-and-disordered `complex crystals'\cite{pazhedath_first-principles_2023}.
To assess how the accuracy in the conductivity prediction  is affected by the precision with which fMLPs describe atomic vibrations, it is sufficient to consider the directionally averaged trace of the conductivity tensor obtained from the  WTE solution in the relaxation-time approximation\cite{simoncelli_wigner_2022}
\begin{equation}
\begin{split}
\kappa(T)&=\kappa_P(T){+}\kappa_C(T)=
\frac{1}{\mathcal{V} N_c }\sum_{\bm{q},s}\bigg[\\
&\hspace*{5mm}C(\bm{q})_s \frac{|\!| \tens{v}(\bm{q})_{s,s}|\!|^2}{3} [\Gamma(\bm{q})_s]^{-1}\\
&{+}\sum_{s'\neq s}\!\frac{C(\bm{q})_s}{C(\bm{q})_s{+}C(\bm{q})_{s'\!}} \hspace*{-1.2pt}\frac{\omega(\bm{q})_s{+}\omega(\bm{q})_{s'\!}}{2}\!\!\left(\frac{C(\bm{q})_s}{\omega(\bm{q})_s}{+}\frac{C(\bm{q})_{s'\!}}{\omega(\bm{q})_{s'\!}}\right)\\
&\hspace*{2mm}{\times} \frac{|\!| \tens{v}(\bm{q})_{s,s'}|\!|^2}{3} \frac{\frac{1}{2} \left[\Gamma(\bm{q})_s{+}\Gamma(\bm{q})_{s'\!}\right]}{[\omega(\bm{q})_{s'\!}{-}\omega(\bm{q})_{s}]^2{+}\frac{1}{4}[\Gamma(\bm{q})_s{+}\Gamma(\bm{q})_{s'\!}]^2}\bigg].
\label{eq:thermal_conductivity_combined}
\end{split}
\raisetag{30mm}
\end{equation}
Here, $C(\bm{q})_s{=}\frac{\hbar^2\omega^2(\bm{q})_s }{k_{\rm B} T^2} \bar{\tenscomp{N}}(\bm{q})_s\big[\bar{\tenscomp{N}}(\bm{q})_s{+}1\big]$ is the specific heat at temperature $T$ of the vibration having: wavevector $\bm{q}$; mode $s$; energy $\omega(\bm{q})_s$; population given by Bose-Einstein distribution $\bar{\tenscomp{N}}(\bm{q})_s{=}[\exp(\hbar \omega(\bm{q})_s/k_{\rm B}T){-}1]^{-1}$; total linewidth $\Gamma(\bm{q})_s{=}\Gamma_a(\bm{q})_s{+}\Gamma_i(\bm{q})_s$, where $\Gamma_a(\bm{q})_s$ depends on $T$  and is determined by anharmonic third-order derivatives of the interatomic potential, while $\Gamma_i(\bm{q})_s$ is $T$-independent and regulated by concentration of isotopes (see Methods for details). $\tenscomp{v}^\beta(\bm{q})_{s,s'}$ is the velocity operator obtained from wavevector differentiation of the solid's dynamical matrix\cite{simoncelli_wigner_2022}, its diagonal elements $s{=}s'$ are the phonon group velocities and its off-diagonal elements describe couplings between modes $s$ and $s'$ at the same $\bm{q}$; $N_{\rm c}$ is the number of $\bm{q}$-points sampling the Brillouin zone and $\mathcal{V}$ is the crystal's unit-cell volume. 

Eq.~(\ref{eq:thermal_conductivity_combined}) shows in the first line that the total macroscopic conductivity is determined by two contributions: the particle-like conductivity ($\kappa_P$), and the wave-like `coherence' conductivity ($\kappa_C$).
In such an equation, the term inside the square brackets is the single-phonon contribution to the total conductivity, $\mathcal{K}(\bm{q})_{s}$, which, as we will show later, is useful for benchmarking fMLPs.
The second line shows that $\kappa_P$ originates from phonons that carry heat $C(\bm{q})_{s}$ by propagating particle-like with velocity ${\tens{v}(\bm{q})_{s,s}}$ over lifetime $[\Gamma(\bm{q})_s]^{-1}$; it can be shown\cite{simoncelli_unified_2019} that $\kappa_P$ is equivalent to the conductivity described by the phonon Boltzmann Transport Equation (BTE).
The term on the third and fourth lines, instead, shows that $\kappa_C$ describes conduction through wave-like tunneling between pairs of phonons $s,s'$ at  wavevector $\bm{q}$. 
It has been shown\cite{simoncelli_unified_2019, simoncelli_wigner_2022} that in simple crystals particle-like propagation dominates over wave-like tunneling ($\kappa_{{\rm P}}{\gg}\kappa_{{\rm C}}$), in complex crystals both these mechanisms co-exist ($\kappa_{_{\rm P}}{\sim}\kappa_{_{\rm C}}$), and in strongly disordered glasses tunneling dominates ($\kappa_{_{\rm P}}{\ll}\kappa_{_{\rm C}}$) \cite{simoncelli_thermal_2023}.

\begin{figure*}[htp]
  \includegraphics[width = \textwidth]{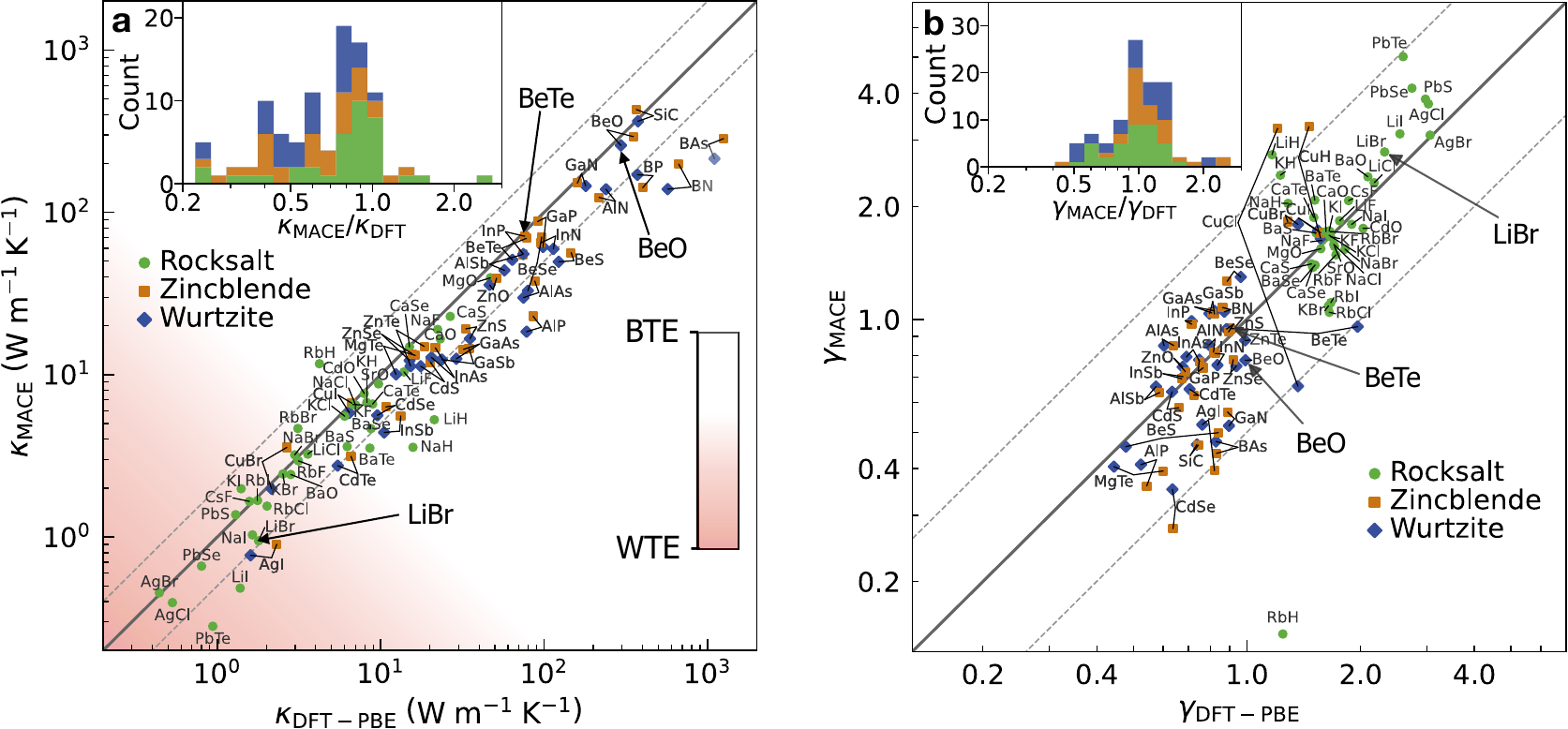}
  \caption{\textbf{Conductivity (left) and Grüneisen parameter (right) at 300 K from DFT-PBE and MACE-MP-0 large,} for 103 compounds taken from the phononDB-PBE database, which have rocksalt (green), zincblende (orange), or wurtzite (blue) structure. The solid line indicates perfect agreement and the dashed lines discrepancies of a factor of 2. 
  The arrows highlight materials discussed in detail later. Inset, distribution of relative deviations between the conductivities predicted by DFT-PBE and MACE-MP-0.
  The color gradient in the left panel shows the transition from the WTE regime where heat transport occurs through both propagation and tunneling mechanisms (red) to the semiclassical propagation-dominated BTE regime (white). }
  \label{fig:mace_pbe}
\end{figure*}

In addition to the conductivity, the thermal-expansion coefficient is also a property of technological interest, e.g., for aerospace applications\cite{padture_advanced_2016}. 
In general, there is no simple correlation between thermal conductivity and thermal expansion coefficients --- materials with similar thermal conductivity can exhibit different coefficients of thermal expansion, and vice-versa\cite{lehmann_thermal_2003,vassen_zirconates_2004}. Therefore, it is important to investigate to which extent accuracy in predicting thermal conductivity correlates with accuracy in predicting thermal expansion. 
Thermal expansion, the change in volume $\mathcal{V}$ driven by a change in temperature $T$ at constant entropy $S$, is quantitatively described by the Grüneisen parameter, $\gamma(T) = -\frac{\partial \ln T }{\partial \ln \mathcal{V} }\big|_S=\frac{\beta B_T \mathcal{V}}{C}$ --- here $\beta$ is the coefficient of volumetric thermal expansion, $B_T$ is the isothermal bulk modulus, and $C=\frac{1}{\mathcal{V}N_C}\sum_{\bm{q},s} C(\bm{q})_S$ is the specific heat at constant volume\cite{ritz_thermal_2019,barron_thermal_1980,born_dynamical_1998,wallace_thermodynamics_1972}.
Within the QHA one can resolve the Grüneisen parameter $\gamma$ in terms of microscopic phonon contributions ($\gamma(\bm{q})_s$),
\begin{equation}
\begin{split}
  \gamma {=}\frac{1}{\mathcal{\mathcal{V}}\hspace{-0.3mm}N_C}\hspace{-1mm}\sum_{\bm{q}s}\hspace{-0.9mm}\bigg[\hspace{-0.2mm}\gamma(\hspace{-0.1mm}\bm{q}\hspace{-0.1mm})\hspace{-0.2mm}_s \hspace{-0.3mm}\frac{C(\hspace{-0.1mm}\bm{q}\hspace{-0.1mm})\hspace{-0.2mm}_s}{C}\hspace{-0.2mm}\bigg],\;\;{\rm with}\;\;
  \gamma(\hspace{-0.1mm}\bm{q}\hspace{-0.1mm})\hspace{-0.2mm}_s {=} {-}\frac{\mathcal{V}}{\omega(\hspace{-0.1mm}\bm{q}\hspace{-0.1mm})\hspace{-0.2mm}_s} \frac{\partial \omega(\hspace{-0.1mm}\bm{q}\hspace{-0.1mm})\hspace{-0.2mm}_s}{\partial \mathcal{V}}  .
\label{eq:mode_gruneisen}
\raisetag{2mm}
\end{split}
\end{equation}
This microscopic decomposition shows that, within the QHA, the microscopic phonon-resolved Grüneisen contributions $\gamma(\bm{q})_s$ are constant over temperature, and the temperature dependence of the macroscopic Grüneisen parameter $\gamma$ entirely derives from the phonon specific heat term $\frac{C(\bm{q})_s}{C}$. In the following, we will refer to the temperature-dependent microscopic contribution to the macroscopic Grüneisen parameter as the quantity in the square brackets in Eq.~(\ref{eq:mode_gruneisen}), i.e., $\mathcal{G}(\bm{q})_s=\gamma(\bm{q})_s\frac{C(\bm{q})_s}{C}$. 

In the Methods, we provide a quantitative explanation about the  different information about the Potential-Energy Surface (PES) provided by constant-volume properties such as the conductivity, and volume-deforming properties such as the Grüneisen parameter. Specifically, within perturbation theory, the mode Grüneisen parameter is linearly related to the third-order derivatives of the (PES)\cite{wallace_thermodynamics_1972}, while the thermal conductivity is related to the inverse of the square of the third-order PES derivatives\cite{simoncelli_wigner_2022}. 
Therefore, macroscopic conductivity and Grüneisen parameter have a very different dependence on microscopic interatomic forces; this implies that conductivity and Grüneisen parameter practically provide complementary information on how accuracy in predicting PES translates into accuracy in describing macroscopic thermomechanical properties. \\

\begin{figure*}[htp]
\vspace*{-3mm}
    \includegraphics[width = \textwidth]{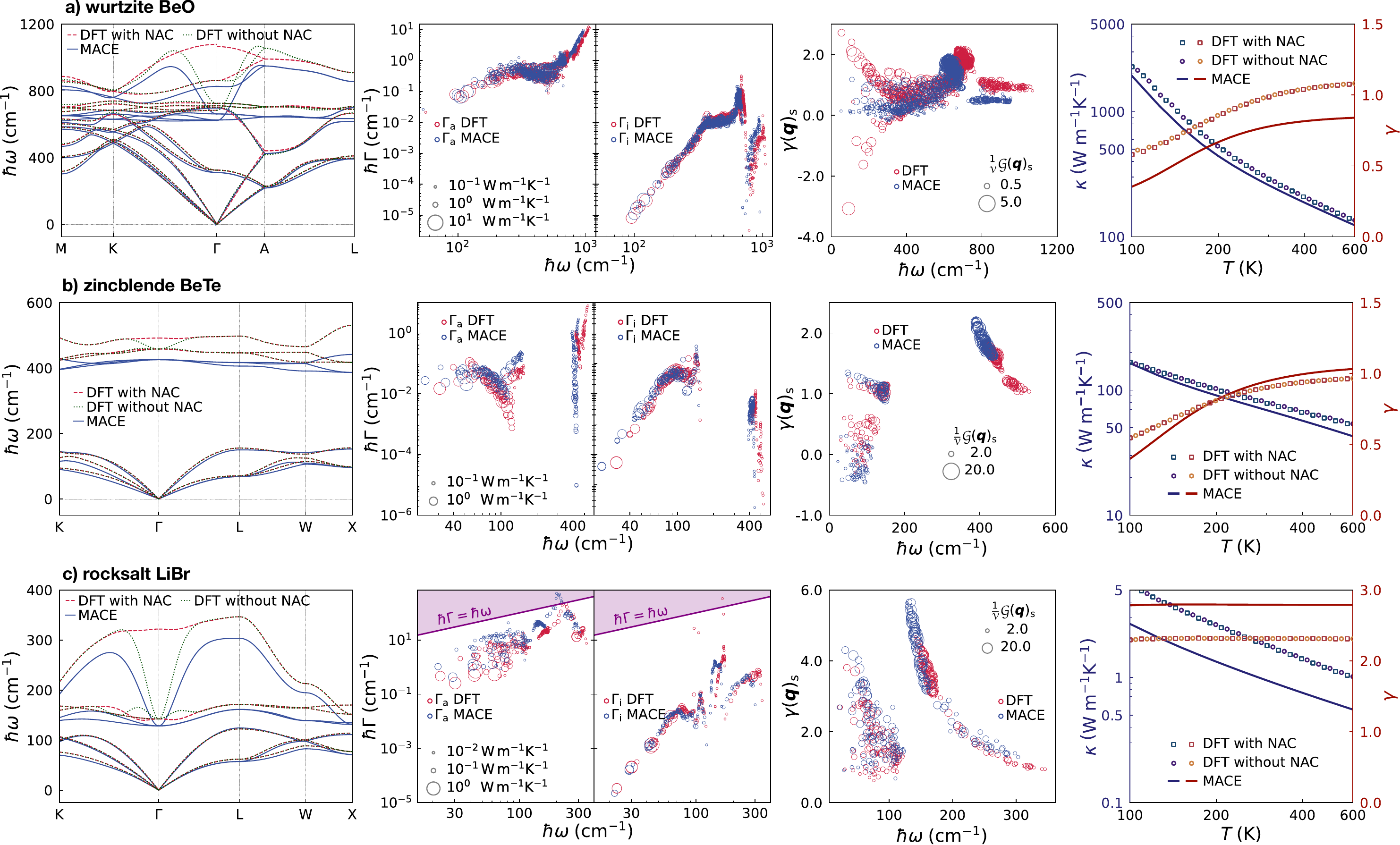}\\[-2mm]
    \caption{\textbf{Phonon dispersion, linewidth \& mode-Grüneisen distributions, along with macroscopic conductivity \& Grüneisen parameter} for wurtzite BeO (panel {\bfseries{a}}), zincblende BeTe ({\bfseries{b}}), and rocksalt LiBr ({\bfseries{c}}). DFT-PBE values are computed using the data from Refs.~\cite{phono3py,seko_prediction_2015}, while `MACE' values are computed using the MACE-MP-0 fMLP\cite{batatia_foundation_2024}. For the DFT-PBE phonons, we show the effect of considering (red) or not (green) the long-range non-analytical contribution (NAC); MACE-MP-0 does not fully account for NAC (see text). In the frequency-linewidths distributions at 300 K, we resolve the anharmonic ($\Gamma_a$) and isotopic ($\Gamma_i$) parts of the linewidths; the markers' areas are proportional to the contribution of the given phonon mode to the conductivity. Similarly, the mode Grüneisen parameter distribution against phonon energy is evaluated at 300K with marker size describing the contribution of the modes to the macroscopic Grüneisen parameter. }\label{fig:detailed_BeO_BeTe_BAs_analysis}
\end{figure*}

\noindent
\textbf{Vibrational and thermal properties from fMLPs}\\
\noindent
Since $\kappa(T)$ and $\gamma(T)$ are observables that can be resolved in terms of microscopic harmonic and anharmonic vibrational properties \cite{plata_efficient_2017,knoop_anharmonicity_2023,bandi_benchmarking_2024,simoncelli_wigner_2022}, it is natural to use them to assess the precision of fMLPs in reproducing reference microscopic and macroscopic vibrational properties obtained from Density-Functional Theory (DFT) or from experiments.
In particular, we rely on the DFT interatomic force data of Refs.~\cite{phono3py,seko_prediction_2015}, which
include 103 diverse binary compounds, involving 34 different chemical species, and rocksalt, zincblende, or wurtzite structures. 
In the following we will refer to this dataset as 
"phononDB-PBE" to highlight that it was computed using the Perdew, Burke, and Ernzerhof (PBE) functional\cite{perdew_generalized_1996} used also to generate the training data for the fMLPs\cite{materialsproject}.
Following the expressions detailed in the Methods, we use these data to calculate the harmonic ($\hbar\omega(\bm{q})_s$, $\tenscomp{v}^\beta(\bm{q})_{s,s'}$, and $\Gamma_i(\bm{q})_s$) and third-order anharmonic ($\Gamma_a(\bm{q})_s$) vibrational properties that determine $\kappa(T)$ (Eq.~\ref{eq:thermal_conductivity_combined}).
To evaluate the Grüneisen parameter~(\ref{eq:mode_gruneisen}), we extend this dataset to include second-order force constants calculations at expanded and contracted volumes, as detailed in the Methods.

\begin{figure*}[ht]
    \centering
    \includegraphics[width=\textwidth]{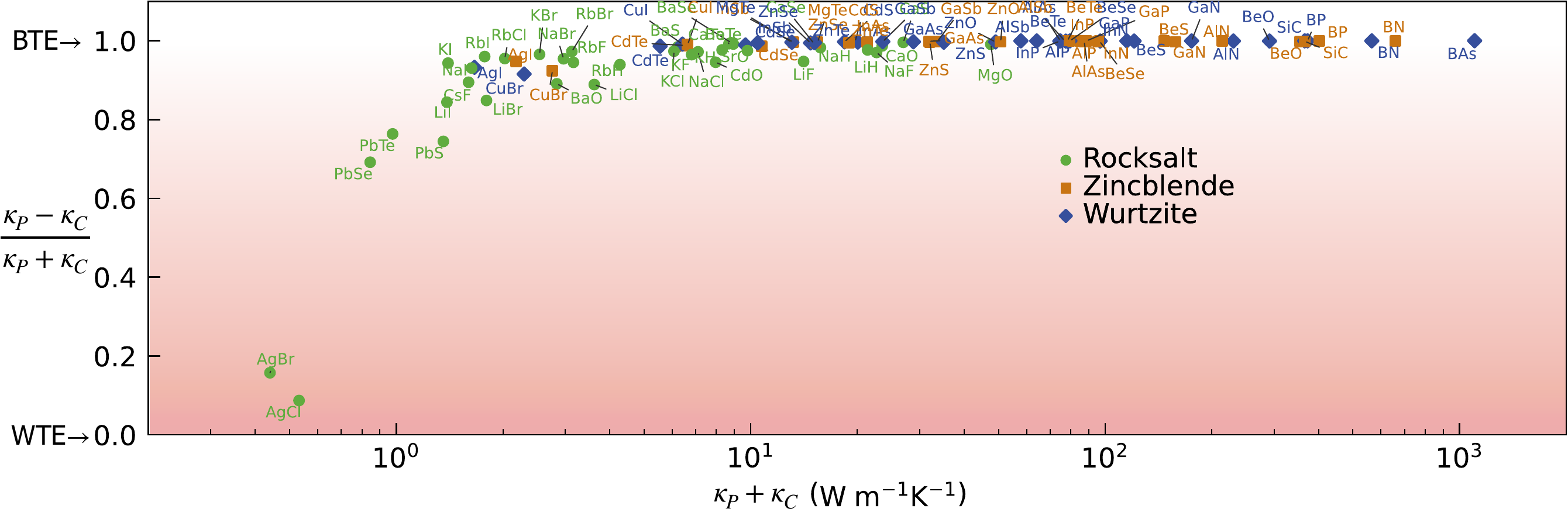}\\[-2mm]
    \caption{\textbf{Failures of semiclassical Boltzmann transport in binary crystals with low conductivity} and having rocksalt (green), zincblende (orange), and wurtzite (blue) structure. These occur when the conductivity described by the BTE in terms of particle-like propagation of phonons ($\kappa_P$) does not dominate over the coherences conductivity due to phonons' wave-like tunneling ($\kappa_C$). The WTE~\cite{simoncelli_wigner_2022} accounts for both  $\kappa_P$ and $\kappa_C$, and failures from semiclassical Boltzmann transport are highlighted by a relative strength of particle-wave transport $(\kappa_P-\kappa_C)/(\kappa_P+\kappa_C)$ appreciably smaller than one (more intense red background, as in Fig.~\ref{fig:mace_pbe}). These conductivities were obtained  from the force constants in the phononDB-PBE dataset\cite{phono3py,seko_prediction_2015}.}
    \label{fig:coherences_vs_populations}
\end{figure*}

{In Fig.~\ref{fig:mace_pbe}{\bfseries{a}} and Fig.~\ref{fig:mace_pbe}{\bfseries{b}} we compare the conductivity and Grüneisen parameter}, respectively, predicted from first principles (DFT-PBE) or from the mp-fMLP MACE-MP-0 large  \cite{batatia_foundation_2024} for the 103 materials in the phononDB-PBE dataset.
We highlight that predictions from DFT or fMLPs are generally compatible within a factor of 2.  Importantly, in about 20\% of the compounds studied, the semiclassical particle-like BTE fails to fully describe heat transport and it is crucial to employ the more general WTE to perform accurate predictions. These cases have very low conductivity ($\kappa_{\rm TOT}\lesssim2$ W/mK), which receives significant contributions from both wave-like tunneling (described by the WTE but missing from the BTE) and particle-like propagation (described by both the WTE and BTE) heat-transport mechanisms, see Methods Fig.~\ref{fig:coherences_vs_populations}. We highlight how having accurate conductivity predictions does not necessarily imply having accuracy in predictions for the Grüneisen parameter, and vice-versa (e.g., in wurtzite BeO $\kappa$ is predicted more accurately than $\gamma$, while in LiBr $\gamma$ is more accurate than $\kappa$). We further observe that the distributions of $\kappa$ and $\gamma$ overlap for zincblende and wurtzite structures, while  rocksalt structures generally exhibit lower $\kappa$ and higher $\gamma$. Moreover, while fMLPs generally tend to systematically underestimate the conductivity, there is no such bias for the Grüneisen parameter.
Overall, our findings highlight that the macroscopic $\kappa$ and $\gamma$ have different sensitivity to atomistic vibrational properties, and in the following we show that they can provide complementary information on the accuracy with which fMLPs describe reference DFT data.\\

\noindent
\textbf{Relating macroscopic and microscopic properties}
\begin{figure*}[t]
\centering\includegraphics[width =\textwidth]{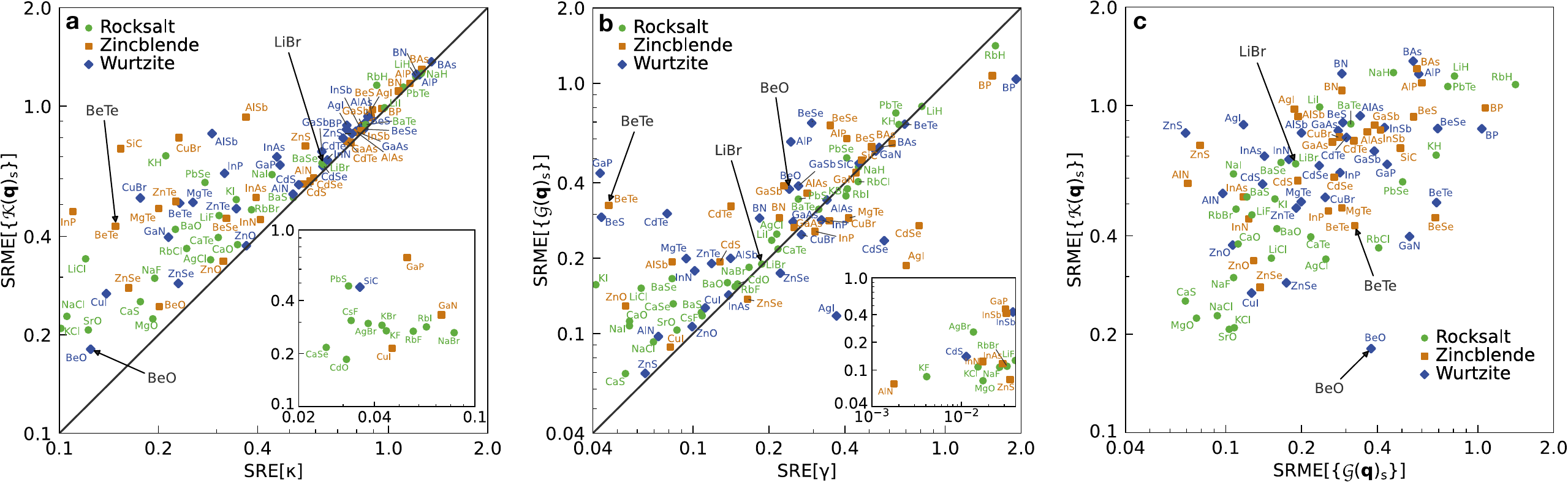}\\[-4mm]
    \caption{\textbf{SRE \& SRME errors on: conductivity (a), Grüneisen parameter (b), and their correlation (c)} in 103 chemically and structurally diverse compounds. 
    For conductivity (panel {\bfseries{a}}), large discrepancies between DFT-PBE and mp-fMLP (MACE-MP-0) on microscopic, single-phonon conductivity  contributions are described by large conductivity SRME~(\ref{eq:SRME_mode_kappa}). The microscopic SRME metric is more informative than the macroscopic SRE metric, since it captures the cases where microscopic (in)accuracy translates into macroscopic (in)accuracy --- e.g., in BeO (LiBr) a low (high) SRME implies a low (high) SRE --- and, because SRME is a sum of non-negative contributions, it detects also cases in which compensation of microscopic errors occurs (e.g., BeTe exhibits high SRME but low SRE).
    {For Grüneisen  parameter (panel {\bfseries{b}}), the behavior of macroscopic (SRE, Eq.~(\ref{eq:SRE_gamma})) and microscopic (SRME, Eq.~(\ref{eq:SRME_mode_gamma})) errors  is more complex due to the possibility of having both positive and negative microscopic contributions, which can lead to both cancellation of single-phonon errors and amplification of  macroscopic relative errors when large negative and positive contributions coexist, as discussed in the text. However, most of the materials studied here (\textbf{b}) display large SRME and positive microscopic contributions to the Grüneisen parameters, implying that cancellation of errors can be detected, as for the conductivity, by large SRME and low SRE.
}
    The insets in panels {\bfseries{a}} and {\bfseries{b}} show materials with very low SRE, while panel {\bfseries{c}} displays the comparison of SRME in thermal conductivity and Grüneisen parameter, demonstrating no clear correlation.
}    \label{fig:errors_ktot}
\end{figure*}
\noindent
To understand the microscopic origin of the discrepancies between the DFT-PBE and MACE-MP-0 conductivities and Grüneisen parameters, we select three representative materials --- wurtzite BeO, zincblende BeTe, and rocksalt LiBr --- and show in Fig.~\ref{fig:detailed_BeO_BeTe_BAs_analysis} their phonon band structures, the macroscopic conductivity and Grüneisen resolved in terms of contributions of microscopic phonon modes.
Examining the phonon band structures in the first column, we observe that MACE-MP-0 tends to underestimate the high-frequency optical modes compared to DFT-PBE. In particular, we plot the DFT-PBE phonons considering or not the long-range non-analytical correction term \cite{gonze_dynamical_1997} (NAC). The NAC term is responsible for the energy splitting between the longitudinal-optical and transverse-optical modes in polar dielectrics \cite{gonze_dynamical_1997}, and is not fully considered in fMLPs trained using a radial cutoff (e.g., for MACE-MP-0\cite{batatia_foundation_2024} such cutoff is 6 \AA, while it is 5 \AA~ for SevenNet\cite{park_scalable_2024}, CHGNet\cite{deng_chgnet_2023}, and M3GNet\cite{chen_universal_2022}).  This explains why the DFT-PBE phonons without NAC are in closer agreement with the MACE-MP-0 phonons. 
Importantly, even without NAC, DFT-PBE energies  for the optical phonon modes tend to be higher than MACE-MP-0 energies, confirming the general tendency of MACE-MP-0 to underestimate phonon energies \cite{deng_overcoming_nodate}.
We show in the Methods that such an underestimation of the optical phonon energies, or considering NAC or not, has negligible effect on the specific heat at constant volume --- this because in these materials the specific heat is dominated by acoustic phonons.
Fig.~\ref{fig:detailed_BeO_BeTe_BAs_analysis} also shows that considering or not the NAC term has a negligible impact on $\kappa(T)$ in BeO, BeTe, and LiBr. This can be understood from the third column of Fig.~\ref{fig:detailed_BeO_BeTe_BAs_analysis}, where we show that 
in these materials a significant amount of heat is carried by low-energy acoustic phonons that are negligibly affected by NAC, as well as by the systematic phonon softening. 
Specifically, we report the energy-linewidth distributions, resolving the anharmonic and isotopic parts of the linewidths ($\Gamma_a(\bm{q})_s$ and $\Gamma_i(\bm{q})_s$, respectively), and quantifying how much each phonon contributes to the conductivity with the single-mode conductivity contribution  ${\mathcal{K}}(\bm{q})_s$ defined in Eq.~(\ref{eq:thermal_conductivity_combined}).
We note, in passing, that while for the 103 materials considered here the NAC term has negligible influence on thermal transport, it could be important in materials where significant heat is carried by optical modes. SOTA fMLPs do not account for NAC, its inclusion will be subject of future work. However, we show in the Methods that, in crystals, this limitation can be bypassed using 
interatomic force constants calculated using supercells in real space (i.e., accessible by finite-difference supercell calculations done with DFT or fMLPs) and the dipole–dipole interactions calculated using density-functional perturbation theory.
As these supercells are smaller than the receptive field of the fMLPs, we use the DFT reference without the NAC contribution to calculate SRME and SRE.

\begin{figure}[b]
  \centering
  \includegraphics[width=\columnwidth]{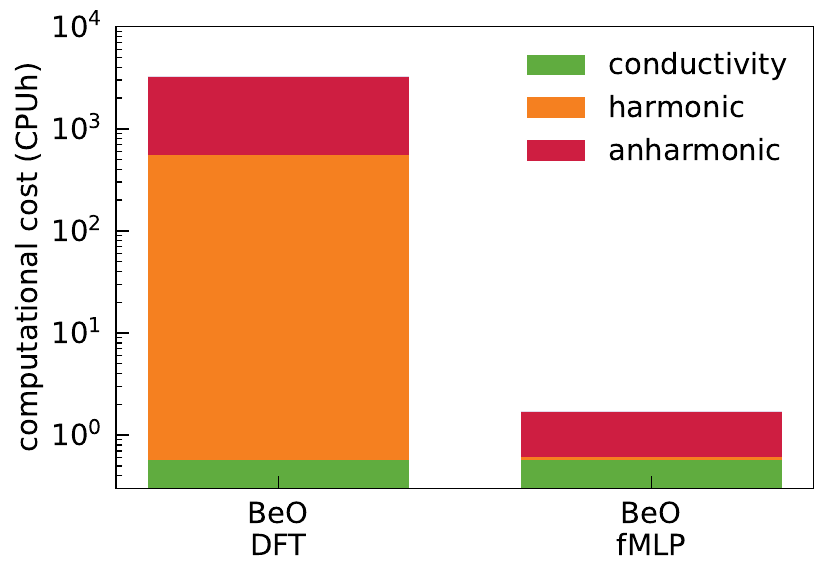}
  \caption{\textbf{Computational cost: DFT vs fMLP} MACE-MP-0-L2 for wurtzite BeO. Orange and red show the cost for computing harmonic and anharmonic force constants, respectively; green is the cost to obtain the conductivity from the force constants through the solution of the  Wigner transport equation. The force constants were computed on CPU for DFT, and on GPU for MACE (see text for details); these costs were compared using the conversion 1 GPUh {=} 55 CPUh \cite{noauthor_csd3_nodate}. }
  \label{fig:comp_cost}
\end{figure}

The most accurate zero-shot predictions for the energy-linewidth distributions (for BeO in Fig.~\ref{fig:detailed_BeO_BeTe_BAs_analysis}{\bfseries{a}}) correspond to a discrepancy of $\sim 7\%$ between $\kappa(T)$ predicted from fMLP (MACE-MP0) and DFT.
We stress that fMLPs allow to obtain this accuracy at a computational cost that is more than 3 orders of magnitude lower than DFT (see Methods Fig.~\ref{fig:comp_cost}).
Fig.~\ref{fig:detailed_BeO_BeTe_BAs_analysis}{\bfseries{b}} shows that
compatibility between $\kappa(T)$ and $\gamma(T)$ predicted from DFT or fMLP can also result from cancellation of microscopic errors. E.g., in BeTe there are visible differences in the energy-linewidth distributions, and these largely cancel out when integrated to determine the conductivity. 
Finally, Fig.~\ref{fig:detailed_BeO_BeTe_BAs_analysis}{\bfseries{c}} illustrates that discrepancies in the energy-linewidth distributions do not always compensate, and can translate into significant differences (a factor of 2) on $\kappa(T)$. We also note that, depending on the chemical composition, the anharmonic linewidths at room temperature can dominate over the isotopic linewidth (e.g., in BeO) or not (e.g., in BeTe). 
Similar cancellation of microscopic errors can emerge also for the Grüneisen parameter, as shown in Fig.~\ref{fig:detailed_BeO_BeTe_BAs_analysis}{\bfseries{b}}.

\begin{figure}[b!]
    \centering\includegraphics[width=\columnwidth]{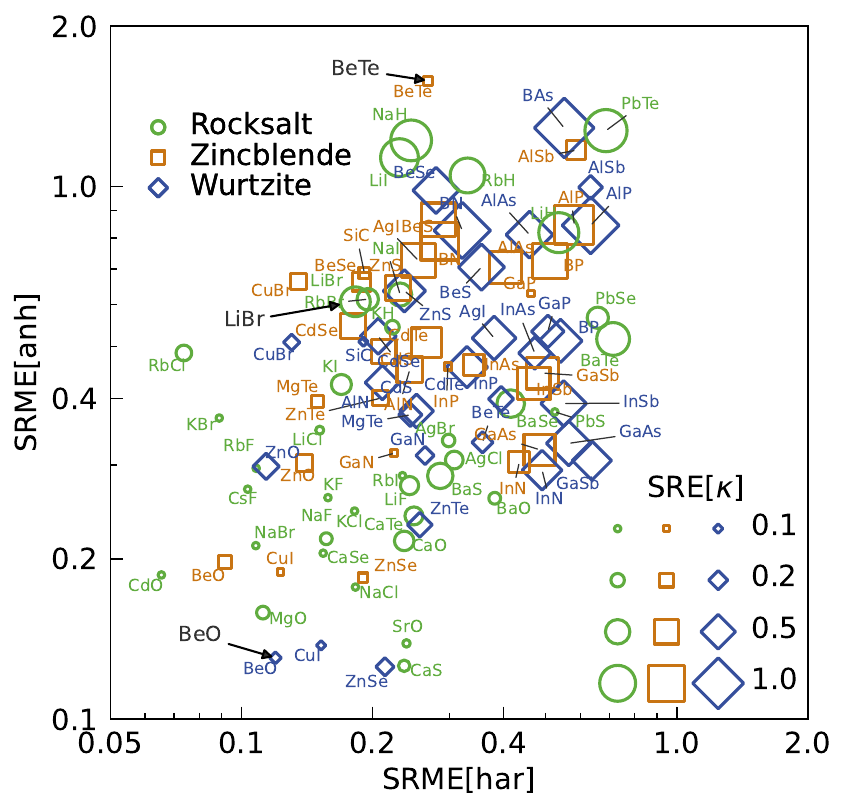}\\[-4mm]
    \caption{\textbf{Relation between harmonic and anharmonic errors, and thermal conductivity,} calculated for 103 compounds using Eq.~(\ref{eq:smape_harmonic_part}) and Eq.~(\ref{eq:smape_anharmonic_part}) considering DFT and MACE-MP-0\cite{batatia_foundation_2024}. The area of the marker is proportional to SRE[$\kappa$], Eq.~(\ref{eq:SAPE_tot_kappa}). We highlight how SRME[har] (Eq.~(\ref{eq:smape_harmonic_part})), SRME[anh] (Eq.~(\ref{eq:smape_anharmonic_part})), and SRE[$\kappa$] are all low (large) in BeO (LiBr); in contrast, BeTe displays large SRME[har], large SRME[anh], but low SRE[$\kappa$] due to compensation of errors.}
    \label{fig:errors}
\end{figure}

The results above motivate us to quantitatively investigate when agreement between DFT and fMLP conductivities and Grüneisen parameters is obtained as a result of accurately described microscopic harmonic and anharmonic vibrational properties, or because of compensating errors.
Starting from conductivity, we quantify the macroscopic discrepancies using the Symmetric Relative Error (SRE), 
\begin{equation}
    \text{SRE}\big[\kappa\big]=2\frac{\left|{\kappa}_{\rm{fMLP}}-{\kappa}_{\rm{DFT}}\right| }{{\kappa}_{\rm{fMLP}}+{\kappa}_{\rm{DFT}}}\,.
    \label{eq:SAPE_tot_kappa}
\end{equation}
Next, to quantify the error on the microscopic phonon properties, we introduce the 
Symmetric Relative Mean Error (SRME) on the single-phonon contribution to $\kappa$:
\begin{equation}
\begin{split}
      \text{SRME}\big[\{\mathcal{K}(\bm{q})_s\}\big]{=}\frac{2}{N_c \mathcal{V}}\frac{\sum_{\bm{q}s}\!\left|{\mathcal{K}}_{\rm{fMLP}}\!(\bm{q})_s{-}{\mathcal{K}}_{\rm{DFT}}\!(\bm{q})_s\right| }{{\kappa}_{\rm{fMLP}}+{\kappa}_{\rm{DFT}}}\hspace{0.1mm},\;\;
\end{split}
\raisetag{6mm}
    \label{eq:SRME_mode_kappa}
\end{equation}
where 
${\mathcal{K}}_{\rm{DFT}}\!(\bm{q})_s$ refers to the single-phonon conductivity (term inside square brackets in Eq.~(\ref{eq:thermal_conductivity_combined})) evaluated using DFT, and 
${\mathcal{K}}_{\rm{fMLP}}\!(\bm{q})_s$ refers to the same expression evaluated using fMLP.
Fig.~\ref{fig:errors_ktot}{\bfseries{a}} illustrates that a large SRME generally implies large SRE, as SRE gives a lower bound on the SRME.  Importantly, knowing both SRE and SRME enables us to identify when microscopic error compensation occurs --- this is indicated by a large SRME but small SRE. 
We note that the SRME[$\mathcal{K}(\bm{q})_s$] error can stem from discrepancies in the harmonic (second-order) or anharmonic (third-order) force constants. In the Methods, Fig.~\ref{fig:errors} we resolve how errors on harmonic and anharmonic vibrational properties contribute to SRME.\\

Analyzing macroscopic and microscopic errors on the Grüneisen parameter requires additional considerations.
In fact, while the positive-definite $\kappa$ can be resolved in terms of microscopic non-negative single-phonon contributions, the single-phonon contributions to the Grüneisen parameter can assume positive or negative sign, and yield a macroscopic  Grüneisen with positive or negative sign depending on their relative strength. To take this into account, we define the SRE for the Grüneisen parameter as
\begin{equation}
  \text{SRE}\big[\gamma\big]=2\frac{\left|{\gamma}_{\rm{fMLP}}-{\gamma}_{\rm{DFT}}\right| }{\left|{\gamma}_{\rm{fMLP}}\right|+\left|{\gamma}_{\rm{DFT}}\right|}\,,
  \label{eq:SRE_gamma}
\end{equation}
and the corresponding SRME as
\begin{equation}
  \begin{split}
        \text{SRME}\big[\{\mathcal{G}(\bm{q})_s\}\big]{=}2\frac{\sum_{\bm{q}s}\!\left|{\mathcal{G}}_{\rm{fMLP}}\!(\bm{q})_s{-}{\mathcal{G}}_{\rm{DFT}}\!(\bm{q})_s\right| }{\sum_{\bm{q}s}\!\left|{\mathcal{G}}_{\rm{fMLP}}\!(\bm{q})_s\right|{+}\left|{\mathcal{G}}_{\rm{DFT}}\!(\bm{q})_s\right|}\,,\;\;
  \end{split}
  \raisetag{6mm}
      \label{eq:SRME_mode_gamma}
  \end{equation}
  where 
${\mathcal{G}}_{\rm{DFT}}\!(\bm{q})_s$ refers to the single-phonon contribution to the Grüneisen parameter (term inside square brackets in Eq.~(\ref{eq:mode_gruneisen})) evaluated using DFT, and 
${\mathcal{G}}_{\rm{fMLP}}\!(\bm{q})_s$ refers to the same expression evaluated using fMLP.
These definitions yield a non-negative Grüneisen SRME~(\ref{eq:SRME_mode_gamma}), with maximal value 2 and minimal value 0, this in analogy to the conductivity SRME~(\ref{eq:SRME_mode_kappa}).

We have seen that because the microscopic conductivity contributions are non-negative, the SRE provides a lower bound on the SRME.
In contrast, microscopic Grüneisen parameter contributions can range from positive to negative, implying four distinct possible scenarios for SRME and SRE:
{(i) when both SRE and SRME are low, the fMLP’s prediction is accurate (e.g., zincblende AlN in Fig.\ref{fig:errors_ktot}{\bfseries{b}});
(ii) when both are high, large microscopic errors add into a large error for the macroscopic Grüneisen parameter, and the prediction is clearly inaccurate (e.g., wurtzite BP in Fig.\ref{fig:errors_ktot}{\bfseries{b}});
(iii) when SRME is high but SRE is low, predictions for the macroscopic Grüneisen parameter appear accurate, but are not reliable because they originate from a fortuitous cancellation of microscopic errors in the underlying phonon physics (e.g., wurtzite GaP in Fig.\ref{fig:errors_ktot}{\bfseries{b}}); and
(iv) when SRE is high and SRME is low, small deviations can induce significant relative errors in cases where the macroscopic Grüneisen parameter vanishes due to large-magnitude and compensating positive and negative microscopic single-phonon contributions (e.g., zincblende AgI in Fig.\ref{fig:errors_ktot}{\bfseries{b}}).}
While cases (i), (ii), and (iii) are totally analogous to those discussed for the positive-definite conductivity, case (iv) emerges only for signed thermomechanical quantities such as the Grüneisen parameter.
For most of the 103 materials in the phononDB-PBE database, the macroscopic Grüneisen parameter at 300 K is positive and SRME exceeds SRE, indicating that case (iv) occurs rarely and that for these materials SRME is overall more informative than SRE in assessing the overall accuracy with which the Grüneisen parameter is described.

Importantly, we note that the SRME in the Grüneisen parameter is not clearly correlated with the SRME in the conductivity, see Fig.~\ref{fig:errors_ktot}{\bfseries{c}}. This demonstrates the importance of analyzing both conductivity and Grüneisen to assess the accuracy and physics-awareness of the fMLPs.\\

\noindent
\textbf{Accuracy of various SOTA foundation models}
\begin{figure}[t]
\includegraphics[width=\columnwidth]{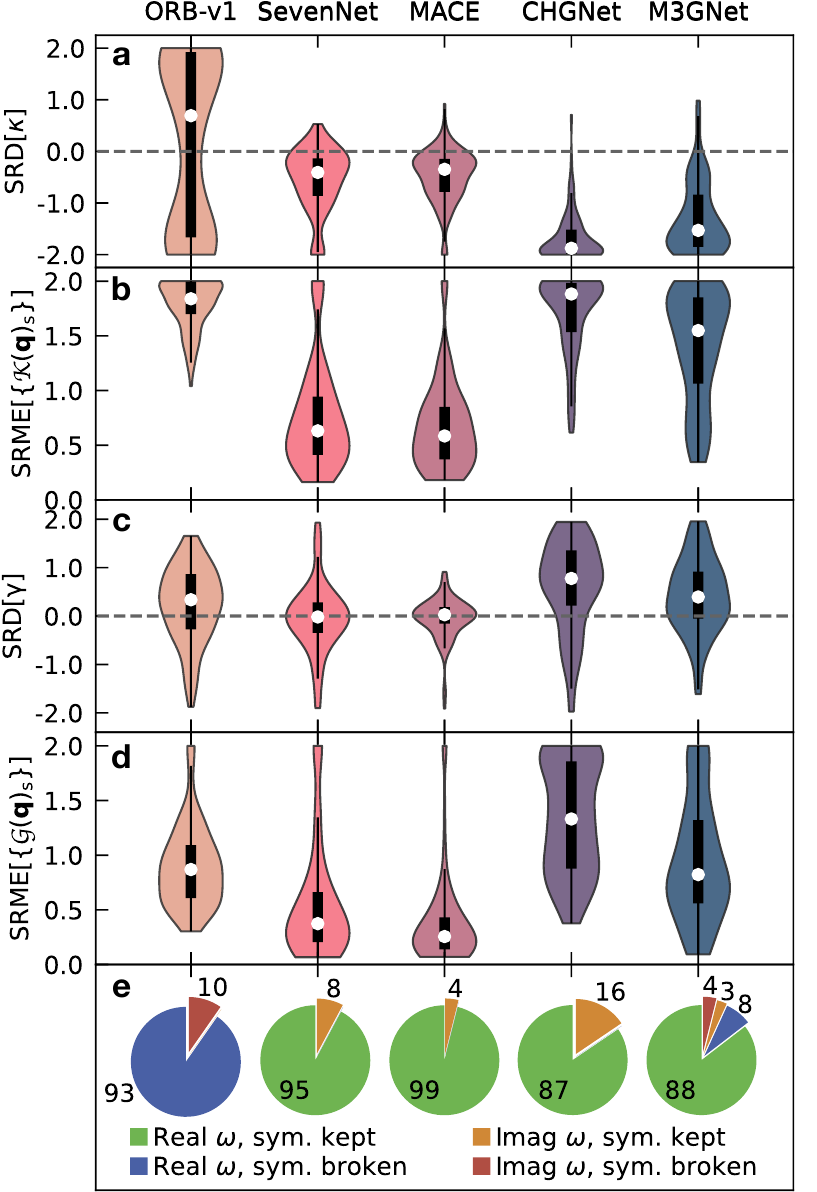}\\[-1mm]
    \caption{\textbf{Performance of mp-fMLPs in predicting thermal conductivity and Grüneisen parameter.} 
    \textbf{a} and \textbf{c} violin plots for the Symmetric Relative Difference (SRD) in the total Wigner conductivity~(\ref{eq:SPD_kappa}) and in the total Grüneisen parameter, respectively, for the materials in phononDB-PBE simulated with ORB-v1-MPtraj\cite{orb_github}, SevenNet\cite{park_scalable_2024}, MACE-MP-0\cite{batatia_foundation_2024}, CHGNet\cite{deng_chgnet_2023}, M3GNet\cite{chen_universal_2022}.
    \textbf{b} and \textbf{d} are $\text{SRME}\big[\{\mathcal{K}(\bm{q})_s\}\big]$ and $\text{SRME}\big[\{\mathcal{G}(\bm{q})_s\}\big]$, respectively, in the same materials database. In all panels, the medians are marked with white scatter points, the widths of the boxes represent the interquartile range, and the whiskers show the range of data points without outliers. The pie charts in \textbf{e}  display the number of compounds with non-negative frequencies that retained or not the correct crystal symmetry after unconstrained relaxation (green and blue, respectively); we also show whether structural instabilities (imaginary phonon frequencies) were observed in relaxations performed with (orange) or without (red) enforcing symmetries. Unstable structures with imaginary phonon frequencies were included considering $\kappa_{\mathrm{fMLP}} = 0$ and $\gamma_{\mathrm{fMLP}} = 0$, i.e., SRD${=}-2$, SRME${=}2$.
    }
 \label{fig:violin}
\end{figure}
\noindent
Here we utilize SRE and SRME for $\kappa$ and $\gamma$ to study the accuracy of non-proprietary mp-fMLPs on the structures contained in the phononDB-PBE database\cite{phono3py,seko_prediction_2015}, comparing DFT predictions against M3GNet\cite{chen_universal_2022}, CHGNet\cite{deng_chgnet_2023}, MACE-MP-0\cite{batatia_foundation_2024}, SevenNet\cite{park_scalable_2024}, and ORB-v1-MPtraj\cite{orb_github}  in  
Fig.~\ref{fig:violin}. 
To determine if a particular mp-fMLP tends to systematically overestimate or underestimate the conductivity and Grüneisen compared to DFT-PBE, we rely on the Symmetric Relative Difference (SRD). For the conductivity this is defined as: 
\begin{equation}
\rm{SRD}\big[\kappa\big]=2\frac{\kappa_{\rm fMLP}-\kappa_{\rm DFT}}{\kappa_{\rm fMLP}+\kappa_{\rm DFT}}\,,
    \label{eq:SPD_kappa}
\end{equation}
and similarly for the Grüneisen parameter,
\begin{equation}
  \rm{SRD}\big[\gamma\big]=2\frac{\gamma_{\rm fMLP}-\gamma_{\rm DFT}}{|\gamma_{\rm fMLP}|+|\gamma_{\rm DFT}|}\,.
\label{eq:SRD_grun}
\end{equation}
In both cases the SRD ranges from -2 to +2, corresponding to fMLP overestimating and underestimating the reference DFT values, respectively. 

Figures ~\ref{fig:violin}\textbf{a,c} summarize in violin plots  the SRD for the compounds in the phononDB-PBE database. 
Unstable structures with imaginary phonon frequencies were included considering $\kappa_{\rm fMLP}{=}0$ and $\gamma_{\rm fMLP}{=}0$, \textit{i.e.}, $\text{SRD}{=}{-}2$, $\text{SRME}{=}2$.
This analysis reveals that even the most accurate mp-fMLP tends to underestimate the thermal conductivity, as a consequence of  underestimating vibrational frequencies~\cite{deng_overcoming_nodate}, and in some cases overestimating anharmonic linewidths. In contrast, such a systematic error is not observed for the Grüneisen parameter, confirming that the analyses of $\kappa$ and $\gamma$ provide complementary information on the accuracy of fMLPs.

To examine how the macroscopic SRD in Figs.~\ref{fig:violin}\textbf{a},\textbf{c} are influenced by microscopic errors, we show in Figs.~\ref{fig:violin}\textbf{b},\textbf{d} violin plots for SRME in the mode contribution to conductivity and Grüneisen parameter. For the conductivity, this directly relates to microscopic error compensation, whereas the relation between SRD (or SRE) and SRME in the Grüneisen parameter is more complex due to the possibility of negative values and amplification of the relative error.

Among the mp-fMLP tested, the one that shows the highest accuracy is MACE-MP-0\cite{batatia_foundation_2024}, which produces zero-shot conductivities compatible within a factor of two of the DFT-PBE values for 69\% of the materials in the phononDB-PBE database\cite{phono3py,seko_prediction_2015}. The zero-shot Grüneisen parameter is compatible within a factor of two for 67\% of the materials.

\begin{table}[t]
    \centering
    \begin{tabular}{m{2.1cm}|c|c|c|c|c}
       & ORB-v1 & 7Net & MACE & CHGN. & M3GN. \\
      \hline 
       \centering mean  $\quad\rm{SRE}[\kappa]$ & $1.555$ &  $0.605$ & $0.528$ & $1.695$ & $1.340$\\
       \hline
        \centering mean ${\rm SRME}[\!\{{\!\mathcal{K}}\!(\bm{q})_{\!s}\!\}\!]$ & $1.786$ & $0.761$ & $0.669$ & $1.717$ & $1.409$ \\ 
        \specialrule{.1em}{.05em}{.05em} 
        \centering mean  $\quad\rm{SRE}[\gamma]$ & $0.667$ &  $0.499$ & $0.282$ & $0.976$ & $0.694$\\
        \hline
         \centering mean ${\rm SRME}[\!\{{\!\mathcal{G}}\!(\bm{q})_{\!s}\!\}\!]$ & {$0.899$} & $0.557$ & $0.374$ & $1.345$ & $0.941$ \\ 
         \specialrule{.1em}{.05em}{.05em}
        \centering F1 (11/9/2024)\cite{riebesell_matbench_2024} & 0.763 & 0.724 & 0.669 & 0.613 & 0.569
    \end{tabular}
    \caption{\textbf{Benchmark metrics for fMLPs.}
    The mean of $\text{SRE}[\kappa]$, $\text{SRME}[\{\mathcal{K}(\bm{q})_s\}]$, $\text{SRE}[\gamma]$ and $\text{SRME}[\{\mathcal{G}(\bm{q})_s\}]$
    introduced here resolve the accuracy of fMLPs in describing interatomic forces and derived structural and thermal properties. The
    Matbench F1\cite{riebesell_matbench_2024} descriptor is based on atomic energies and resolves thermodynamic stability.
    Different (similar) values for SRE and SRME indicate presence (absence) of cancellation of errors on microscopic vibrational properties.
  The lower the SRE is, the higher the accuracy is in predicting macroscopic $\kappa(300 \rm{K})$ and $\gamma(300 \rm{K})$. The lower the SRME is, the higher the accuracy is in predicting both microscopic vibrational properties, and macroscopic $\kappa(300 \rm{K})$ and $\gamma(300 \rm{K})$. 
  The higher the F1 score is, the higher the accuracy is in describing thermodynamic stability.
    }
\label{tab:umlp_performance}
\end{table}

To quantify the overall accuracy of a fMLPs in predicting the macroscopic conductivity and Grüneisen parameter over a materials' database, without resolving the possible compensation of microscopic errors, it is informative to consider the mean of the modulus of the deviations, \textit{i.e.}, the mean of the distribution of SRE~(\ref{eq:SAPE_tot_kappa}) --- we prefer this over the mean of the SRD distribution, as the latter can be close to zero in the presence of very broad but sign-symmetric distribution.

Starting from the conductivity, the mean for SRE[$\kappa$] and mean for SRME[$\{\mathcal{K}(\bm{q})_s\}$] (Table~\ref{tab:umlp_performance}) are comparable in the absence of cancellation of microscopic errors,  while the mean SRE[$\kappa$] is significantly lower than mean SRME[$\{\mathcal{K}(\bm{q})_s\}$] when cancellation of microscopic errors occurs. 
{As anticipated, this relation is not necessarily respected for SRE[$\gamma$] and SRME[$\{\mathcal{G}(\bm{q})_s\}$], however the materials in focus here have non-negligible Grüneisen parameter with dominating positive single-phonon contributions, and therefore mostly display a relation between SRE[$\gamma$] and SRME[$\{\mathcal{G}(\bm{q})_s\}$] analogous to the one for the conductivity.}

Importantly, we note that our benchmarks on {SRME}[$\{\mathcal{K}(\bm{q})_s\}$], {SRME}[$\{\mathcal{G}(\bm{q})_s\}$] and SRE[$\kappa$], SRE[$\gamma$] evaluate the accuracy of fMLPs in predicting experimentally observable conductivity and Grüneisen parameter, which are macroscopic properties determined not only by single (local-energy-minima) point of the potential energy surface (PES), but also by its second- and third-order derivatives (i.e., harmonic and anharmonic interatomic forces\cite{ziman1960electrons}).
Because these derivatives expose even subtle discontinuities in the PES, our metrics provide information that is
physically richer and more robust compared to established energy-based metrics, and overall stricter in measuring both the utility of fMLPs and the physics encoded in the smoothness of the PES\cite{matbench_discovery_url}. \\

Table~\ref{tab:umlp_performance} highlights how fMLPs that are highly accurate according to tests based on interaction energies can be inaccurate in predicting crystal structures, interatomic forces, and thermomechanical properties. 
Conversely, for the materials studied so far, fMLPs that are highly accurate according to the {SRME}[$\{\mathcal{K}(\bm{q})_s\}$] and {SRME}[$\{\mathcal{G}(\bm{q})_s\}$] metrics are also highly ranked by energy-based tests, and Fig.~\ref{fig:violin}\textbf{c} also shows that low {SRME}[$\{\mathcal{K}(\bm{q})_s\}$] is correlated with correctly describing crystal symmetries. \\

Importantly, the MPtrj dataset, used to train the fMLPs discussed in this work, includes multiple DFT simulations for the same structure computed with DFT settings of different accuracy (see, e.g., BeTe mp-252, which contains tasks mp-657299 and mp-1057046 that were computed with different k-point grids); this may introduce noise into the training data. In this study, to ensure consistency, we adopted the same DFT parameters as those used for the phononDB-PBE database \cite{togo_2015_distribution,seko_prediction_2015} in both conductivity and Grüneisen calculations.
Therefore, even if the MPtrj and phononDB-PBE datasets were computed with the same DFT functional, the DFT thresholds in these two datasets are not always exactly the same, and the MPtrj dataset may contain noise related to variations in the stringency of the DFT thresholds.
The strong improvement in the accuracy of fMLPs upon fine-tuning shows that these discrepancies are unimportant for assessing the performance of the fMLPs discussed in this work. However, as fMLPs continue to improve in accuracy, discrepancies between DFT and fMLP predictions—especially the residual few-percent differences after fine-tuning—may become comparable to the noise introduced by variations in the stringency of the DFT thresholds in the different datasets.
Future work should therefore focus on generating more accurate training and reference DFT data for fMLPs, with higher accuracy, comprehensiveness, and quality \cite{kaplanFoundationalPotentialEnergy2025}.\\

\begin{figure*}
    \centering
    \includegraphics[width=\textwidth]{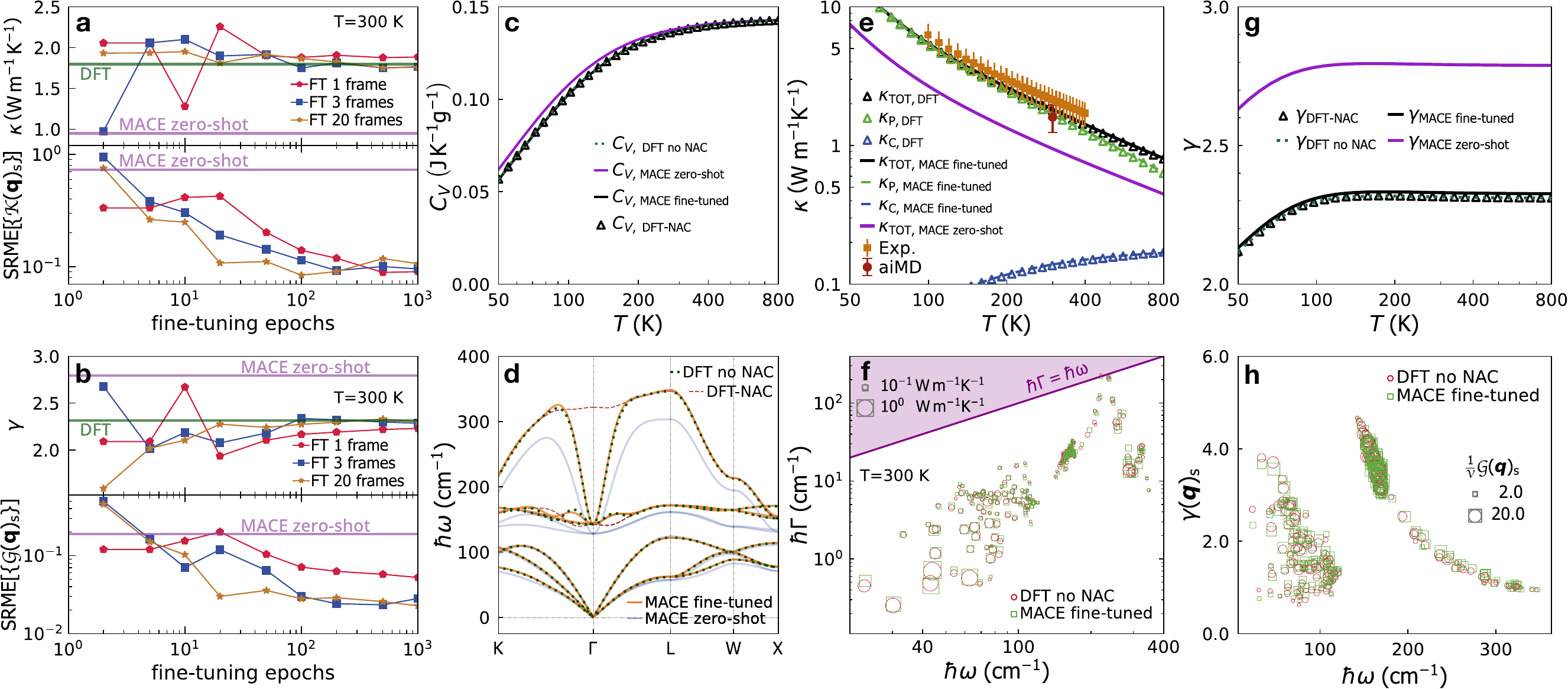}\\[-2mm]
    \caption{\textbf{Achieving first-principles accuracy on LiBr's vibrational \& thermal properties through fine-tuning.} Top of panel \textbf{a} (\textbf{b}), conductivity (Grüneisen parameter) as a function of number of DFT-PBE supercell frames (see text) in the fine-tuning dataset, 
    and number of fine-tuning epochs. The MACE zero-shot prediction for conductivity (Grüneisen parameter) is purple; performing fine-tuning on one single DFT frame and 100 fine-tuning epochs yields compatibility within 7\% for conductivity (5.5\% for Grüneisen) with the reference DFT-PBE value (green), while with datasets containing 3 frames with rattling and volume variations yields compatibility within 2\% for conductivity (0.5\% for Grüneisen) at 200 epochs.
    The bottom of panel \textbf{a} (\textbf{b}) show that SRME$\big[\{\mathcal{K}(\bm{q})_s\}\big]$ (SRME$\big[\{\mathcal{G}(\bm{q})_s\}\big]$) converges to a minimum upon increasing the number of epochs, and larger datasets yield faster convergence. 
    After fine-tuning on 3 frames for 200 epochs, remarkable agreement between fine-tuned MACE and DFT-PBE is observed for: specific heat  (\textbf{c}), phonons (\textbf{d}), microscopic phonon energy-linewidth (\textbf{f}) and  energy-mode Grüneisen parameter distributions  (\textbf{h}), macroscopic $\kappa(T)$ (\textbf{e}) and $\gamma(T)$  (\textbf{g}). We note that after fine-tuning, our WTE  $\kappa(T)$ predictions agree with  experiments\cite{hakansson_thermal_1989}, as well as with predictions from \textit{ab-initio} (PBE) molecular-dynamics (aiMD) at 300 K\cite{knoop_anharmonicity_2023}.
    }
    \label{fig:fine-tuning}
\end{figure*}

\noindent 
\textbf{Achieving first-principles accuracy via fine-tuning}\\
In the previous section we discussed LiBr as paradigmatic example where the zero-shot conductivity predictions from MACE-MP-0 are in strong disagreement --- about a factor of two --- with DFT-PBE reference data. Predictions for the Grüneisen parameter also show a $20\%$ discrepancy. These discrepancies might be due to lack of relevant reference data in the MP dataset\cite{materialsproject} used to train MACE-MP-0, and in this section we introduce a fine-tuning  protocol to correct them and achieve first-principles accuracy.

Employing the same DFT settings used to generate reference DFT data (see Methods for details), we prepared three distinct datasets for fine-tuning. Each dataset contains a set of different DFT-PBE frames --- with 'frame' we mean a set comprising forces, total energy, and stress computed from DFT for an  atomic configuration perturbed from equilibrium.
Perturbations consist of rattling (random atomic displacements drawn from Gaussian distribution with standard deviation of 0.03\AA)\cite{eriksson_hiphive_2019}, and in some cases also a few-percent isotropic rescaling of volume. DFT frames have the same size as those used to compute the reference phononDB-PBE phonons, $4\times4\times4$ supercell of LiBr (512 atoms). 

\begin{figure}[t]
  \centering
  \includegraphics[width=\columnwidth]{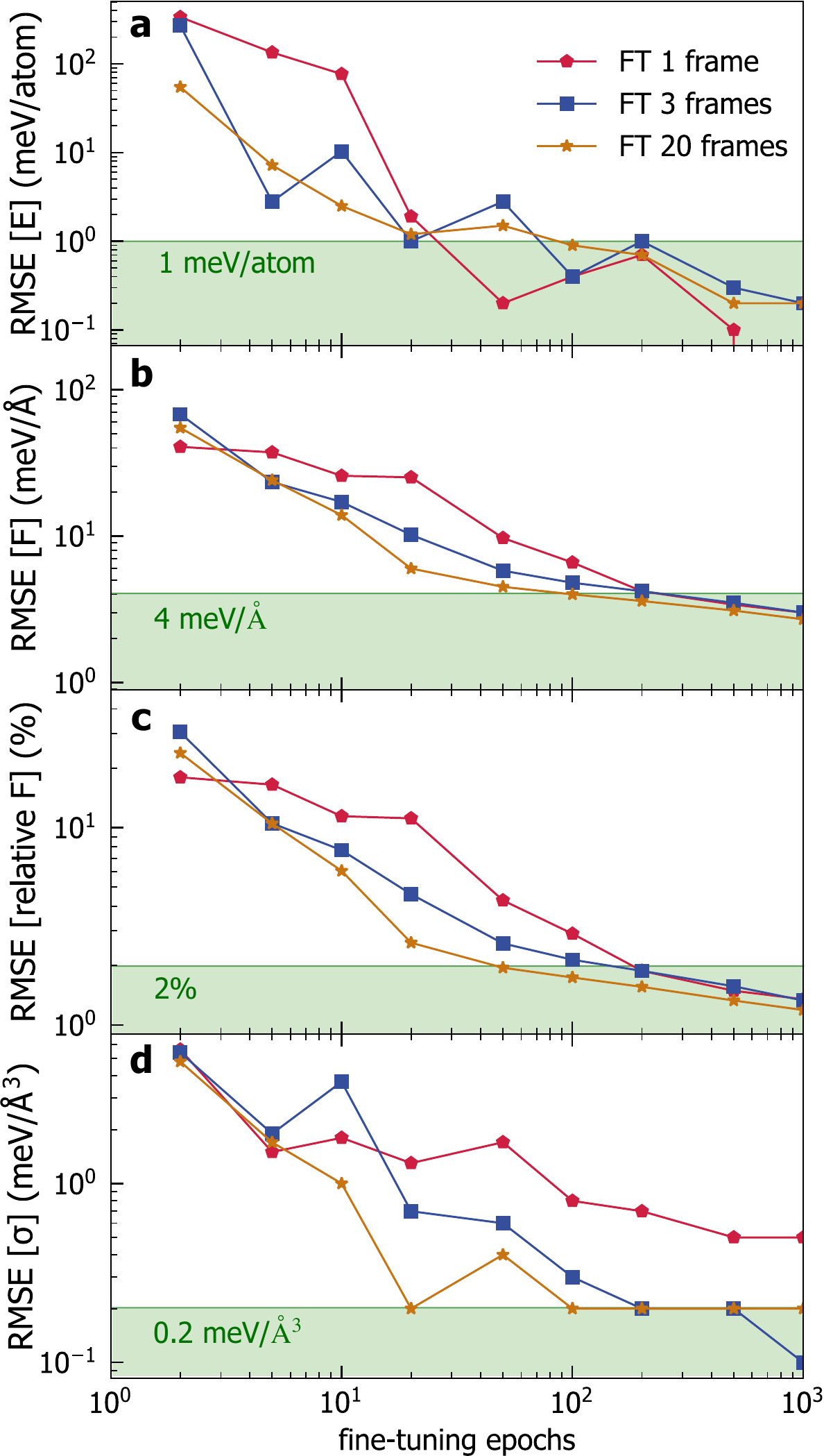}
  \caption{\textbf{Improving fMLP accuracy in LiBr via fine-tuning.} Upon increasing the number of fine-tuning epochs, the root mean square training errors on energies (panel \textbf{a}), forces (\textbf{b} and \textbf{c}), and stresses (\textbf{d}) decrease. 
  The green regions highlight the errors at the end of the 3-frame fine-tuning for MACE-MP-0, which was then used to produce Fig.~\ref{fig:fine-tuning}\textbf{b-d}. These errors approximately correspond to a validation error at 300 K of 2\% on SRE~(\ref{eq:SAPE_tot_kappa}), and of 0.1 on SRME~(\ref{eq:SRME_mode_kappa}).  }
\label{fig:finetuning_methods}
\end{figure}

We generated three training datasets for fine-tuning, which contain: (i) one single frame at equilibrium volume and with atomic rattling; (ii) 3 frames, out of which 2 feature \textpm 1\% isotropic volume rescaling and rattling, plus the frame in (i) at equilibrium volume and with rattling; (iii) 20 frames, out of which 10 are at equilibrium volume with different rattling, and 10 frames with rattling and volume variations of \textpm0.25 \%, \textpm0.50\%, \textpm1\%, \textpm2\%, and \textpm5\%. Using these datasets, we performed fine-tuning training with number of epochs ranging from 1 to 1000. 
Training root-mean-square errors (RMSE) on energy, forces, and stresses, are discussed in the Methods (Fig.~\ref{fig:finetuning_methods}).
Validation was performed relying on reference data for LiBr in the phononDB-PBE database, in particular Fig.~\ref{fig:fine-tuning}\textbf{a} and Fig.~\ref{fig:fine-tuning}\textbf{b} show the validation error on SRME[$\{\mathcal{K}(\bm{q})_s\}$] and SRME[$\{\mathcal{G}(\bm{q})_s\}$] respectively. 
Fig.~\ref{fig:fine-tuning}\textbf{a} highlights how fine-tuning relying on the 3-frame dataset (ii) and for 200 epochs yields a validation error on SRME[$\{\mathcal{K}(\bm{q})_s\}$]<0.1, which corresponds to compatibility within 2\% with the reference DFT-PBE conductivity $\kappa(300 \rm{K})$. 
In the case of the Grüneisen parameter, in Fig.~\ref{fig:fine-tuning}\textbf{b}, the SRME[$\{\mathcal{G}(\bm{q})_s\}$] is <0.03, which corresponds to compatibility within 0.5\% with the reference DFT-PBE Grüneisen parameter $\gamma(300 \rm{K})$.  
We note that low ${\rm SRME}[\!\{{\mathcal{K}}(\!\bm{q}\!)_s\}\!]$ and ${\rm SRME}[\!\{{\mathcal{G}}(\!\bm{q}\!)_s\}\!]$ imply also remarkable compatibility with reference phononDB-PBE data for specific heat (Fig.~\ref{fig:fine-tuning}\textbf{d}), phonons (Fig.~\ref{fig:fine-tuning}\textbf{d}), microscopic energy-linewidth distribution (Fig.~\ref{fig:fine-tuning}\textbf{f})  and mode Grüneisen parameter (Fig.~\ref{fig:fine-tuning}\textbf{h}).
The accuracy of the fine-tuned fMLP in predicting the macroscopic, experimentally measurable conductivity  was evaluated over a test set including: \textit{ab-initio} molecular dynamics (aiMD) \cite{knoop_anharmonicity_2023} performed at 300 K using the same PBE functional used in our work;
experimental data~\cite{hakansson_thermal_1989} in the temperature range 100-800 K.
Fig.~\ref{fig:fine-tuning}\textbf{e} highlights how our predictions agree, within error bars, both in trend and magnitude with experiments (error bars show the compatibility threshold of 20\% discussed in Ref.~\cite{hakansson_thermal_1989}), as well as with independent aiMD predictions \cite{knoop_anharmonicity_2023}.
We also note that using the larger fine-tuning dataset (iii) does not show significant improvements in accuracy, and the single-frame dataset (i) allows to reduce conductivity discrepancies from 47\% to 7\% in 100 fine-tuning epochs. 
Fig.~\ref{fig:fine-tuning}\textbf{g} shows that the obtained Grüneisen parameter is compatible with the reference DFT-PBE data (the maximum deviation in the temperature range 50-800 K is smaller than 0.7\%); comparisons against experiments on thermal expansion or Grüneisen parameter  are not reported because, to the best of our knowledge, these are not available in the literature.

Finally, we emphasize that employing the WTE is essential for accurately describing heat transport in LiBr over a broad temperature range.
In fact, with increasing temperature the tunneling conductivity $\kappa_C$ becomes more important and non-negligible compared to the propagation conductivity $\kappa_P$, and the total WTE conductivity $\kappa_T(T){=}\kappa_P(T){+}\kappa_C(T)$ is in better agreement with the experimental trend compared to the BTE (propagation-only) conductivity $\kappa_P(T)$. 

We highlight how with one single fine-tuning procedure we obtained quantitatively accurate predictions for both thermal conductivity and Grüneisen parameter (evaluated with the volume-derivative method, see Methods for workflow). 
More generally, additional properties like temperature-dependent bulk modulus and thermal-expansion coefficient can be directly obtained from fine-tuned fMLPs by direct free energy minimization within QHA\cite{ritz_thermal_2019,togo_first-principles_2023} (see Methods). This represents an advantage compared to established methods such as: (i) compressive sensing, which requires different DFT datasets at different volumes to compute these properties (see, e.g., Refs. ~\cite{eriksson_hiphive_2019} for a thorough discussion compressive sensing methods applied to simulations of vibrational properties); (ii) training a conventional (non-foundation) machine learning potential from scratch, which requires a large training dataset to obtain accurate results.
{Importantly, fMLPs enable zero-shot predictions that are rapidly becoming increasingly more accurate\cite{matbench_discovery_url}; 
this is an advantage compared to  established (compressive-sensing or non-foundation methods), which  require supercell DFT calculations for direct predictions or training.}\\

Another important application of fMLPs is that of high-throughput materials screening, which aims at identifying optimal materials for target applications by systematically calculating and comparing the properties of a large number of materials with different compositions and structures\cite{mounetTwodimensionalMaterialsHighthroughput2018,lee_accelerating_2024,liang_high-throughput_2019,bagheri_high-throughput_2023,Uhrin2021,ojih2024graph,roy_machine_2021,jain_commentary_2013}.
In the Methods, we demonstrate that SOTA fMLPs can be simultaneously 
fine-tuned for multiple compositions (rather than independently fine-tuning different compositions), a property that may be highly beneficial to minimize the computational cost of high-throughput screening applications.
In particular, we show that simultaneous fine-tuning in three compositions yields conductivity and Grüneisen parameter compatible within 5\% with DFT reference.
This fine-tuning was particularly data efficient, as it was performed using only three atomic configurations: a single rattled supercell at equilibrium volume per material; two unit-cell-sized configurations at expanded and contracted volume, respectively, to describe stresses.\\

\noindent
{{\textbf{Conclusions}}}\\
\noindent
We introduced a framework that leverages foundation models for atomistic materials chemistry\cite{chen_universal_2022,deng_chgnet_2023,batatia_foundation_2024,park_scalable_2024,yang2024mattersimdeeplearningatomistic,merchant_scaling_2023, ALIGNN, MEGNET, Voronoi_RF, BOWSR, CGCNN, CGCNN+P, Wrenformer}, the Wigner formulation of heat transport\cite{simoncelli_unified_2019,simoncelli_wigner_2022}, and the quasiharmonic approach for thermal expansion \cite{ritz_thermal_2019,wallace_thermodynamics_1972} to address the major challenges of current techniques for designing heat-management materials: high computational cost, limited transferability, or lack of physics awareness. 
We leverage this framework to quantify how the accuracy of fMLPs in describing many-body interatomic forces connects to their accuracy in predicting experimentally observable thermomechanical properties. We leverage these insights to define physically interpretable, robust, and stringent benchmark metrics that expose how subtle artifacts or inaccuracies in the smoothness of the Born-Oppenheimer potential energy surface (PES) encoded by fMLPs affect the prediction of constant-volume observables such as the thermal conductivity, and volume-deforming properties such as the Grüneisen parameter.
Because our metrics test not only single-energy values of the PES but also its derivatives (smoothness), they are physically richer, more robust, and stricter compared to established single-energy metrics \cite{matbench_discovery_url}, as recognized by several works \cite{fu_learning_2025,rhodes_orb-v3_2025,matbench_discovery_url,egip_nodate} released after the first preprint version of this manuscript. 
For these reasons, shortly after the preprint version of this work, our $\kappa_{\rm SRME}$ conductivity metric~(Eq.~(\ref{eq:SRME_mode_kappa})) has been included on the Matbench Discovery platform \cite{matbench_discovery_url} and recognized as a key metric to assess both the utility and physical accuracy of the potential energy surface described by fMLPs\cite{fu_learning_2025,rhodes_orb-v3_2025,matbench_discovery_url,egip_nodate}.
These statements are supported by very recent work from Orbital Materials \cite{rhodes_orb-v3_2025}: 
they employed the metrics discussed here (in the first preprint version of this manuscript) to develop the Orb-v3 model, which displays significantly improved zero-shot accuracy compared to the model Orb-v1 discussed in this manuscript.

We also demonstrated that zero-shot foundation machine-learning potentials (fMLPs) predictions for conductivity and Grüneisen parameter can be within a factor of 2 accuracy from DFT reference data, and near DFT accuracy (within 2\%) can be obtained through fine-tuning fMLPs. Most importantly, we showed how fine-tuned fMLPs also enable the prediction of several thermomechanical properties at significantly reduced computational cost compared to current DFT-based methods (e.g., based on systematic finite-differences\cite{togo_2015_distribution}, compressive sensing\cite{eriksson_hiphive_2019}, or density-functional perturbation theory\cite{baroni_phonons_2001}), as well as compared to conventional (composition-specific, not-foundation) machine-learning potentials that require a computationally expensive training from scratch.
Importantly, our physics-aware framework also overcomes the reliability issues (unknown unknowns) of current end-to-end machine-learning methods for predicting vibrational or thermal properties of materials with arbitrary composition and structure.

Finally, we have shown that our framework enables the screening or prediction of materials that violate semiclassical Boltzmann transport, which are crucial for applications ranging from thermal insulation\cite{qian_phonon-engineered_2021,simoncelli_wigner_2022} to neuromorphic computing\cite{nataf_using_2024}.
It is worth mentioning that after posting the preprint of this manuscript on arXiv, the effectiveness of fMLP-based computational frameworks in predicting thermomechanical properties has been discussed also by Microsoft Research for AI, which employed the fMLP MatterSim\cite{yang2024mattersimdeeplearningatomistic} to perform high-throughput screening of materials with extreme lattice thermal conductivities\cite{li_probing_2025}.
Ultimately, we have shown that combining the Wigner Transport Equation with fMLP enables the theory-driven optimization, design, and discovery of materials for next-generation energy and information-management technologies.

\vspace*{3mm}
\noindent
{\Large{\textbf{Methods}}}\\ 
\textbf{Vibrational properties relevant to compute $\kappa(T)$ and $\gamma(T)$}\\
In this section we discuss the relation between interatomic vibrational energies and the quantities that appear in the WTE thermal conductivity expression~(\ref{eq:thermal_conductivity_combined}) and in the formula for the Grüneisen parameter~(\ref{eq:mode_gruneisen}). 
The starting point is the Born-Oppenheimer Hamiltonian for atomic vibrations \cite{simoncelli_wigner_2022} expanded up to anharmonic third-order\cite{carrete2017almabte,alamode,kaldo,paulatto_anharmonic_2013,fugallo2013ab,cepellotti_phoebe_2022,plata_efficient_2017,savic_dimensionality_2013} in the atomic displacements from equilibrium, and accounting for energy perturbations due to presence of isotopes \cite{tamura_isotope,foss_effects_2020,hahn_engineering_2021,abou_el_kheir_unraveling_2024,stewart_first-principles_2009}
 \begin{equation}
 \begin{split}
  &\hat{H}=\sum_{\bm{R},b,\alpha}{\frac{\hat{p}^2_{\bm{R}b\alpha}}{2 M_{b}}} + \frac{1}{2}\!\!\!\sum_{\substack{\scriptscriptstyle{\bm{R},b,\alpha}\\ \scriptscriptstyle{\bm{R'}\!,b'\!,\alpha'}} } \!\!
\frac{\partial^2 {V} }{\partial {u}_{\bm{R}b\alpha} \partial {u}_{\bm{R'}b'\hspace*{-0.5mm}\alpha'}\! }\bigg\lvert_{\!\rm eq}\!\!
\hat{u}_{\bm{R}b\alpha}
\hat{u}_{\bm{R'}b'\hspace*{-0.5mm}\alpha'}\\
&+\frac{1}{3!}\!\!\!\sum_{\substack{\scriptscriptstyle{\bm{R},b,\alpha}\\ \scriptscriptstyle{\bm{R'}\!,b'\!,\alpha'}} } \!\!
\frac{\partial^3 {V} }{\partial {u}_{\bm{R}b\alpha} \partial {u}_{\bm{R'}b'\hspace*{-0.5mm}\alpha'}\!\partial {u}_{\bm{R}''\hspace*{-0.5mm}b''\hspace*{-0.5mm}\alpha''}\! }\bigg\lvert_{\!\rm eq}\!\!
\hat{u}_{\bm{R}b\alpha}
\hat{u}_{\bm{R'}b'\hspace*{-0.5mm}\alpha'}
\hat{u}_{\bm{R}''\hspace*{-0.5mm}b''\hspace*{-0.5mm}\alpha''} \, \\
&+\sum_{\scriptscriptstyle{\bm{R},b,\alpha}}\left(\frac{M_{b} }{m_b}-1\right){\frac{\hat{p}^2_{\bm{R}b\alpha}}{2 M_{b}}}.   
  \label{eq:Born_Oppenheimer}
   \end{split}
   \raisetag{6mm}
\end{equation}
Here, $\hat{p}_{\bm{R}b\alpha}$ and $\hat{u}_{\bm{R}b\alpha}$ are the momentum and position-displacement operators  for the atom $b$ having isotope-averaged mass $M_{b}$, position $\bm{R}{+}\bm{\tau}_b$ ($\bm{R}$ is the Bravais-lattice vector and $\bm{\tau}_b$ the position in the crystal's unit cell), and along the Cartesian direction $\alpha$; the last term  describes kinetic-energy perturbations induced by isotopes ($m_{b}$ is the exact mass of atom at $\bm{\tau}_b$, we adopt the common approximation\cite{tamura_isotope,fugallo2013ab} that the mass $m_{b}$ depends only on the index $b$ and not on the lattice vector $\bm{R}$).
The leading (harmonic) term in Eq.~(\ref{eq:Born_Oppenheimer}) determines the vibrational frequencies. In  particular,  the Fourier transform of the mass-rescaled Hessian of the interatomic potential yields the dynamical matrix at wavevector $\bm{q}$,
\begin{equation}
  \tenscomp{D}(\bm{q})_{b\alpha,b'\!\alpha'}{=}\sum_{\bm{R}}
\frac{\partial^2 {V} }{\partial {u}_{\bm{R}b\alpha} \partial {u}_{\bm{R}'\hspace*{-0.5mm}b'\hspace*{-0.5mm}\alpha'} }\Big\lvert_{\rm eq}\frac{e^{-i\bm{q}\cdot(\bm{R}{+}\bm{\tau}_b{-}\bm{\tau}_{b'})}}{\sqrt{M_bM_{b'}}},
\end{equation}
and by diagonalizing the dynamical matrix,
\begin{equation}
  \textstyle\sum_{b'\alpha'}\tenscomp{D}(\bm{q})_{b\alpha,b'\!\alpha'}\mathcal{E}(\bm{q})_{s,b'\!\alpha'}=\omega^2(\bm{q})_s\mathcal{E}(\bm{q})_{s,b\alpha},
  \label{eq:diagonalization_dynamical_matrix}
\end{equation}
one obtains from the eigenvalues the phonon energies of the solid $\hbar\omega(\bm{q})_s$  ($s$ is a  band index ranging from 1 to 3$N_{\rm at}$, where $N_{\rm at}$ is the number of atoms in the unit cell), and from the eigenvectors $\mathcal{E}(\bm{q})_{s,b\alpha}$ the displacement patterns of atom $b$ in direction $\alpha$ for the phonon having wavevector $\bm{q}$ and mode $s$.

The vibrational frequencies $\hbar\omega(\bm{q})_s$ are sensitive to the thermally induced volume expansion, as quantified by mode Grüneisen parameter, Eq.~(\ref{eq:mode_gruneisen})\cite{wallace_thermodynamics_1972}. 
In practice, the volume derivative is calculated using the Hellman-Feynman theorem\cite{togo_first-principles_2023},
\begin{equation}
\begin{split}
    \gamma(\bm{q})_s{=}{-}\frac{\mathcal{V}}{2\omega(\bm{q})_s^2} \hspace{-1mm}\sum_{b\alpha,b'\alpha'}\hspace{-1.7mm}\mathcal{E}^*\hspace{-0.3mm}(\bm{q})_{s,b\alpha}\!\frac{\partial \tenscomp{D}(\bm{q})_{b\alpha,b'\!\alpha'\!}}{\partial \mathcal{V}} \mathcal{E}(\bm{q})_{s,b'\!\alpha'\!}\hspace{0.1mm},\hspace{5.5mm}
    \raisetag{7mm}
  \end{split}
\end{equation}
where the partial derivative of the dynamical matrix is obtained by computing the numerical differences at volumes expanded or contracted by $1\%$.

As discussed in detail in Ref.~\cite{simoncelli_wigner_2022}, the velocity operator appearing in the WTE conductivity expression~(\ref{eq:thermal_conductivity_combined}) is determined by the wavevector derivative of the square root of the dynamical matrix:
\begin{equation}
\begin{split}
    \tenscomp{v}^\beta(\bm{q})_{s,s'}{=}\!\!\sum_{\substack{b,\alpha,b'\!,\alpha'}}\!\!\!\mathcal{E}^\star(\bm{q})_{{s},b\alpha}[{{\nabla^{\beta}_{\bm{q}} \sqrt{\tenscomp{D}(\bm{q})}} }_{b\alpha,b'\hspace*{-0.5mm}\alpha'} ]\mathcal{E}(\bm{q})_{s'\hspace*{-0.5mm},b'\hspace*{-0.5mm}\alpha'}.\;\;\;\;\;\;\;
  \raisetag{8mm}
  \end{split}
\end{equation}

The anharmonic linewidths $\hbar\Gamma_{\rm a}(\bm{q})_s$, which contribute to the total linewidths $\hbar\Gamma(\bm{q})_s=\hbar\Gamma_{\rm a}(\bm{q})_s+\hbar\Gamma_{\rm i}(\bm{q})_s$ appearing in Eq.~(\ref{eq:thermal_conductivity_combined}), are the energy broadenings due to anharmonic three-phonon interactions \cite{fugallo2013ab,togo_--fly_2024,tadano_anharmonic_2014,carrete2017almabte,li_shengbte_2014}. It can be shown\cite{paulatto_anharmonic_2013} that these are determined by the third derivative of the interatomic potential (see Eq.~(\ref{eq:Born_Oppenheimer})),
\begin{equation}
\begin{split}
&\hbar\Gamma_{\!\!\rm a}\!(\bm{q})_{\!s}{=}\!\frac{\pi}{N_c^2}
\!{\sum_{\substack{\bm{q'}\!\bm{q'\!'\!}\\s'\!,s''\!}}} \!\Big\{\!2\!\big[\bar{\tenscomp{N}}(\bm{q'\!})_{s'\!}{-}\bar{\tenscomp{N}}(\bm{q'\!'\!})_{s'\!'\!}\big]\delta\big(\omega(\bm{q})_s\!{+}\omega(\bm{q'\!})_{s'\!}{-}\omega(\bm{q'\!'\!})_{s'\!'\!}\big) \\
 &+ \big[1{+}\bar{\tenscomp{N}}(\bm{q'\!})_{s'\!}{+}\bar{\tenscomp{N}}(\bm{q'\!'\!})_{s'\!'\!}\big]\delta\big[\omega(\bm{q})_{s}{-}\omega(\bm{q'\!})_{s'\!}{-}\omega(\bm{q'\!'\!})_{s'\!'\!}\big]\Big \}\times\\
&\Bigg|\hspace*{-2mm}
\sum_{\substack{\alpha,\alpha'\!,\alpha'\!'\\b,b'\!,b'\!'\!,\bm{R'}\! \bm{R'\!'}}}\hspace*{-4mm}\frac{\partial^3 V}{\partial u_{\bm{0}b\alpha}\partial u_{\bm{R'\!}b'\!\alpha'\!}\partial u_{\bm{R'\!'}\!b'\!'\!\alpha'\!'}\!\!\!}
\mathcal{E}(\bm{q})_{s,b\alpha}
\mathcal{E}(\bm{q'})_{s'\!,b'\!\alpha'}\mathcal{E}(\bm{q'\!'})_{s'\!'\!,b'\!'\!\alpha'\!'\!} \\
&
\sqrt{\frac{\hbar^{3}}{8}}\frac{\Delta(\bm{q}{+}\bm{q'\!}{+}\bm{q'\!'\!})e^{-i[\bm{q}\cdot \bm{\tau}_b{+}\bm{q'\!}\cdot (\bm{R'\!}{+}\bm{\tau}_{b'\!}){+}\bm{q'\!'\!}\cdot (\bm{R'\!'\!}{+}\bm{\tau}_{b'\!'\!})]}}{\sqrt{M_b M_{b'\!} M_{b'\!'\!}\omega(\bm{q})_s\omega(\bm{q'\!})_{s'}\omega(\bm{q'\!'\!})_{s'\!'\!}}}
\Bigg|^2,
  \label{eq:anh_linewidth}
\end{split}
\raisetag{6mm}
\end{equation}
where $\bar{\tenscomp{N}}(\bm{q})_s{=}[\exp(\hbar \omega(\bm{q})_s/k_{\rm B}T){-}1]^{-1}$ is the Bose-Einstein distribution, $\Delta(\bm{q}{+}\bm{q'\!}{+}\bm{q'\!'\!})$ is the Kronecker delta (equal to 1 if $\bm{q}{+}\bm{q'\!}{+}\bm{q'\!'\!}$ is a reciprocal lattice vector, 0 otherwise), $\delta$ is the Dirac delta. 
The linewidth due to isotopic-mass disorder is determined exclusively by harmonic properties and concentration of isotopes\cite{tamura_isotope}:
\begin{equation}
\begin{split}
 \hbar\Gamma_{\rm{i}}(\bm{q})_s{=}&\frac{\hbar\pi}{2 N_c}[\omega(\bm{q})_s]^2
{\textstyle\sum_{\bm{q'},s'}} 
\delta\big[\omega(\bm{q})_{s}{-}\omega(\bm{q'})_{s'}\big]\\
&\times\textstyle\sum_b g_2^b \Big|\textstyle\sum_\alpha \mathcal{E}(\bm{q})^{\star}_{s,b\alpha} \mathcal{E}(\bm{q'})_{s',b\alpha} \Big|^2,
  \label{eq:linewidts_iso}
\end{split}
\end{equation}
where $g_2^b=\sum_i f_{i,b}\big(\frac{M_b -m_{i,b}}{M_b}\big)^2$ describes the variance of the isotopic masses of atom $b$ ($f_{i,b}$ and $m_{i,b}$ are the mole fraction and mass, respectively, of the $i$th isotope of atom $b$; $M_b=\sum_i f_{i,b} m_{i,b}$ is the weighted average mass).\\

\noindent
\textbf{Relation between Grüneisen parameter and third-order force constants}\\
In the quasi-harmonic approximation (QHA), expanding the Hamiltonian in the first-order of applied strain allows one to evaluate the mode Grüneisen parameter\cite{hellman_temperature-dependent_2013,wallace_thermodynamics_1972},
which depends linearly on the third-order derivatives of the potential energy surface (PES).
In contrast, the thermal conductivity is, in the BTE relaxation-time approximation regime, inversely proportional to the anharmonic linewidths, and the latter depend on the square of the third-order derivatives of the PES. These distinct dependencies of conductivity and Grüneisen parameter from PES derivatives intuitively shows why our SRME$[\{\mathcal{K}(\bm{q})_s\}\big]$ and SRME$[\{\mathcal{G}(\bm{q})_s\}\big]$ provide complementary information: errors in the PES derivative that are very important for conductivity may have negligible impact on Grüneisen parameter, and vice-versa. Therefore, jointly analyzing thermal conductivity and Grüneisen parameters enables a more comprehensive benchmark of fMLP accuracy, helping in minimizing the risks of overlooking errors that could arise from subtle inaccuracies in modeling the physical PES.\\

\noindent
\textbf{Failures of semiclassical Boltzmann transport} \label{sec:population_and_coherence_conductivity}\\
As anticipated in the main text, our automated framework can be used to identify materials in which the semiclassical particle-like BTE fails. This happens when particle-like propagation mechanisms --- described by both the BTE and WTE\cite{simoncelli_unified_2019,PhysRevX.10.011019,dragasevic_viscous_2023}, and determining $\kappa_P$ in Eq.~(\ref{eq:thermal_conductivity_combined}) --- do not dominate over wave-like tunneling mechanisms  --- missing from the BTE but described by the WTE and determining $\kappa_C$ in Eq.~(\ref{eq:thermal_conductivity_combined}).
Failures of the BTE have been discussed to appear in, e.g., strongly anharmonic complex crystals for energy harvesting\cite{di_lucente_crossover_2023,jain_multichannel_2020,xia_microscopic_2020,shen_amorphous-like_2024,pandey_origin_2022,fiorentino_green-kubo_2023,zhou_phonon_2024,zheng_unravelling_2024,jia_cu3bis3_2023,tadano_first-principles_2022,bernges_considering_2022,yang_reduced_2022,zeng_pushing_2024},  thermal barrier coatings \cite{caldarelli_many-body_2022,xia_unified_2023,luo_vibrational_2020,pazhedath_first-principles_2023}, as well as in several disordered functional materials \cite{harper_vibrational_2023,liu2023unraveling,ndour_practical_2023,lundgren_mode_2021,kielar_anomalous_2024,simoncelli_temperature-invariant_2024}.
Our automated framework computes the total Wigner thermal conductivity $\kappa_P+\kappa_C$ \cite{simoncelli_wigner_2022}, and therefore can be used to find materials that violate semiclassical Boltzmann thermal transport.
In Fig.~\ref{fig:coherences_vs_populations} we show that among the 103 compounds analyzed at 300 K, the BTE fails in materials with low thermal conductivity ($\kappa_{\rm TOT}\lesssim 2$ W/mK), such as AgBr and AgCl, since the wave-like tunneling conductivity becomes comparable to the particle-like propagation  conductivity. We also note that in simple crystals with large conductivity, the BTE is accurate as the particle-like propagation conductivity dominates over the wave-like tunneling conductivity.\\

\noindent
\textbf{Symmetric Relative Mean Error on harmonic and anharmonic vibrational properties}\\
The analyses reported in the main text discuss SRME$[\{\mathcal{K}(\bm{q})_s\}\big]$ as a descriptor that is informative of the accuracy of fMLPs in describing microscopic harmonic and anharmonic properties. In this section we provide the tools to resolve whether a large SRME$[\{\mathcal{K}(\bm{q})_s\}\big]$ originates from large errors in microscopic harmonic properties, or large errors in anharmonic properties, or from both.
To this aim, we employ the expression for the microscopic, single-phonon contributions to the thermal conductivity~(see terms inside the square brackets in Eq.~(\ref{eq:thermal_conductivity_combined})) to define descriptors that quantify the accuracy with which harmonic and anharmonic properties are predicted.
In particular, we note that the conductivity contribution of a phonon $(\bm{q})_{s}$ is a function of: (i) all the phonon frequencies $\{\omega(\bm{q})_{s'}\}$ at a fixed wavevector $\bm{q}$ and variable mode $s'=1,\dots,N_{\rm s}$ ($N_{\rm s}$ is the number of phonon bands, equal to 3 times the number of atoms in the crystal's unit cell \cite{ziman1960electrons}); (ii) all velocity-operator elements $\{\tens{v}(\bm{q})_{s,s'}\}$ at fixed $\bm{q}$, $s$ and variable $s'$; (iii) all isotopic linewidths at fixed $\bm{q}$ and variable $s'$, $\{\Gamma_{\!\rm i}(\bm{q})_{s'}\}$, which depend solely on harmonic properties, see Eq.~\ref{eq:linewidts_iso}; (iv) the anharmonic linewidths $\{\Gamma_{\!\rm a}(\bm{q})_{s'}\}$ at fixed $\bm{q}$ and variable $s'$. This can be summarized using the following notation: 
${\mathcal{K}}(\bm{q})_s={\mathcal{K}}(\bm{q})_s[\{\omega(\bm{q})_{s'}\},\{\tens{v}(\bm{q})_{s,s'} \},\{\Gamma_{\rm i}(\bm{q})_{s'}\},\{\Gamma_{\rm a}(\bm{q})_{s'}\}]$
where the curly brackets denote the set of values of a certain quantity over all the bands at fixed $\bm{q}$ (e.g., $\{\omega(\bm{q})_{s'}\}=\{\omega(\bm{q})_{1},\omega(\bm{q})_{2},\dots,\omega(\bm{q})_{N_s}\}$). 

To quantify the impact of errors on the harmonic (har) and anharmonic (anh) properties on the single-phonon  conductivity contributions, we define the Symmetric Relative Mean Error (SRME) on these properties as follows:  
\begin{widetext}
    \begin{equation}
\text{SRME}[{\rm{har}}]{=}2\frac{\sum_{\bm{q}s}\left|
    {\mathcal{K}}(\bm{q})_s\big[\{\omega_{{\rm fMLP}}(\bm{q})_{s'}\},\{\tens{v}_{{\rm fMLP}}(\bm{q})_{s,s'} \},\{\Gamma_{{\!\rm i, fMLP}}(\bm{q})_{s'}\},\{\Gamma_{{\!\rm a, DFT}}(\bm{q})_{s'}\}\big]-
    {\mathcal{K}}_{\rm DFT}(\bm{q})_s
    \right| }{\sum_{\bm{q}s}\left(
    {\mathcal{K}}(\bm{q})_s\big[\{\omega_{{\rm fMLP}}(\bm{q})_{s'}\},\{\tens{v}_{{\rm fMLP}}(\bm{q})_{s,s'} \},\{\Gamma_{{\!\rm i, fMLP}}(\bm{q})_{s'}\},\{\Gamma_{{\!\rm a, DFT}}(\bm{q})_{s'}\}\big]
    +
    {\mathcal{K}}_{\rm DFT}(\bm{q})_s
    \right)} ,
    \label{eq:smape_harmonic_part}
\end{equation}
\begin{equation}
\text{SRME}[{\rm{anh}}]{=}2\frac{\sum_{\bm{q}s}\left|
    {\mathcal{K}}(\bm{q})_s\big[\{\omega_{{\rm DFT}}(\bm{q})_{s'}\},\{\tens{v}_{{\rm DFT}}(\bm{q})_{s,s'} \},\{\Gamma_{{\!\rm i, DFT}}(\bm{q})_{s'}\},\{\Gamma_{{\!\rm a, fMLP}}(\bm{q})_{s'}\}\big]-
    {\mathcal{K}}_{\rm DFT}(\bm{q})_s
    \right| }{\sum_{\bm{q}s}\left(
    {\mathcal{K}}(\bm{q})_s\big[\{\omega_{{\rm DFT}}(\bm{q})_{s'}\},\{\tens{v}_{{\rm DFT}}(\bm{q})_{s,s'} \},\{\Gamma_{{\!\rm i, DFT}}(\bm{q})_{s'}\},\{\Gamma_{{\!\rm a, fMLP}}(\bm{q})_{s'}\}\big]
    +
    {\mathcal{K}}_{\rm DFT}(\bm{q})_s
    \right)},
    \label{eq:smape_anharmonic_part}
\end{equation}
\end{widetext}

where we have used the shorthand notation
$\mathcal{K}_{{\!\rm DFT}}\!(\bm{q})_{\!s}=\mathcal{K}_{_{\!\rm DFT}}\!(\bm{q})_{\!s}\big[\{\!\omega_{_{\!\rm DFT}}\!(\bm{q})_{\!s'}\!\}{,}\{\!\tens{v}_{_{\!\rm DFT}}\!(\bm{q})_{\!s{,}s'} \!\}{,}\{\!\Gamma_{{\rm i{,} DFT}}\!(\bm{q})_{\!s'}\!\}{,}\{\!\Gamma_{{\rm a{,} DFT}}\!(\bm{q})_{\!s'}\!\}\big]$.
Intuitively, SRME[har]~(\ref{eq:smape_harmonic_part}) is large when harmonic vibrational properties 
($\omega(\bm{q})_{s}$, or $\tens{v}(\bm{q})_{s,s'}$, or $\Gamma_{{\!\rm i}}(\bm{q})_{s}$) differ between DFT and fMLP, while SRME[anh]~(\ref{eq:smape_anharmonic_part}) is large when the anharmonic linewidths 
($\Gamma_{{\!\rm a}}(\bm{q})_s$) differ between DFT and fMLP. 
We show in Fig.~\ref{fig:errors} that small SRME[har] and small SRME[anh] (e.g., as in BeO), imply a small SRE[$\kappa$]
Fig.~\ref{fig:errors} also shows that often large SRME[har] and large SRME[anh] (e.g. LiBr) translate into large  $\rm{SRE}\big[\kappa\big]$; however there are also cases (e.g. BeTe) in which large SRME[har] and large SRME[anh] can also compensate each other and result in a small $\rm{SRE}\big[\kappa\big]$. This confirms that  $\rm{SRE}\big[\kappa\big]$  is not a reliable descriptor for the capability of fMLPs to capture the harmonic and anharmonic physics underlying heat conduction. In contrast, having both small SRME[har] and small SRME[anh] is a sufficient condition to accurately capture harmonic and anharmonic vibrational properties, as well as conductivity.

Overall, these tests illustrate the importance of benchmarking the accuracy of fMLPs in predicting both microscopic harmonic and anharmonic vibrational properties, further motivating the introduction of the SRME[$\{\mathcal{K}(\bm{q})_s\}$].\\

\noindent
\textbf{Influence of isotopic scattering on conductivity}\label{sec:effect_of_isotope_scattering}\\
To analyze the influence of isotope scattering on the conductivity, we compare thermal conductivity values with and without considering the linewidths from isotope mass-disorder (Eq.~(\ref{eq:linewidts_iso})). 
The results are summarized in Table~\ref{tab:iso_effect}. In wurtzite BeO and rocksalt LiBr, the impact of isotope scattering is minimal or small, respectively. However, in zincblende BeTe isotopic and anharmonic linewidths have comparable values (Fig.~\ref{fig:detailed_BeO_BeTe_BAs_analysis}\textbf{b}), and therefore both affect the conductivity. In this material, MACE-MP-0 significantly overestimates some linewidths compared to DFT-PBE, and this compensates for the overestimation in other regions. Consequently, MACE-MP-0 shows a significantly larger error in thermal conductivity when isotope scattering is not considered.
\begin{table}[htb]
    \centering
    \begin{tabular}{m{1.1cm}|m{1.2cm}|m{1.2cm}|m{1.1cm}|m{1.2cm}|m{0.8cm}|m{1.0cm}}
         & \multicolumn{2}{c|}{wurtzite BeO} & \multicolumn{2}{c|}{zincblende BeTe} & \multicolumn{2}{c}{rocksalt LiBr} \\
          &  \centering DFT &\centering MACE & \centering DFT &\centering MACE & \centering DFT & \centering MACE\cr
         \hline
         \centering with $\Gamma_{\rm{i}}(\bm{q})_s$ & 286.391 & 259.443 & 78.246 & 69.138 & 1.789 & 0.948  \\
         \hline
         \centering without $\Gamma_{\rm{i}}(\bm{q})_s$ & 291.963 & 263.956 & 289.374 & 402.030 & 1.803 & 0.951 
    \end{tabular}
    \caption{\textbf{Influence of isotopic scattering on the thermal conductivity at 300 K} for BeO, BeTe, and LiBr. }
    \label{tab:iso_effect}
\end{table}

\begin{figure}[htb]
    \centering
    \includegraphics[width=0.95\columnwidth]{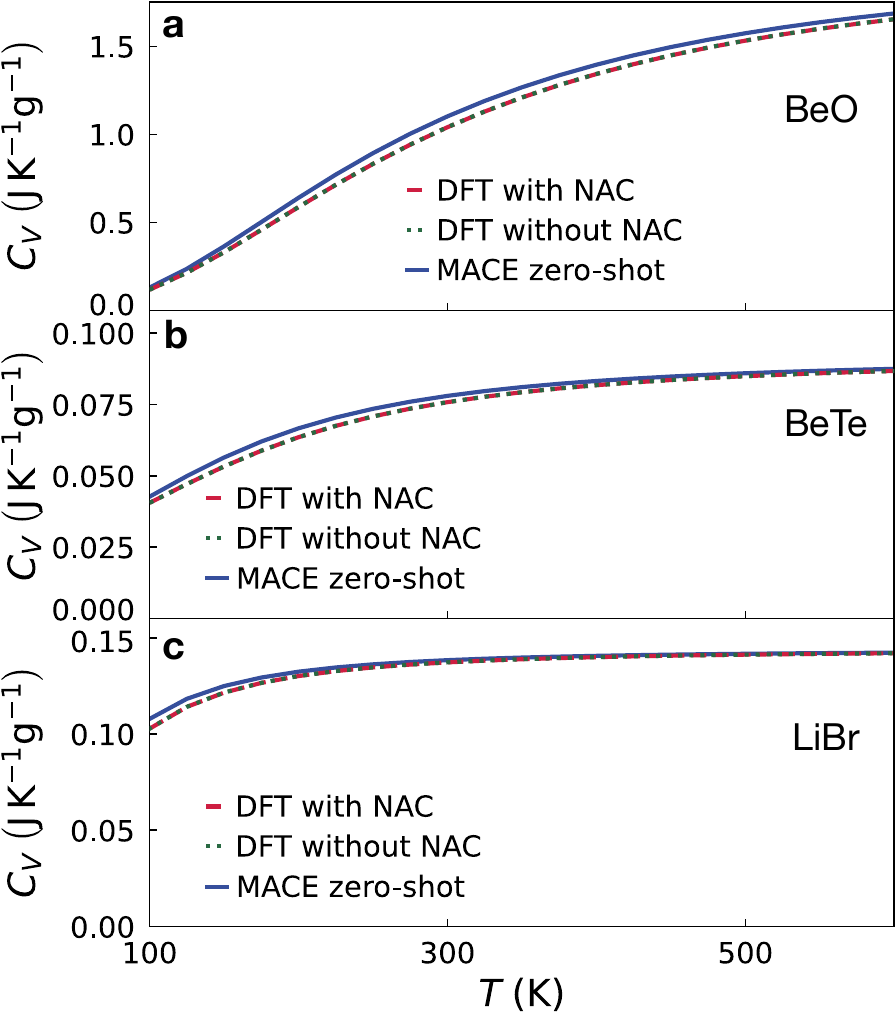}
    \caption{\textbf{Specific heat of a) wurtzite BeO, b) zincblende BeTe c) rocksalt LiBr.} The zero-shot MACE-MP-0 predictions are in good agreement with DFT confirming that optical phonon mode softening has small (unimportant)effect on the  specific heat.}
    \label{fig:specific_heats}
\end{figure}

\noindent
\textbf{Sensitivity of specific heat to fMLP accuracy}\\
To show that inaccuracies in the PES described by SOTA fMLPs have a small (almost negligible) impact on the specific heat, we consider as paradigmatic cases the materials discussed in Fig.~\ref{fig:detailed_BeO_BeTe_BAs_analysis}, namely wurtzite BeO, zincblende BeTe and rocksalt LiBr. 
Fig.~\ref{fig:specific_heats} presents a comparison of the specific heat computed from DFT and zero-shot MACE-MP-0 for these materials, illustrating that specific heat --- and more broadly, phonon spectrum–derived observables --- exhibit reduced sensitivity to phonon band structure and hence to the accuracy of the model.
This can be intuitively understood from the expression for the specific heat:
\begin{equation}
  C =\frac{1}{V N_c}\sum_{\bm{q}s}\frac{\hbar^2\omega^2(\bm{q})_s }{k_{\rm B} T^2} \bar{\tenscomp{N}}(\bm{q})_s\big[\bar{\tenscomp{N}}(\bm{q})_s{+}1\big]
  \label{eq:spec_heat_1}
\end{equation}
where $\bm{q}$ and $s$ are the phonon  wavevector  and mode index, 
$V$ is the unit cell volume and  $N_c$ the number of wavevectors appearing in the sum; $\omega(\bm{q})_s$ is the phonon energy, and $\bar{\tenscomp{N}}(\bm{q})_s{=}[\exp(\hbar \omega(\bm{q})_s/k_{\rm B}T){-}1]^{-1}$ the Bose-Einstein distribution at temperature $T$.
Eq.~(\ref{eq:spec_heat_1}) shows that at finite $T$ the quantum specific heat is more sensitive to low-energy (acoustic) phonon modes and becomes decreasingly less sensitive to modes with increasingly higher frequency (e.g., optical phonon modes). SOTA fMLPs describe relatively well the low-energy phonons and suffer from a systematic softening of optical phonons \cite{deng_overcoming_nodate}, and because the latter inaccuracy involves predominantly high-energy phonons that are unimportant for the specific heat, the specific heat is reasonably well described by SOTA fMLPs already at the zero-shot level.
\\

\noindent
\textbf{Automated Wigner conductivity workflow}\\
\begin{figure}[t]
    \centering
    \includegraphics[width=0.8\columnwidth]{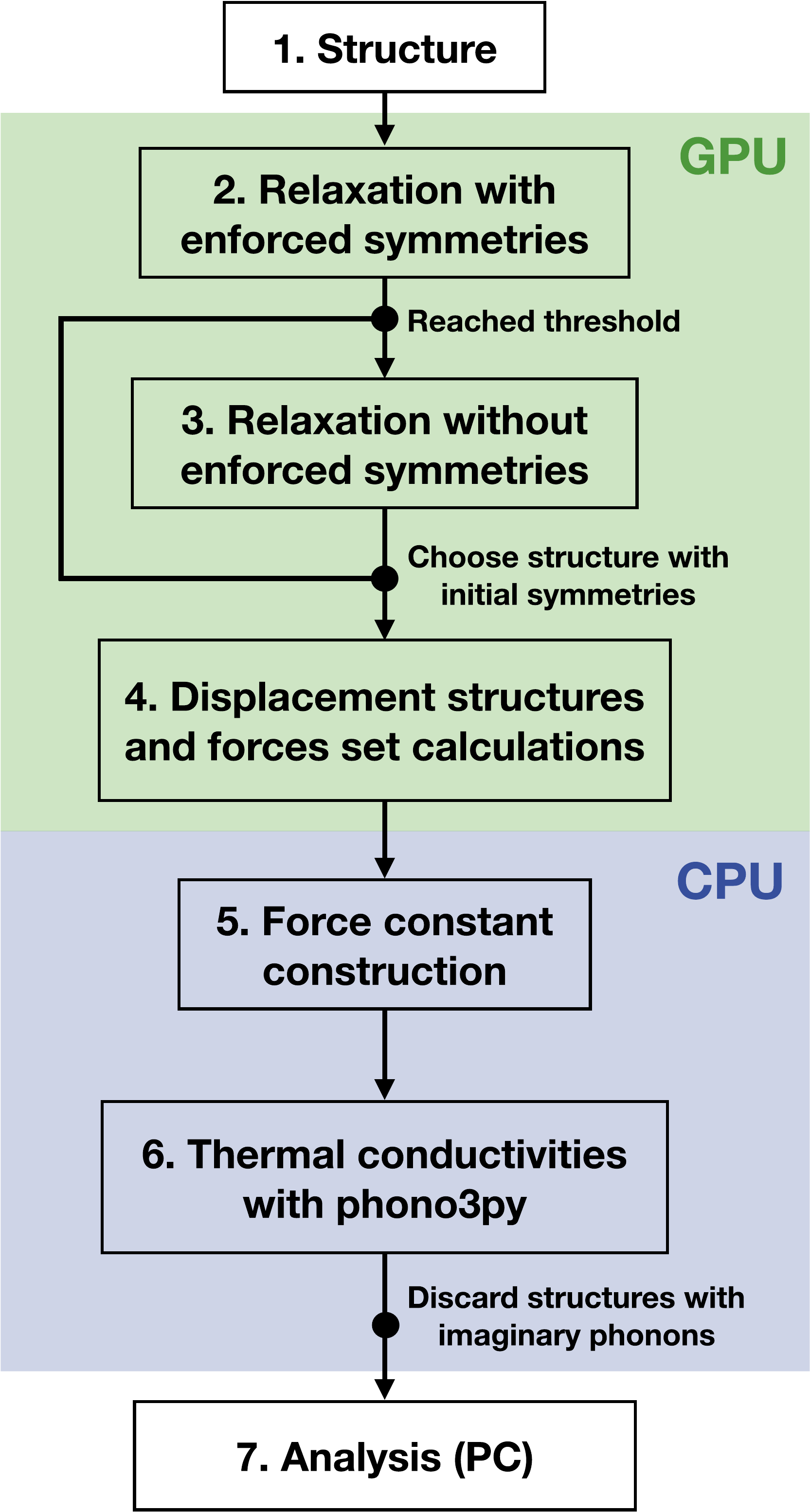}
    \caption{\textbf{Automated Wigner Conductivity Workflow.} The first part of the workflow, involving force evaluation, is performed on a GPU. The input structures undergo relaxation in two stages. In the first stage, initial symmetries are enforced while simultaneously relaxing atomic positions and cell parameters. In the second stage, the structure is further relaxed by removing the symmetry constraint. If the symmetry group of the final structure changes, the structure with enforced symmetry relaxation is used instead. After initializing the displacement structures, the force sets are calculated for each displacement. CPUs handle the remainder of the workflow. Force constants are constructed from the displacement force sets, and thermal conductivity is calculated using \texttt{phono3py}. Structures with imaginary phonon frequencies are discarded from the final analysis (conducted on a PC).}
    \label{fig:workflow}
\end{figure}
The workflow, outlined in Fig.~\ref{fig:workflow}, includes structure relaxation, force constant computation, thermal conductivity evaluation and analysis. 
To automate the calculation of interatomic forces, we developed an interface between the packages \texttt{ase}\cite{ase-paper} (version 3.23.0) and \texttt{phono3py}\cite{phono3py,togo_implementation_2023} (version 3.2.0).
Atomic positions and unit-cell lattice parameters were simultaneously relaxed to minimize energy, forces, and stresses. We note that the determination of irreducible $\bm{q}$ points needed for thermal conductivity calculations depends on the crystal symmetry, and to evaluate $\text{SRME}[\{\mathcal{K}(\bm{q})_s\}]$  (Eq.~(\ref{eq:SRME_mode_kappa})) it is essential to compare structures that have the same crystal symmetry when studied using DFT or fMLPs.

We constructed a two-stage relaxation protocol, employing the FretchetCellFilter and the FIRE algorithms in both stages to simultaneously optimize atomic positions and cell parameters. We used force convergence threshold of $10^{-4}$eV/\AA\ and a maximum number of steps of $300$.
During the first stage, symmetries are explicitly enforced as constraints, whereas in the second stage, instead, the constraint on the crystal symmetry is removed, to test whether the fMLP correctly preserves the physical crystal symmetry or not. 
If the symmetry is preserved, the final relaxed structure is used. If symmetry is broken, the structure from the first stage, with enforced symmetry, is retained. We detect symmetries utilizing the \texttt{spglib} package\cite{spglib} with the default precision parameters. 
As discussed in Fig.~\ref{fig:violin}\textbf{c}, in some cases retaining the symmetric structure yielded structural instabilities (imaginary phonon frequencies), which disappeared when the structure was allowed to break the symmetry and relax.
While these structurally stable and broken-symmetry structures are not useful for evaluating $\text{SRME}[\{\mathcal{K}(\bm{q})_s\}]$, they provide information about shortcomings of fMLPs that could addressed in future studies or used as benchmark metric for accuracy in the description of crystal structures.

In the supercell force-constant calculations, we used the same parameters used in the DFT reference data\cite{phono3py,seko_prediction_2015}.

Thermal conductivities were computed using \texttt{phono3py} following Refs.~\cite{phono3py,togo_implementation_2023}. A $19\times19\times19$ $\bm{q}$-mesh was used for rocksalt and zincblende structures, and a $19\times19\times15$ $\bm{q}$-mesh for wurtzite structures. The collision operator was computed using the tetrahedron method. For Fig.~\ref{fig:detailed_BeO_BeTe_BAs_analysis}, linewidths were calculated on a $9\times9\times9$ $\bm{q}$-mesh, for graphical clarity. {We used version 3 of \texttt{phono3py} for anharmonic linewidth calculations and employed the Gonze method \cite{Gonze1997} for the NAC term, as implemented in the \texttt{phonopy}\cite{togo_implementation_2023,togo_first-principles_2023} package (version 2.26.6). For consistency with the phononDB-PBE dataset, in our data release we also include the conductivity results computed with the Wang method\cite{Wang_2010}.} When structures with imaginary phonon frequencies were found with mp-fMLPs, these were discarded from subsequent thermal-conductivity analysis.\\

\noindent
\textbf{Automated Grüneisen parameter workflow} \\
{The workflow to compute Grüneisen consists of structure relaxation, force constant computation for three structures (equilibrium volume, volume expanded $1\%$, and volume contracted $1\%$), Grüneisen parameter calculation, and analysis.
The structure relaxation is carried out using the same two-stage protocol described in the previous section.
To generate the configurations at volume expanded or contracted by $1\%$, we scale the unit cells while keeping the relative atomic positions fixed, then relax the structures using the same procedure, with the cell volume held fixed at each step. Harmonic force constants for the three volumes are computed using $0.03$~\AA\ displacements and the same supercells used in the conductivity workflow.
The \texttt{phonopy} package \cite{togo_first-principles_2023,togo_implementation_2023} is then used to calculate the mode Grüneisen parameters over the same $\bm{q}$-mesh as in the conductivity calculations.}\\

\noindent
\textbf{DFT reference and zero-shot mp-fMLP calculations}\\
Computational details of the DFT calculations underlying the phononDB-PBE database can be found in Refs.~\cite{phono3py,seko_prediction_2015}. Applying the same settings on the same set of materials, we computed the Grüneisen parameters by calculating the phonon band structures the same way as detailed in the previous section. 
These calculations employed the Perdew, Burke, and Ernzerhof (PBE) exchange-correlation functional  \cite{perdew_generalized_1996}, consistent with the approach used for generating the MP dataset \cite{materialsproject}, used to train mp-fMLPs.

Fig.~\ref{fig:comp_cost} shows that in wurtzite BeO, the zero-shot MACE-MP-0-L2 calculation has a computational cost that is more than 3 orders of magnitude lower than the conventional DFT calculation, and yields a conductivity compatible within 7\% with DFT reference values (Fig.~\ref{fig:detailed_BeO_BeTe_BAs_analysis}\textbf{a}). Importantly, we note that speedups much stronger than those highlighted in Fig.~\ref{fig:comp_cost} are expected in disordered materials without symmetries\cite{simoncelli_thermal_2023,harper_vibrational_2023}. In fact,  BeO has a simple unit cell containing just 4 atoms and with 12 crystal symmetries\cite{togo_spglib_nodate}, which allow to reduce the number of displacements\cite{togo_implementation_2023}. In contrast, in disordered materials symmetries are broken and consequently a larger number of finite-displacement calculations are needed. Moreover, standard DFT approaches scale with the cube of the number of atoms, while fMLPs have a linear scaling. Based on these considerations, we expect even stronger speedups in disordered materials\cite{reicht_designing_2024,PhysRevB.102.024108,de_araujo_oliveira_tuning_2024,bosoni_atomistic_2020,deangelis_thermal_2019,guerboub_impact_2023}.

We note that the DFT parameters used to generate the phononDB-PBE database are similar but not always exactly equal to those used for DFT calculations in the MP database. In the regime of computational convergence,
variations of the DFT parameters at fixed functional are expected to yield unimportant effects on the conductivity. In particular, Ref.~\cite{jain_effect_2015} shows that for the prototypical case of silicon, computing force constants using DFT parameters within the range of `computationally converged values' found in the literature produces conductivity variations that are within 7 $\%$, and also shows that in ideal cases varying DFT parameters within the domain of computational convergence should yield conductivity variations smaller than 2 $\%$. As such, the significant discrepancies shown in Fig.~\ref{fig:errors_ktot} are expected to be not caused by possible small differences in the DFT parameters.

The computational details on the mp-fMLPs can be found in the following references: M3GNet\cite{chen_universal_2022},  CHGNet\cite{deng_chgnet_2023}, MACE-MP-0\cite{batatia_foundation_2024}, SevenNet\cite{park_scalable_2024}, and ORB-v1-MPtraj\cite{orb_github}. M3GNet was used through the \texttt{matgl}\cite{ko_2023_8025189_matgl} (version 1.1.2) package with the {M3GNet-MP-2021.2.8-PES} model. The CHGNet version 0.3.8 was used. The MACE-MP-0 {2024-01-07-mace-128-L2} model was run through LAMMPS\cite{plimpton_lammps_1995}. SevenNet was used through the \texttt{sevenn} package (version 0.9.2) with the {SevenNet-0\_11July2024} model. ORB-v1-MPtraj was used through the \texttt{orb-models} package (version 0.3.0, commit d142854) with the {ORB-v1-MPtraj-only} model.The details on the violin plots presented in Fig.~\ref{fig:violin} can be found in Ref.~\cite{hintze_violin_plots_1998}.\\

The ORB-v1-MPtraj, SevenNet, MACE, and CHGNet calculations were performed on an Nvidia Tesla A100 SXM4 80GB GPU, with the relaxation and force calculations for all 103 structures costing approximately $\sim 3$ GPU hours per model. M3GNet was executed on AMD EPYC 7702 CPUs, with relaxation and force calculations costing around $\sim 30$ CPU hours. Thermal conductivity calculations for all 103 structures required about $\sim 50$ CPU hours per model, and similar resources were needed for evaluating thermal conductivity from the DFT force constants. Overall, the analysis of these five models required approximately $\sim 12$ GPU hours and $\sim 330$ CPU hours.\\

\noindent
\textbf{Details on fine-tuning}\\
In this section we discuss how the size of the dataset and the number of fine-tuning epochs influence the training error for energies, forces, and stresses in LiBr. Results are shown for different finite-tuning datasets in Fig.~\ref{fig:finetuning_methods}.
The three datasets used for fine-tuning contain DFT single-point calculations in $4\times4\times4$ supercells of LiBr (same size used for the phonon calculation in the phononDB-PBE\cite{seko_prediction_2015,phono3py} database) perturbed from equilibrium with atomic rattling and in some cases also volume variations from 0.25\% to 5\%.
All DFT calculations are performed employing the Perdew, Burke, and Ernzerhof (PBE) exchange-correlation functional \cite{perdew_generalized_1996}, and the projector augmented wave (PAW) formalism for electronic minimization, implemented within the VASP (Vienna ab initio simulation package) code~\cite{kresse_ab_1993, kresse_efficiency_1996}. 
Following Ref.~\cite{seko_prediction_2015}, we use a 600 eV plane-wave-basis cutoff (20\% higher than the default one), and the Brillouin zone is sampled at the Gamma point.
Using these DFT settings, the LiBr structure in the phononDB-PBE database shows a residual negative stress of 960 Bar. Therefore, to make the most accurate possible comparison between fine-tuned MACE-MP-0 and reference DFT data, we relaxed the fine-tuned MACE-MP-0 structure exactly to the same residual stress present in the reference DFT-PBE data.\\

The MACE-MP-0 {2024-01-07-mace-128-L2} model \url{https://github.com/ACEsuit/mace-mp/commits/mace_mp_0} was fine-tuned using the \texttt{mace\_run\_train} script in the main branch of \url{https://github.com/ACEsuit/mace}. 
The training started from the last checkpoint of MACE-MP-0 L2. The three datasets discussed above were used to perform three independent trainings. 
For trainings on the 1-frame and 3-frame datasets we employed a batch size of 1, and for trainings on the 20-frame dataset we used a batch of size 4.
Since our fine-tuning datasets were computed with the DFT-PBE functional, which was also employed to generate the MP database\cite{materialsproject} used to train MACE-MP-0, in the fine-tuning we kept the E0s fixed at their original MACE-MP-0 value. 
Representative examples of training scripts are available at: \url{https://github.com/MSimoncelli/fMLP_conductivity.git}. The fine-tuning was performed on Nvidia Tesla A100 SXM4 80GB GPU.\\  

\begin{figure}[htb]
  \centering
  \includegraphics[width=\columnwidth]{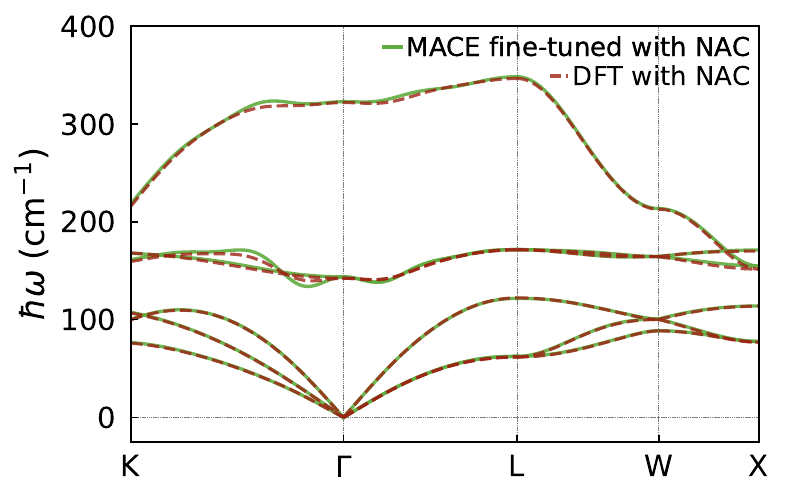}
  \caption{\textbf{Hybrid fMLP-DFT phonon band structure including non-analytical correction (NAC) term.} The hybrid fMLP-DFT approach shows good agreement with  reference band structure obtained from DFT. As an example, we used the MACE-MP-0 model fine-tuned on 3 frames trained over 200 epochs.}
\label{fig:fMLP-NAC}
\end{figure}

\noindent
\textbf{Hybrid fMLP-DFT approach to include non-analytical correction term.}\\
We propose a hybrid fMLP-DFT method to incorporate the non-analytical correction (NAC) term\cite{gonze_dynamical_1997} computed from density functional perturbation theory (DFPT)\cite{ghosez_dynamical_1998,baroni_phonons_2001} into phonon band structures predicted by fMLPs. 
To limit the computational cost, we neglect changes in the Born effective charges that may originate from small structural differences between the DFT-relaxed and fMLP-relaxed primitive cell; the accuracy of this approximation will be discussed a posteriori. 
We demonstrate the effectiveness of this hybrid approach in LiBr, using MACE-MP-0 potential fine-tuned on 3 frames for over 200 epochs, and combining them with DFT-quality Born effective charges from the phononDB-PBE database \cite{seko_prediction_2015}. As shown in Fig.~\ref{fig:fMLP-NAC}, the hybrid approach with fine-tuned fMLPs and DFT Born effective charges is in good agreement with the DFT reference calculation, confirming the accuracy of our hybrid fMLP-DFT approach, and, a posteriori, the validity of the approximations performed therein.\\

\begin{figure*}[htb]
  \centering
  \includegraphics[width=\textwidth]{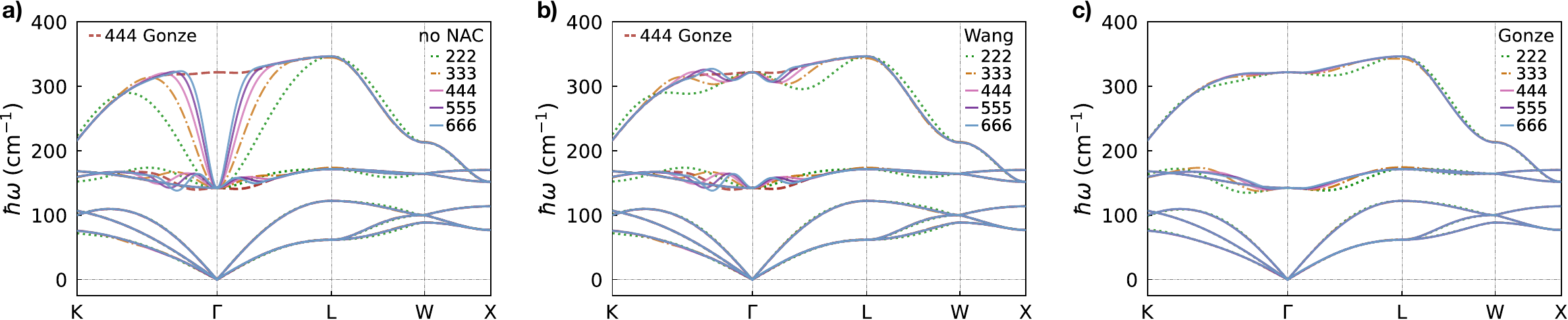}
  \caption{
  \textbf{Numerical performance of different approaches to account for the effect of the long-range NAC term on phonon dispersion.}
  {We show the phonon band structure of LiBr without non-analytical correction (NAC) (a), considering NAC using Wang's\cite{Wang_2010} approach (b) or Gonze's\cite{gonze_dynamical_1997} method (c). These three cases are compared in simulations that use supercells of different sizes.} }
\label{fig:nac-convergence}
\end{figure*}

\noindent
\textbf{Influence of Born effective charges on phonon dispersion, conductivity, and Grüneisen parameter.}\\
The long-range non-analytical correction (NAC) term to the phonon band structure can be described using two methodologies: (i) the one proposed by Gonze and Lee\cite{gonze_dynamical_1997} using density-functional perturbation theory in reciprocal space, or using the mixed-space approach by Wang et al.\cite{Wang_2010}. To evaluate the numerical difference between these two methods, we employed both of them to calculate the phonon band structures of LiBr, also checking how these two methods numerically converge with the size of the supercell used (all the other parameters were kept fixed and consistent parameters to those of the phononDB-PBE dataset). Fig.~\ref{fig:nac-convergence} shows that increasing the supercell size from $2\times 2\times2$ to $6\times 6\times6$ yields a strong effect in the band structures that do not include the NAC term (this was expected and is explained in detail in Ref.~\cite{Wang_2010}), and a noticeable effect in the band structure computed accounting for NAC with Wang's method. In contrast, Gonze's method shows significantly smaller variations with respect to varying supercell size, displaying acceptable numerical convergence already in supercells of size $4\times 4\times4$. Therefore, in the plots of this work we account for NAC using the approach by Gonze and Lee\cite{born_dynamical_1998}.

\begin{figure}[htb]
  \centering
  \includegraphics[width=0.9\columnwidth]{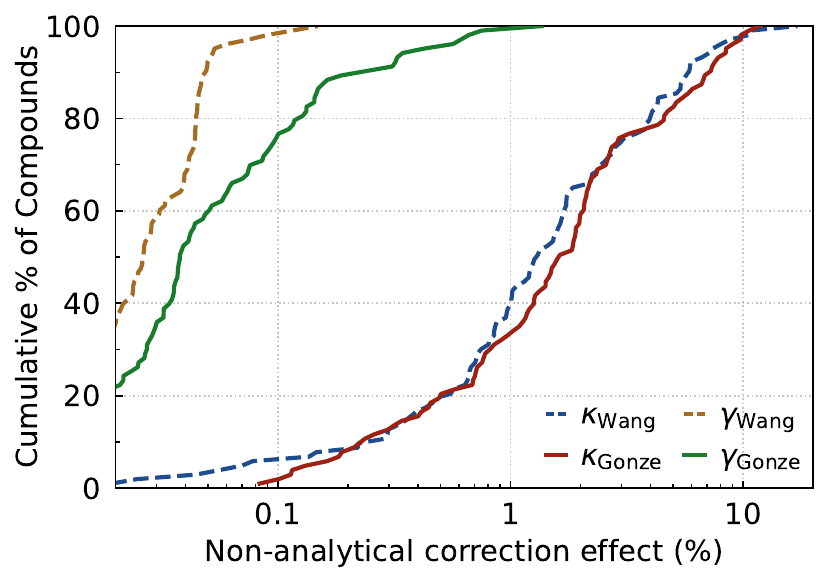}
  \caption{\textbf{Cumulative distribution  of non-analytical correction term's effect on conductivity and Grüneisen parameter,} and comparison between the methods by Gonze and Lee\cite{gonze_dynamical_1997},  and Wang \textit{et al.}\cite{Wang_2010}}
\label{fig:nac_cdf}
\end{figure}

In Fig.\ref{fig:nac_cdf} we show how thermal conductivity and Grüneisen parameter change considering or not the NAC term, reporting the cumulative distribution of the relative variation for all the 103 materials analyzed.
Overall, the materials analyzed do not show a significant dependence on considering or not the NAC term. 
In particular, the maximum variation for the Grüneisen parameter is $1.4\%$ in rocksalt AgCl, while for conductivity is $11.8\%$ in wurtzite CdSe. The cumulative distribution of the relative NAC effect on conductivity and Grüneisen parameter is presented in Fig.\ref{fig:nac_cdf}. \\

\noindent
\textbf{fMLPs to describe thermal expansion 
and bulk modulus within the quasi harmonic approximation.} \\
To demonstrate that fine-tuned fMLPs can accurately describe not only thermal conductivity and Grüneisen parameter, but also other technologically relevant thermomechanical properties, we employed the fine-tuned fMLP to compute the volumetric thermal expansion and temperature-dependent bulk modulus of LiBr using free energy minimization within the quasi-harmonic approximation (QHA), as implemented in the \texttt{phonopy} package\cite{togo_first-principles_2023}.
The capability of fine-tuned fMLPs to describe multiple different thermomechanical properties provides an advantage in terms of computational cost over the following established simulation approaches: (i) compressive sensing\cite{eriksson_hiphive_2019,di_lucente_crossover_2023}, which require separate DFT datasets at contracted and expanded unit-cell volumes to describe both conductivity and Grüneisen parameter; (ii) traditional non-foundation machine learning potentials, which rely on much larger training datasets to reach accuracy comparable to fine-tuned fMLPs \cite{harper_vibrational_2023}. 

\begin{figure}[htb]
  \centering
  \includegraphics[width=\columnwidth]{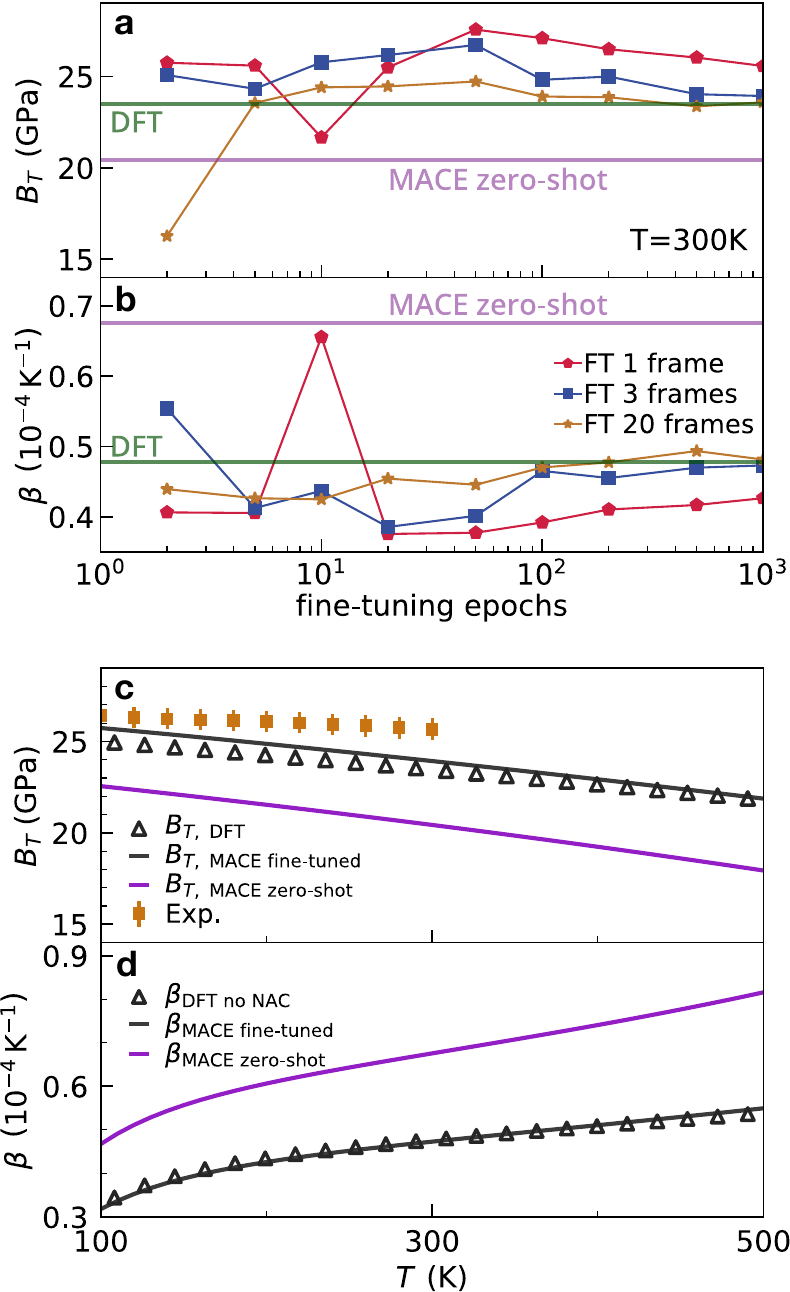}
  \caption{\textbf{Volumetric thermal expansion 
  and temperature-dependent bulk modulus 
  of LiBr within the quasi-harmonic approximation (QHA) from free energy minimization.} Fine-tuning the MACE-MP-0 model on three supercell frames for 500 epochs yields volumetric thermal expansion coefficients in close agreement with DFT reference values across the full temperature range considered.}
\label{fig:volume_expansion}
\end{figure}

The DFT reference data were generated with the same computational settings as the fine-tuning dataset, applying volume variations of $\pm1\%$, $\pm2\%$, $\pm3\%$, $\pm4\%$, and $\pm5\%$, alongside $0.03\AA$ displacements for interatomic force constants and the default Vinet equation of state. The MACE-MP-0 model was evaluated both in a zero-shot setting and after fine-tuning on three frames over 500 epochs. Fig.~\ref{fig:volume_expansion} presents the resulting volumetric thermal expansion coefficients 
and temperature-dependent bulk modulus compared to the DFT reference. The fine-tuned MACE-MP-0 model closely reproduces the DFT results for thermal expansion within $1.5\%$ and for bulk modulus within $3\%$ across the full temperature range, whereas the zero-shot model significantly overestimates the thermal expansion coefficient
and underestimates the bulk modulus. Experimental bulk modulus values\cite{marshall_elastic_1969} were plotted for comparison. We note that the bulk modulus exhibits a clear temperature dependence over the plotted range, emphasizing that thermal expansion cannot be divided into independent single-phonon contributions. This is due to the temperature dependence of both the bulk modulus\cite{wallace_thermodynamics_1972} and the Grüneisen parameter, the phonon-dependent terms appearing in the expression for the volumetric thermal expansion, $\beta= \frac{\gamma C}{B_T \mathcal{V}}$, as discussed in the main text.\\

\noindent
\textbf{Simultaneous fine-tuning for multiple materials}\\ 
To demonstrate the applicability and generalizability of fine-tuning fMLPs for thermomechanical properties, we simultaneously fine-tuned models for three distinct materials: wurtzite AlP, zincblende BN, and rocksalt LiBr, targeting both thermal conductivity and the Grüneisen parameter. For each material, three structural configurations were used: a rattled supercell (atomic displacements sampled with a standard deviation of $0.1\ \text{\AA}$), and two unit cells uniformly scaled by $\pm1\%$ in volume. This setup enhances the accuracy of stress predictions and yields relaxed structures that more closely align with DFT references.

\begin{figure*}[htb]
  \centering
  \includegraphics[width=\textwidth]{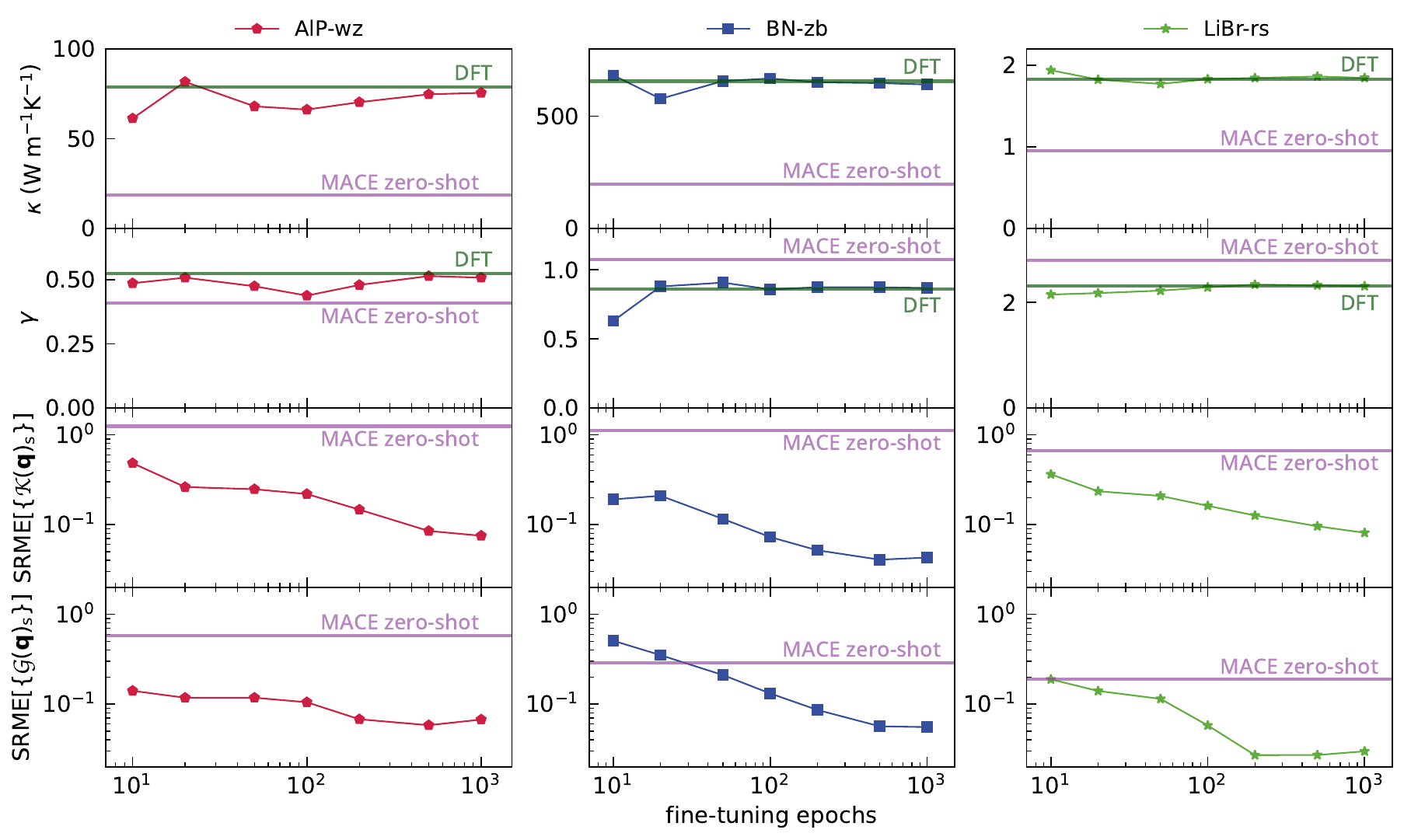}
  \caption{\textbf{Improving fMLP accuracy via simultaneous fine-tuning of three materials.} The thermal conductivity, Grüneisen parameter and the corresponding SRME values for wurtzite AlP, zincblende BN and rocksalt LiBr after fine-tuning simultaneously on 3 frames (1 rattled supercell and 2 isotropically scaled unit cell) against the number of epochs. }
\label{fig:finetuning_3mat}
\end{figure*}
Fig.~\ref{fig:finetuning_3mat} shows that the fMLP effectively learns across multiple materials in parallel, improving accuracy while reducing the per-observable training cost. Fine-tuning was performed using the same settings as for LiBr, on a Nvidia Tesla A100 SXM4 80GB GPU.\\

\noindent
\textbf{Data availability.}
The Matbench Discovery results are available at \url{https://matbench-discovery.materialsproject.org/}. The phononDB-PBE dataset including the displacements generated by \texttt{phono3py} and the corresponding force sets are available \url{https://github.com/atztogo/phonondb} \cite{phono3py,seko_prediction_2015}. 
The dataset needed to reproduce the findings of this study is available on the Materials Cloud Archive\cite{talirz_materials_2020}: \url{https://doi.org/10.24435/materialscloud:2d-4b}; such dataset contains thermal conductivity resolved for each phonon mode, as well as data for fine-tuning.\\

\noindent
\textbf{Code availability.}
The \text{phono3py} and \texttt{phonopy} packages are available at \url{https://github.com/phonopy/}; the \texttt{ase} package is available at \url{https://gitlab.com/ase/ase};
the \texttt{spglib} package is available at \url{https://github.com/spglib/spglib}. \texttt{matgl} containing the M3GNet model is available at \url{https://github.com/materialsvirtuallab/matgl}; the CHGNet model and package is available at \url{https://github.com/CederGroupHub/chgnet}; the SevenNet model and package is available at \url{https://github.com/MDIL-SNU/SevenNet}; the ORB-v1-MPtraj model and package is available at \url{https://github.com/orbital-materials/orb-models}. The MACE used through LAMMPS\cite{plimpton_lammps_1995} is available at \url{https://github.com/ACEsuit/lammps}, and the MACE-MP-0 model is available at \url{https://github.com/ACEsuit/mace-mp}. The Hiphive package is available at \url{https://gitlab.com/materials-modeling/hiphive}.
The source code used to reproduce the results presented in this work is publicly available. The \texttt{k\_SRME} code, which enables the thermal conductivity benchmarks, is accessible at \url{https://github.com/MPA2suite/k_SRME}. The \texttt{gamma\_SRME} code, used for computing Grüneisen parameter benchmarks, is available at \url{https://github.com/MPA2suite/gamma_SRME}.

\vspace*{2mm}
\noindent
\textbf{Acknowledgements.}
The computational resources were provided by: (i) the Kelvin2 HPC platform at the NI-HPC Centre (funded by EPSRC and jointly managed by Queen’s University Belfast and Ulster University); (ii) the UK National Supercomputing Service ARCHER2, for which access was obtained via the UKCP consortium and funded by EPSRC [EP/X035891/1].
M. S. acknowledges support from Gonville and Caius College. P.A. acknowledges support from SNSF through Post.Doc mobility fellowship P500PN\_206693. 
We thank William C. Witt, Ilyes Batatia, Andrew D. P. Smith, and Janosh Riebesell for the useful discussions. We gratefully acknowledge Atsushi Togo for having provided harmonic and anharmonic force constants computed using density functional theory (PBE functional) from the phononDB-PBE database~\cite{phono3py,seko_prediction_2015}.\\

\noindent
{\textbf{Author contributions.}
M.S. conceived and supervised the project. B.P. developed the automated computational framework and performed the numerical calculations with inputs from P.A., G.C., and M.S. The results were analyzed and organized by B.P. and M.S., with inputs from P.A. All authors contributed to discussing the results, writing and editing the manuscript.}\\

\noindent
\textbf{Competing interests.}
C.G. has equity interest of Symmetric Group LLP that licenses force fields commercially and also in \AA ngstrom AI. The other authors declare that they have no competing interest.\\

\noindent
\textbf{Inclusion \& ethics statement.}
All the authors of this study have fulfilled the criteria for authorship required by Nature Portfolio journals, and their participation was essential for the design and implementation of the study. Roles and responsibilities were discussed and agreed among collaborators. Local and regional research relevant to our study has been considered in the citations.


\begin{thebibliography}{137}%
\makeatletter
\providecommand \@ifxundefined [1]{%
 \@ifx{#1\undefined}
}%
\providecommand \@ifnum [1]{%
 \ifnum #1\expandafter \@firstoftwo
 \else \expandafter \@secondoftwo
 \fi
}%
\providecommand \@ifx [1]{%
 \ifx #1\expandafter \@firstoftwo
 \else \expandafter \@secondoftwo
 \fi
}%
\providecommand \natexlab [1]{#1}%
\providecommand \enquote  [1]{``#1''}%
\providecommand \bibnamefont  [1]{#1}%
\providecommand \bibfnamefont [1]{#1}%
\providecommand \citenamefont [1]{#1}%
\providecommand \href@noop [0]{\@secondoftwo}%
\providecommand \href [0]{\begingroup \@sanitize@url \@href}%
\providecommand \@href[1]{\@@startlink{#1}\@@href}%
\providecommand \@@href[1]{\endgroup#1\@@endlink}%
\providecommand \@sanitize@url [0]{\catcode `\\12\catcode `\$12\catcode `\&12\catcode `\#12\catcode `\^12\catcode `\_12\catcode `\%12\relax}%
\providecommand \@@startlink[1]{}%
\providecommand \@@endlink[0]{}%
\providecommand \url  [0]{\begingroup\@sanitize@url \@url }%
\providecommand \@url [1]{\endgroup\@href {#1}{\urlprefix }}%
\providecommand \urlprefix  [0]{URL }%
\providecommand \Eprint [0]{\href }%
\providecommand \doibase [0]{https://doi.org/}%
\providecommand \selectlanguage [0]{\@gobble}%
\providecommand \bibinfo  [0]{\@secondoftwo}%
\providecommand \bibfield  [0]{\@secondoftwo}%
\providecommand \translation [1]{[#1]}%
\providecommand \BibitemOpen [0]{}%
\providecommand \bibitemStop [0]{}%
\providecommand \bibitemNoStop [0]{.\EOS\space}%
\providecommand \EOS [0]{\spacefactor3000\relax}%
\providecommand \BibitemShut  [1]{\csname bibitem#1\endcsname}%
\let\auto@bib@innerbib\@empty
\bibitem [{\citenamefont {Blank}\ \emph {et~al.}(1995)\citenamefont {Blank}, \citenamefont {Brown}, \citenamefont {Calhoun},\ and\ \citenamefont {Doren}}]{blank_neural_1995}%
  \BibitemOpen
  \bibfield  {author} {\bibinfo {author} {\bibfnamefont {T.~B.}\ \bibnamefont {Blank}}, \bibinfo {author} {\bibfnamefont {S.~D.}\ \bibnamefont {Brown}}, \bibinfo {author} {\bibfnamefont {A.~W.}\ \bibnamefont {Calhoun}},\ and\ \bibinfo {author} {\bibfnamefont {D.~J.}\ \bibnamefont {Doren}},\ }\href {https://doi.org/10.1063/1.469597} {\bibfield  {journal} {\bibinfo  {journal} {The Journal of Chemical Physics}\ }\textbf {\bibinfo {volume} {103}},\ \bibinfo {pages} {4129} (\bibinfo {year} {1995})}\BibitemShut {NoStop}%
\bibitem [{\citenamefont {Behler}\ and\ \citenamefont {Parrinello}(2007)}]{behler_generalized_2007}%
  \BibitemOpen
  \bibfield  {author} {\bibinfo {author} {\bibfnamefont {J.}~\bibnamefont {Behler}}\ and\ \bibinfo {author} {\bibfnamefont {M.}~\bibnamefont {Parrinello}},\ }\href {https://doi.org/10.1103/PhysRevLett.98.146401} {\bibfield  {journal} {\bibinfo  {journal} {Physical Review Letters}\ }\textbf {\bibinfo {volume} {98}},\ \bibinfo {pages} {146401} (\bibinfo {year} {2007})}\BibitemShut {NoStop}%
\bibitem [{\citenamefont {Bartók}\ \emph {et~al.}(2010)\citenamefont {Bartók}, \citenamefont {Payne}, \citenamefont {Kondor},\ and\ \citenamefont {Csányi}}]{bartok_gaussian_2010}%
  \BibitemOpen
  \bibfield  {author} {\bibinfo {author} {\bibfnamefont {A.~P.}\ \bibnamefont {Bartók}}, \bibinfo {author} {\bibfnamefont {M.~C.}\ \bibnamefont {Payne}}, \bibinfo {author} {\bibfnamefont {R.}~\bibnamefont {Kondor}},\ and\ \bibinfo {author} {\bibfnamefont {G.}~\bibnamefont {Csányi}},\ }\href {https://doi.org/10.1103/PhysRevLett.104.136403} {\bibfield  {journal} {\bibinfo  {journal} {Physical Review Letters}\ }\textbf {\bibinfo {volume} {104}},\ \bibinfo {pages} {136403} (\bibinfo {year} {2010})}\BibitemShut {NoStop}%
\bibitem [{\citenamefont {Rupp}\ \emph {et~al.}(2012)\citenamefont {Rupp}, \citenamefont {Tkatchenko}, \citenamefont {Müller},\ and\ \citenamefont {von Lilienfeld}}]{rupp_fast_2012}%
  \BibitemOpen
  \bibfield  {author} {\bibinfo {author} {\bibfnamefont {M.}~\bibnamefont {Rupp}}, \bibinfo {author} {\bibfnamefont {A.}~\bibnamefont {Tkatchenko}}, \bibinfo {author} {\bibfnamefont {K.-R.}\ \bibnamefont {Müller}},\ and\ \bibinfo {author} {\bibfnamefont {O.~A.}\ \bibnamefont {von Lilienfeld}},\ }\href {https://doi.org/10.1103/PhysRevLett.108.058301} {\bibfield  {journal} {\bibinfo  {journal} {Physical Review Letters}\ }\textbf {\bibinfo {volume} {108}},\ \bibinfo {pages} {058301} (\bibinfo {year} {2012})}\BibitemShut {NoStop}%
\bibitem [{\citenamefont {Zhang}\ \emph {et~al.}(2018)\citenamefont {Zhang}, \citenamefont {Han}, \citenamefont {Wang}, \citenamefont {Car},\ and\ \citenamefont {E}}]{zhang_deep_2018}%
  \BibitemOpen
  \bibfield  {author} {\bibinfo {author} {\bibfnamefont {L.}~\bibnamefont {Zhang}}, \bibinfo {author} {\bibfnamefont {J.}~\bibnamefont {Han}}, \bibinfo {author} {\bibfnamefont {H.}~\bibnamefont {Wang}}, \bibinfo {author} {\bibfnamefont {R.}~\bibnamefont {Car}},\ and\ \bibinfo {author} {\bibfnamefont {W.}~\bibnamefont {E}},\ }\href {https://doi.org/10.1103/PhysRevLett.120.143001} {\bibfield  {journal} {\bibinfo  {journal} {Physical Review Letters}\ }\textbf {\bibinfo {volume} {120}},\ \bibinfo {pages} {143001} (\bibinfo {year} {2018})}\BibitemShut {NoStop}%
\bibitem [{\citenamefont {Drautz}(2019)}]{drautz_atomic_2019}%
  \BibitemOpen
  \bibfield  {author} {\bibinfo {author} {\bibfnamefont {R.}~\bibnamefont {Drautz}},\ }\href {https://doi.org/10.1103/PhysRevB.99.014104} {\bibfield  {journal} {\bibinfo  {journal} {Physical Review B}\ }\textbf {\bibinfo {volume} {99}},\ \bibinfo {pages} {014104} (\bibinfo {year} {2019})}\BibitemShut {NoStop}%
\bibitem [{\citenamefont {Seko}\ \emph {et~al.}(2019)\citenamefont {Seko}, \citenamefont {Togo},\ and\ \citenamefont {Tanaka}}]{seko_group-theoretical_2019}%
  \BibitemOpen
  \bibfield  {author} {\bibinfo {author} {\bibfnamefont {A.}~\bibnamefont {Seko}}, \bibinfo {author} {\bibfnamefont {A.}~\bibnamefont {Togo}},\ and\ \bibinfo {author} {\bibfnamefont {I.}~\bibnamefont {Tanaka}},\ }\href {https://doi.org/10.1103/PhysRevB.99.214108} {\bibfield  {journal} {\bibinfo  {journal} {Physical Review B}\ }\textbf {\bibinfo {volume} {99}},\ \bibinfo {pages} {214108} (\bibinfo {year} {2019})}\BibitemShut {NoStop}%
\bibitem [{\citenamefont {Deringer}\ \emph {et~al.}(2021)\citenamefont {Deringer}, \citenamefont {Bartók}, \citenamefont {Bernstein}, \citenamefont {Wilkins}, \citenamefont {Ceriotti},\ and\ \citenamefont {Csányi}}]{deringer_gaussian_2021}%
  \BibitemOpen
  \bibfield  {author} {\bibinfo {author} {\bibfnamefont {V.~L.}\ \bibnamefont {Deringer}}, \bibinfo {author} {\bibfnamefont {A.~P.}\ \bibnamefont {Bartók}}, \bibinfo {author} {\bibfnamefont {N.}~\bibnamefont {Bernstein}}, \bibinfo {author} {\bibfnamefont {D.~M.}\ \bibnamefont {Wilkins}}, \bibinfo {author} {\bibfnamefont {M.}~\bibnamefont {Ceriotti}},\ and\ \bibinfo {author} {\bibfnamefont {G.}~\bibnamefont {Csányi}},\ }\href {https://doi.org/10.1021/acs.chemrev.1c00022} {\bibfield  {journal} {\bibinfo  {journal} {Chemical Reviews}\ }\textbf {\bibinfo {volume} {121}},\ \bibinfo {pages} {10073} (\bibinfo {year} {2021})}\BibitemShut {NoStop}%
\bibitem [{\citenamefont {Batzner}\ \emph {et~al.}(2022)\citenamefont {Batzner}, \citenamefont {Musaelian}, \citenamefont {Sun}, \citenamefont {Geiger}, \citenamefont {Mailoa}, \citenamefont {Kornbluth}, \citenamefont {Molinari}, \citenamefont {Smidt},\ and\ \citenamefont {Kozinsky}}]{batzner_e3-equivariant_2022}%
  \BibitemOpen
  \bibfield  {author} {\bibinfo {author} {\bibfnamefont {S.}~\bibnamefont {Batzner}}, \bibinfo {author} {\bibfnamefont {A.}~\bibnamefont {Musaelian}}, \bibinfo {author} {\bibfnamefont {L.}~\bibnamefont {Sun}}, \bibinfo {author} {\bibfnamefont {M.}~\bibnamefont {Geiger}}, \bibinfo {author} {\bibfnamefont {J.~P.}\ \bibnamefont {Mailoa}}, \bibinfo {author} {\bibfnamefont {M.}~\bibnamefont {Kornbluth}}, \bibinfo {author} {\bibfnamefont {N.}~\bibnamefont {Molinari}}, \bibinfo {author} {\bibfnamefont {T.~E.}\ \bibnamefont {Smidt}},\ and\ \bibinfo {author} {\bibfnamefont {B.}~\bibnamefont {Kozinsky}},\ }\href {https://doi.org/10.1038/s41467-022-29939-5} {\bibfield  {journal} {\bibinfo  {journal} {Nature Communications}\ }\textbf {\bibinfo {volume} {13}},\ \bibinfo {pages} {2453} (\bibinfo {year} {2022})}\BibitemShut {NoStop}%
\bibitem [{\citenamefont {Kocer}\ \emph {et~al.}(2022)\citenamefont {Kocer}, \citenamefont {Ko},\ and\ \citenamefont {Behler}}]{kocer_neural_2022}%
  \BibitemOpen
  \bibfield  {author} {\bibinfo {author} {\bibfnamefont {E.}~\bibnamefont {Kocer}}, \bibinfo {author} {\bibfnamefont {T.~W.}\ \bibnamefont {Ko}},\ and\ \bibinfo {author} {\bibfnamefont {J.}~\bibnamefont {Behler}},\ }\href {https://doi.org/10.1146/annurev-physchem-082720-034254} {\bibfield  {journal} {\bibinfo  {journal} {Annual Review of Physical Chemistry}\ }\textbf {\bibinfo {volume} {73}},\ \bibinfo {pages} {163} (\bibinfo {year} {2022})}\BibitemShut {NoStop}%
\bibitem [{\citenamefont {Qian}\ \emph {et~al.}(2021)\citenamefont {Qian}, \citenamefont {Zhou},\ and\ \citenamefont {Chen}}]{qian_phonon-engineered_2021}%
  \BibitemOpen
  \bibfield  {author} {\bibinfo {author} {\bibfnamefont {X.}~\bibnamefont {Qian}}, \bibinfo {author} {\bibfnamefont {J.}~\bibnamefont {Zhou}},\ and\ \bibinfo {author} {\bibfnamefont {G.}~\bibnamefont {Chen}},\ }\href {https://doi.org/10.1038/s41563-021-00918-3} {\bibfield  {journal} {\bibinfo  {journal} {Nature Materials}\ }\textbf {\bibinfo {volume} {20}},\ \bibinfo {pages} {1188} (\bibinfo {year} {2021})}\BibitemShut {NoStop}%
\bibitem [{\citenamefont {Nataf}\ \emph {et~al.}(2024)\citenamefont {Nataf}, \citenamefont {Volz}, \citenamefont {Ordonez-Miranda}, \citenamefont {Íñiguez González}, \citenamefont {Rurali},\ and\ \citenamefont {Dkhil}}]{nataf_using_2024}%
  \BibitemOpen
  \bibfield  {author} {\bibinfo {author} {\bibfnamefont {G.~F.}\ \bibnamefont {Nataf}}, \bibinfo {author} {\bibfnamefont {S.}~\bibnamefont {Volz}}, \bibinfo {author} {\bibfnamefont {J.}~\bibnamefont {Ordonez-Miranda}}, \bibinfo {author} {\bibfnamefont {J.}~\bibnamefont {Íñiguez González}}, \bibinfo {author} {\bibfnamefont {R.}~\bibnamefont {Rurali}},\ and\ \bibinfo {author} {\bibfnamefont {B.}~\bibnamefont {Dkhil}},\ }\href {https://doi.org/10.1038/s41578-024-00690-1} {\bibfield  {journal} {\bibinfo  {journal} {Nature Reviews Materials}\ ,\ \bibinfo {pages} {1}} (\bibinfo {year} {2024})}\BibitemShut {NoStop}%
\bibitem [{\citenamefont {Okabe}\ \emph {et~al.}(2024)\citenamefont {Okabe}, \citenamefont {Chotrattanapituk}, \citenamefont {Boonkird}, \citenamefont {Andrejevic}, \citenamefont {Fu}, \citenamefont {Jaakkola}, \citenamefont {Song}, \citenamefont {Nguyen}, \citenamefont {Drucker}, \citenamefont {Mu}, \citenamefont {Wang}, \citenamefont {Liao}, \citenamefont {Cheng},\ and\ \citenamefont {Li}}]{okabe_virtual_2024}%
  \BibitemOpen
  \bibfield  {author} {\bibinfo {author} {\bibfnamefont {R.}~\bibnamefont {Okabe}}, \bibinfo {author} {\bibfnamefont {A.}~\bibnamefont {Chotrattanapituk}}, \bibinfo {author} {\bibfnamefont {A.}~\bibnamefont {Boonkird}}, \bibinfo {author} {\bibfnamefont {N.}~\bibnamefont {Andrejevic}}, \bibinfo {author} {\bibfnamefont {X.}~\bibnamefont {Fu}}, \bibinfo {author} {\bibfnamefont {T.~S.}\ \bibnamefont {Jaakkola}}, \bibinfo {author} {\bibfnamefont {Q.}~\bibnamefont {Song}}, \bibinfo {author} {\bibfnamefont {T.}~\bibnamefont {Nguyen}}, \bibinfo {author} {\bibfnamefont {N.}~\bibnamefont {Drucker}}, \bibinfo {author} {\bibfnamefont {S.}~\bibnamefont {Mu}}, \bibinfo {author} {\bibfnamefont {Y.}~\bibnamefont {Wang}}, \bibinfo {author} {\bibfnamefont {B.}~\bibnamefont {Liao}}, \bibinfo {author} {\bibfnamefont {Y.}~\bibnamefont {Cheng}},\ and\ \bibinfo {author} {\bibfnamefont {M.}~\bibnamefont {Li}},\ }\href {https://www.nature.com/articles/s43588-024-00661-0} {\bibfield  {journal} {\bibinfo  {journal} {Nature Computational Science}\ }\textbf {\bibinfo {volume} {4}} (\bibinfo {year} {2024})}\BibitemShut {NoStop}%
\bibitem [{\citenamefont {Guo}\ \emph {et~al.}(2023)\citenamefont {Guo}, \citenamefont {Roy~Chowdhury}, \citenamefont {Han}, \citenamefont {Sun}, \citenamefont {Feng}, \citenamefont {Lin},\ and\ \citenamefont {Ruan}}]{Guo2023-pm}%
  \BibitemOpen
  \bibfield  {author} {\bibinfo {author} {\bibfnamefont {Z.}~\bibnamefont {Guo}}, \bibinfo {author} {\bibfnamefont {P.}~\bibnamefont {Roy~Chowdhury}}, \bibinfo {author} {\bibfnamefont {Z.}~\bibnamefont {Han}}, \bibinfo {author} {\bibfnamefont {Y.}~\bibnamefont {Sun}}, \bibinfo {author} {\bibfnamefont {D.}~\bibnamefont {Feng}}, \bibinfo {author} {\bibfnamefont {G.}~\bibnamefont {Lin}},\ and\ \bibinfo {author} {\bibfnamefont {X.}~\bibnamefont {Ruan}},\ }\href {https://doi.org/10.1038/s41524-023-01020-9} {\bibfield  {journal} {\bibinfo  {journal} {npj Computational Materials}\ }\textbf {\bibinfo {volume} {9}},\ \bibinfo {pages} {1} (\bibinfo {year} {2023})}\BibitemShut {NoStop}%
\bibitem [{\citenamefont {Ojih}\ \emph {et~al.}(2024)\citenamefont {Ojih}, \citenamefont {Al-Fahdi}, \citenamefont {Yao}, \citenamefont {Hu},\ and\ \citenamefont {Hu}}]{ojih2024graph}%
  \BibitemOpen
  \bibfield  {author} {\bibinfo {author} {\bibfnamefont {J.}~\bibnamefont {Ojih}}, \bibinfo {author} {\bibfnamefont {M.}~\bibnamefont {Al-Fahdi}}, \bibinfo {author} {\bibfnamefont {Y.}~\bibnamefont {Yao}}, \bibinfo {author} {\bibfnamefont {J.}~\bibnamefont {Hu}},\ and\ \bibinfo {author} {\bibfnamefont {M.}~\bibnamefont {Hu}},\ }\href {https://doi.org/10.1039/D3TA06190F} {\bibfield  {journal} {\bibinfo  {journal} {Journal of Materials Chemistry A}\ }\textbf {\bibinfo {volume} {12}},\ \bibinfo {pages} {8502} (\bibinfo {year} {2024})}\BibitemShut {NoStop}%
\bibitem [{\citenamefont {Rodriguez}\ \emph {et~al.}(2023)\citenamefont {Rodriguez}, \citenamefont {Lin}, \citenamefont {Yang}, \citenamefont {Al-Fahdi}, \citenamefont {Shen}, \citenamefont {Choudhary}, \citenamefont {Zhao}, \citenamefont {Hu}, \citenamefont {Cao}, \citenamefont {Zhang} \emph {et~al.}}]{rodriguez2023million}%
  \BibitemOpen
  \bibfield  {author} {\bibinfo {author} {\bibfnamefont {A.}~\bibnamefont {Rodriguez}}, \bibinfo {author} {\bibfnamefont {C.}~\bibnamefont {Lin}}, \bibinfo {author} {\bibfnamefont {H.}~\bibnamefont {Yang}}, \bibinfo {author} {\bibfnamefont {M.}~\bibnamefont {Al-Fahdi}}, \bibinfo {author} {\bibfnamefont {C.}~\bibnamefont {Shen}}, \bibinfo {author} {\bibfnamefont {K.}~\bibnamefont {Choudhary}}, \bibinfo {author} {\bibfnamefont {Y.}~\bibnamefont {Zhao}}, \bibinfo {author} {\bibfnamefont {J.}~\bibnamefont {Hu}}, \bibinfo {author} {\bibfnamefont {B.}~\bibnamefont {Cao}}, \bibinfo {author} {\bibfnamefont {H.}~\bibnamefont {Zhang}}, \emph {et~al.},\ }\href {https://doi.org/10.1038/s41524-023-00974-0} {\bibfield  {journal} {\bibinfo  {journal} {npj Computational Materials}\ }\textbf {\bibinfo {volume} {9}},\ \bibinfo {pages} {20} (\bibinfo {year} {2023})}\BibitemShut {NoStop}%
\bibitem [{\citenamefont {Srivastava}\ and\ \citenamefont {Jain}(2023)}]{srivastava2023end}%
  \BibitemOpen
  \bibfield  {author} {\bibinfo {author} {\bibfnamefont {Y.}~\bibnamefont {Srivastava}}\ and\ \bibinfo {author} {\bibfnamefont {A.}~\bibnamefont {Jain}},\ }\href {https://doi.org/10.1063/5.0183513} {\bibfield  {journal} {\bibinfo  {journal} {Journal of Applied Physics}\ }\textbf {\bibinfo {volume} {134}} (\bibinfo {year} {2023})}\BibitemShut {NoStop}%
\bibitem [{\citenamefont {Rose}\ \emph {et~al.}(2023)\citenamefont {Rose}, \citenamefont {Kozachinskiy}, \citenamefont {Rojas}, \citenamefont {Petrache},\ and\ \citenamefont {Barceló}}]{rose_three_2023}%
  \BibitemOpen
  \bibfield  {author} {\bibinfo {author} {\bibfnamefont {V.~D.}\ \bibnamefont {Rose}}, \bibinfo {author} {\bibfnamefont {A.}~\bibnamefont {Kozachinskiy}}, \bibinfo {author} {\bibfnamefont {C.}~\bibnamefont {Rojas}}, \bibinfo {author} {\bibfnamefont {M.}~\bibnamefont {Petrache}},\ and\ \bibinfo {author} {\bibfnamefont {P.}~\bibnamefont {Barceló}},\ }\href {http://arxiv.org/abs/2303.12853} {\bibinfo {title} {Three iterations of \$(d-1)\$-{WL} test distinguish non isometric clouds of \$d\$-dimensional points}} (\bibinfo {year} {2023})\BibitemShut {NoStop}%
\bibitem [{\citenamefont {Gilmer}\ \emph {et~al.}(2017)\citenamefont {Gilmer}, \citenamefont {Schoenholz}, \citenamefont {Riley}, \citenamefont {Vinyals},\ and\ \citenamefont {Dahl}}]{gilmer_neural_2017}%
  \BibitemOpen
  \bibfield  {author} {\bibinfo {author} {\bibfnamefont {J.}~\bibnamefont {Gilmer}}, \bibinfo {author} {\bibfnamefont {S.~S.}\ \bibnamefont {Schoenholz}}, \bibinfo {author} {\bibfnamefont {P.~F.}\ \bibnamefont {Riley}}, \bibinfo {author} {\bibfnamefont {O.}~\bibnamefont {Vinyals}},\ and\ \bibinfo {author} {\bibfnamefont {G.~E.}\ \bibnamefont {Dahl}},\ }in\ \href {https://proceedings.mlr.press/v70/gilmer17a.html} {\emph {\bibinfo {booktitle} {Proceedings of the 34th {International} {Conference} on {Machine} {Learning}}}}\ (\bibinfo  {publisher} {PMLR},\ \bibinfo {year} {2017})\ pp.\ \bibinfo {pages} {1263--1272}\BibitemShut {NoStop}%
\bibitem [{\citenamefont {Batatia}\ \emph {et~al.}(2023)\citenamefont {Batatia}, \citenamefont {Kovács}, \citenamefont {Simm}, \citenamefont {Ortner},\ and\ \citenamefont {Csányi}}]{batatia_mace_2023}%
  \BibitemOpen
  \bibfield  {author} {\bibinfo {author} {\bibfnamefont {I.}~\bibnamefont {Batatia}}, \bibinfo {author} {\bibfnamefont {D.~P.}\ \bibnamefont {Kovács}}, \bibinfo {author} {\bibfnamefont {G.~N.~C.}\ \bibnamefont {Simm}}, \bibinfo {author} {\bibfnamefont {C.}~\bibnamefont {Ortner}},\ and\ \bibinfo {author} {\bibfnamefont {G.}~\bibnamefont {Csányi}},\ }\href {http://arxiv.org/abs/2206.07697} {\bibinfo {title} {{MACE}: {Higher} {Order} {Equivariant} {Message} {Passing} {Neural} {Networks} for {Fast} and {Accurate} {Force} {Fields}}} (\bibinfo {year} {2023})\BibitemShut {NoStop}%
\bibitem [{\citenamefont {Chen}\ and\ \citenamefont {Ong}(2022)}]{chen_universal_2022}%
  \BibitemOpen
  \bibfield  {author} {\bibinfo {author} {\bibfnamefont {C.}~\bibnamefont {Chen}}\ and\ \bibinfo {author} {\bibfnamefont {S.~P.}\ \bibnamefont {Ong}},\ }\href {https://www.nature.com/articles/s43588-022-00349-3} {\bibfield  {journal} {\bibinfo  {journal} {Nature Computational Science}\ }\textbf {\bibinfo {volume} {2}} (\bibinfo {year} {2022})}\BibitemShut {NoStop}%
\bibitem [{\citenamefont {Deng}\ \emph {et~al.}(2023)\citenamefont {Deng}, \citenamefont {Zhong}, \citenamefont {Jun}, \citenamefont {Riebesell}, \citenamefont {Han}, \citenamefont {Bartel},\ and\ \citenamefont {Ceder}}]{deng_chgnet_2023}%
  \BibitemOpen
  \bibfield  {author} {\bibinfo {author} {\bibfnamefont {B.}~\bibnamefont {Deng}}, \bibinfo {author} {\bibfnamefont {P.}~\bibnamefont {Zhong}}, \bibinfo {author} {\bibfnamefont {K.}~\bibnamefont {Jun}}, \bibinfo {author} {\bibfnamefont {J.}~\bibnamefont {Riebesell}}, \bibinfo {author} {\bibfnamefont {K.}~\bibnamefont {Han}}, \bibinfo {author} {\bibfnamefont {C.~J.}\ \bibnamefont {Bartel}},\ and\ \bibinfo {author} {\bibfnamefont {G.}~\bibnamefont {Ceder}},\ }\href {https://doi.org/10.1038/s42256-023-00716-3} {\bibfield  {journal} {\bibinfo  {journal} {Nature Machine Intelligence}\ }\textbf {\bibinfo {volume} {5}},\ \bibinfo {pages} {1031} (\bibinfo {year} {2023})}\BibitemShut {NoStop}%
\bibitem [{\citenamefont {Batatia}\ \emph {et~al.}(2024)\citenamefont {Batatia}, \citenamefont {Benner}, \citenamefont {Chiang}, \citenamefont {Elena}, \citenamefont {Kovács}, \citenamefont {Riebesell}, \citenamefont {Advincula}, \citenamefont {Asta}, \citenamefont {Avaylon}, \citenamefont {Baldwin}, \citenamefont {Berger}, \citenamefont {Bernstein}, \citenamefont {Bhowmik}, \citenamefont {Blau}, \citenamefont {Cărare}, \citenamefont {Darby}, \citenamefont {De}, \citenamefont {Della~Pia}, \citenamefont {Deringer}, \citenamefont {Elijošius}, \citenamefont {El-Machachi}, \citenamefont {Falcioni}, \citenamefont {Fako}, \citenamefont {Ferrari}, \citenamefont {Genreith-Schriever}, \citenamefont {George}, \citenamefont {Goodall}, \citenamefont {Grey}, \citenamefont {Grigorev}, \citenamefont {Han}, \citenamefont {Handley}, \citenamefont {Heenen}, \citenamefont {Hermansson}, \citenamefont {Holm}, \citenamefont {Jaafar}, \citenamefont {Hofmann}, \citenamefont {Jakob}, \citenamefont {Jung}, \citenamefont {Kapil}, \citenamefont {Kaplan}, \citenamefont {Karimitari}, \citenamefont {Kermode}, \citenamefont {Kroupa}, \citenamefont {Kullgren}, \citenamefont {Kuner}, \citenamefont {Kuryla}, \citenamefont {Liepuoniute}, \citenamefont {Margraf}, \citenamefont {Magdău}, \citenamefont {Michaelides}, \citenamefont {Moore}, \citenamefont {Naik}, \citenamefont {Niblett}, \citenamefont {Norwood}, \citenamefont {O'Neill}, \citenamefont {Ortner}, \citenamefont {Persson}, \citenamefont {Reuter}, \citenamefont {Rosen}, \citenamefont {Schaaf}, \citenamefont {Schran}, \citenamefont {Shi}, \citenamefont {Sivonxay}, \citenamefont {Stenczel}, \citenamefont {Svahn}, \citenamefont {Sutton}, \citenamefont {Swinburne}, \citenamefont {Tilly}, \citenamefont {van~der Oord}, \citenamefont {Varga-Umbrich}, \citenamefont {Vegge}, \citenamefont {Vondrák}, \citenamefont {Wang}, \citenamefont {Witt}, \citenamefont {Zills},\ and\ \citenamefont {Csányi}}]{batatia_foundation_2024}%
  \BibitemOpen
  \bibfield  {author} {\bibinfo {author} {\bibfnamefont {I.}~\bibnamefont {Batatia}}, \bibinfo {author} {\bibfnamefont {P.}~\bibnamefont {Benner}}, \bibinfo {author} {\bibfnamefont {Y.}~\bibnamefont {Chiang}}, \bibinfo {author} {\bibfnamefont {A.~M.}\ \bibnamefont {Elena}}, \bibinfo {author} {\bibfnamefont {D.~P.}\ \bibnamefont {Kovács}}, \bibinfo {author} {\bibfnamefont {J.}~\bibnamefont {Riebesell}}, \bibinfo {author} {\bibfnamefont {X.~R.}\ \bibnamefont {Advincula}}, \bibinfo {author} {\bibfnamefont {M.}~\bibnamefont {Asta}}, \bibinfo {author} {\bibfnamefont {M.}~\bibnamefont {Avaylon}}, \bibinfo {author} {\bibfnamefont {W.~J.}\ \bibnamefont {Baldwin}}, \bibinfo {author} {\bibfnamefont {F.}~\bibnamefont {Berger}}, \bibinfo {author} {\bibfnamefont {N.}~\bibnamefont {Bernstein}}, \bibinfo {author} {\bibfnamefont {A.}~\bibnamefont {Bhowmik}}, \bibinfo {author} {\bibfnamefont {S.~M.}\ \bibnamefont {Blau}}, \bibinfo {author} {\bibfnamefont {V.}~\bibnamefont {Cărare}}, \bibinfo {author} {\bibfnamefont {J.~P.}\ \bibnamefont {Darby}}, \bibinfo {author} {\bibfnamefont {S.}~\bibnamefont {De}}, \bibinfo {author} {\bibfnamefont {F.}~\bibnamefont {Della~Pia}}, \bibinfo {author} {\bibfnamefont {V.~L.}\ \bibnamefont {Deringer}}, \bibinfo {author} {\bibfnamefont {R.}~\bibnamefont {Elijošius}}, \bibinfo {author} {\bibfnamefont {Z.}~\bibnamefont {El-Machachi}}, \bibinfo {author} {\bibfnamefont {F.}~\bibnamefont {Falcioni}}, \bibinfo {author} {\bibfnamefont {E.}~\bibnamefont {Fako}}, \bibinfo {author} {\bibfnamefont {A.~C.}\ \bibnamefont {Ferrari}}, \bibinfo {author} {\bibfnamefont {A.}~\bibnamefont {Genreith-Schriever}}, \bibinfo {author} {\bibfnamefont {J.}~\bibnamefont {George}}, \bibinfo {author} {\bibfnamefont {R.~E.~A.}\ \bibnamefont {Goodall}}, \bibinfo {author} {\bibfnamefont {C.~P.}\ \bibnamefont {Grey}}, \bibinfo {author} {\bibfnamefont {P.}~\bibnamefont {Grigorev}}, \bibinfo {author} {\bibfnamefont {S.}~\bibnamefont {Han}}, \bibinfo {author} {\bibfnamefont {W.}~\bibnamefont {Handley}}, \bibinfo {author} {\bibfnamefont {H.~H.}\ \bibnamefont {Heenen}}, \bibinfo {author} {\bibfnamefont {K.}~\bibnamefont {Hermansson}}, \bibinfo {author} {\bibfnamefont {C.}~\bibnamefont {Holm}}, \bibinfo {author} {\bibfnamefont {J.}~\bibnamefont {Jaafar}}, \bibinfo {author} {\bibfnamefont {S.}~\bibnamefont {Hofmann}}, \bibinfo {author} {\bibfnamefont {K.~S.}\ \bibnamefont {Jakob}}, \bibinfo {author} {\bibfnamefont {H.}~\bibnamefont {Jung}}, \bibinfo {author} {\bibfnamefont {V.}~\bibnamefont {Kapil}}, \bibinfo {author} {\bibfnamefont {A.~D.}\ \bibnamefont {Kaplan}}, \bibinfo {author} {\bibfnamefont {N.}~\bibnamefont {Karimitari}}, \bibinfo {author} {\bibfnamefont {J.~R.}\ \bibnamefont {Kermode}}, \bibinfo {author} {\bibfnamefont {N.}~\bibnamefont {Kroupa}}, \bibinfo {author} {\bibfnamefont {J.}~\bibnamefont {Kullgren}}, \bibinfo {author} {\bibfnamefont {M.~C.}\ \bibnamefont {Kuner}}, \bibinfo {author} {\bibfnamefont {D.}~\bibnamefont {Kuryla}}, \bibinfo {author} {\bibfnamefont {G.}~\bibnamefont {Liepuoniute}}, \bibinfo {author} {\bibfnamefont {J.~T.}\ \bibnamefont {Margraf}}, \bibinfo {author} {\bibfnamefont {I.-B.}\ \bibnamefont {Magdău}}, \bibinfo {author} {\bibfnamefont {A.}~\bibnamefont {Michaelides}}, \bibinfo {author} {\bibfnamefont {J.~H.}\ \bibnamefont {Moore}}, \bibinfo {author} {\bibfnamefont {A.~A.}\ \bibnamefont {Naik}}, \bibinfo {author} {\bibfnamefont {S.~P.}\ \bibnamefont {Niblett}}, \bibinfo {author} {\bibfnamefont {S.~W.}\ \bibnamefont {Norwood}}, \bibinfo {author} {\bibfnamefont {N.}~\bibnamefont {O'Neill}}, \bibinfo {author} {\bibfnamefont {C.}~\bibnamefont {Ortner}}, \bibinfo {author} {\bibfnamefont {K.~A.}\ \bibnamefont {Persson}}, \bibinfo {author} {\bibfnamefont {K.}~\bibnamefont {Reuter}}, \bibinfo {author} {\bibfnamefont {A.~S.}\ \bibnamefont {Rosen}}, \bibinfo {author} {\bibfnamefont {L.~L.}\ \bibnamefont {Schaaf}}, \bibinfo {author} {\bibfnamefont {C.}~\bibnamefont {Schran}}, \bibinfo {author} {\bibfnamefont {B.~X.}\ \bibnamefont {Shi}}, \bibinfo {author} {\bibfnamefont {E.}~\bibnamefont {Sivonxay}}, \bibinfo {author} {\bibfnamefont {T.~K.}\ \bibnamefont {Stenczel}}, \bibinfo {author} {\bibfnamefont {V.}~\bibnamefont {Svahn}}, \bibinfo {author} {\bibfnamefont {C.}~\bibnamefont {Sutton}}, \bibinfo {author} {\bibfnamefont {T.~D.}\ \bibnamefont {Swinburne}}, \bibinfo {author} {\bibfnamefont {J.}~\bibnamefont {Tilly}}, \bibinfo {author} {\bibfnamefont {C.}~\bibnamefont {van~der Oord}}, \bibinfo {author} {\bibfnamefont {E.}~\bibnamefont {Varga-Umbrich}}, \bibinfo {author} {\bibfnamefont {T.}~\bibnamefont {Vegge}}, \bibinfo {author} {\bibfnamefont {M.}~\bibnamefont {Vondrák}}, \bibinfo {author} {\bibfnamefont {Y.}~\bibnamefont {Wang}}, \bibinfo {author} {\bibfnamefont {W.~C.}\ \bibnamefont {Witt}}, \bibinfo {author} {\bibfnamefont {F.}~\bibnamefont {Zills}},\ and\ \bibinfo {author} {\bibfnamefont {G.}~\bibnamefont {Csányi}},\ }\href {https://doi.org/10.48550/arXiv.2401.00096} {\bibinfo {title} {A foundation model for atomistic materials chemistry}} (\bibinfo {year} {2024}),\ \bibinfo {note} {arXiv:2401.00096}\BibitemShut {NoStop}%
\bibitem [{\citenamefont {Park}\ \emph {et~al.}(2024)\citenamefont {Park}, \citenamefont {Kim}, \citenamefont {Hwang},\ and\ \citenamefont {Han}}]{park_scalable_2024}%
  \BibitemOpen
  \bibfield  {author} {\bibinfo {author} {\bibfnamefont {Y.}~\bibnamefont {Park}}, \bibinfo {author} {\bibfnamefont {J.}~\bibnamefont {Kim}}, \bibinfo {author} {\bibfnamefont {S.}~\bibnamefont {Hwang}},\ and\ \bibinfo {author} {\bibfnamefont {S.}~\bibnamefont {Han}},\ }\href {https://doi.org/10.1021/acs.jctc.4c00190} {\bibfield  {journal} {\bibinfo  {journal} {Journal of Chemical Theory and Computation}\ }\textbf {\bibinfo {volume} {20}},\ \bibinfo {pages} {4857} (\bibinfo {year} {2024})}\BibitemShut {NoStop}%
\bibitem [{\citenamefont {Yang}\ \emph {et~al.}(2024)\citenamefont {Yang}, \citenamefont {Hu}, \citenamefont {Zhou}, \citenamefont {Liu}, \citenamefont {Shi}, \citenamefont {Li}, \citenamefont {Li}, \citenamefont {Chen}, \citenamefont {Chen}, \citenamefont {Zeni}, \citenamefont {Horton}, \citenamefont {Pinsler}, \citenamefont {Fowler}, \citenamefont {Zügner}, \citenamefont {Xie}, \citenamefont {Smith}, \citenamefont {Sun}, \citenamefont {Wang}, \citenamefont {Kong}, \citenamefont {Liu}, \citenamefont {Hao},\ and\ \citenamefont {Lu}}]{yang2024mattersimdeeplearningatomistic}%
  \BibitemOpen
  \bibfield  {author} {\bibinfo {author} {\bibfnamefont {H.}~\bibnamefont {Yang}}, \bibinfo {author} {\bibfnamefont {C.}~\bibnamefont {Hu}}, \bibinfo {author} {\bibfnamefont {Y.}~\bibnamefont {Zhou}}, \bibinfo {author} {\bibfnamefont {X.}~\bibnamefont {Liu}}, \bibinfo {author} {\bibfnamefont {Y.}~\bibnamefont {Shi}}, \bibinfo {author} {\bibfnamefont {J.}~\bibnamefont {Li}}, \bibinfo {author} {\bibfnamefont {G.}~\bibnamefont {Li}}, \bibinfo {author} {\bibfnamefont {Z.}~\bibnamefont {Chen}}, \bibinfo {author} {\bibfnamefont {S.}~\bibnamefont {Chen}}, \bibinfo {author} {\bibfnamefont {C.}~\bibnamefont {Zeni}}, \bibinfo {author} {\bibfnamefont {M.}~\bibnamefont {Horton}}, \bibinfo {author} {\bibfnamefont {R.}~\bibnamefont {Pinsler}}, \bibinfo {author} {\bibfnamefont {A.}~\bibnamefont {Fowler}}, \bibinfo {author} {\bibfnamefont {D.}~\bibnamefont {Zügner}}, \bibinfo {author} {\bibfnamefont {T.}~\bibnamefont {Xie}}, \bibinfo {author} {\bibfnamefont {J.}~\bibnamefont {Smith}}, \bibinfo {author} {\bibfnamefont {L.}~\bibnamefont {Sun}}, \bibinfo {author} {\bibfnamefont {Q.}~\bibnamefont {Wang}}, \bibinfo {author} {\bibfnamefont {L.}~\bibnamefont {Kong}}, \bibinfo {author} {\bibfnamefont {C.}~\bibnamefont {Liu}}, \bibinfo {author} {\bibfnamefont {H.}~\bibnamefont {Hao}},\ and\ \bibinfo {author} {\bibfnamefont {Z.}~\bibnamefont {Lu}},\ }\href {https://arxiv.org/abs/2405.04967} {\bibinfo {title} {Mattersim: A deep learning atomistic model across elements, temperatures and pressures}} (\bibinfo {year} {2024}),\ \Eprint {https://arxiv.org/abs/2405.04967} {arXiv:2405.04967 [cond-mat.mtrl-sci]} \BibitemShut {NoStop}%
\bibitem [{\citenamefont {Merchant}\ \emph {et~al.}(2023)\citenamefont {Merchant}, \citenamefont {Batzner}, \citenamefont {Schoenholz}, \citenamefont {Aykol}, \citenamefont {Cheon},\ and\ \citenamefont {Cubuk}}]{merchant_scaling_2023}%
  \BibitemOpen
  \bibfield  {author} {\bibinfo {author} {\bibfnamefont {A.}~\bibnamefont {Merchant}}, \bibinfo {author} {\bibfnamefont {S.}~\bibnamefont {Batzner}}, \bibinfo {author} {\bibfnamefont {S.~S.}\ \bibnamefont {Schoenholz}}, \bibinfo {author} {\bibfnamefont {M.}~\bibnamefont {Aykol}}, \bibinfo {author} {\bibfnamefont {G.}~\bibnamefont {Cheon}},\ and\ \bibinfo {author} {\bibfnamefont {E.~D.}\ \bibnamefont {Cubuk}},\ }\href {https://doi.org/10.1038/s41586-023-06735-9} {\bibfield  {journal} {\bibinfo  {journal} {Nature}\ }\textbf {\bibinfo {volume} {624}},\ \bibinfo {pages} {80} (\bibinfo {year} {2023})}\BibitemShut {NoStop}%
\bibitem [{\citenamefont {Choudhary}\ and\ \citenamefont {DeCost}(2021)}]{ALIGNN}%
  \BibitemOpen
  \bibfield  {author} {\bibinfo {author} {\bibfnamefont {K.}~\bibnamefont {Choudhary}}\ and\ \bibinfo {author} {\bibfnamefont {B.}~\bibnamefont {DeCost}},\ }\href {https://doi.org/10.1038/s41524-021-00650-1} {\bibfield  {journal} {\bibinfo  {journal} {npj Computational Materials}\ }\textbf {\bibinfo {volume} {7}},\ \bibinfo {pages} {185} (\bibinfo {year} {2021})}\BibitemShut {NoStop}%
\bibitem [{\citenamefont {Chen}\ \emph {et~al.}(2019)\citenamefont {Chen}, \citenamefont {Ye}, \citenamefont {Zuo}, \citenamefont {Zheng},\ and\ \citenamefont {Ong}}]{MEGNET}%
  \BibitemOpen
  \bibfield  {author} {\bibinfo {author} {\bibfnamefont {C.}~\bibnamefont {Chen}}, \bibinfo {author} {\bibfnamefont {W.}~\bibnamefont {Ye}}, \bibinfo {author} {\bibfnamefont {Y.}~\bibnamefont {Zuo}}, \bibinfo {author} {\bibfnamefont {C.}~\bibnamefont {Zheng}},\ and\ \bibinfo {author} {\bibfnamefont {S.~P.}\ \bibnamefont {Ong}},\ }\href {https://doi.org/10.1021/acs.chemmater.9b01294} {\bibfield  {journal} {\bibinfo  {journal} {Chemistry of Materials}\ }\textbf {\bibinfo {volume} {31}},\ \bibinfo {pages} {3564} (\bibinfo {year} {2019})}\BibitemShut {NoStop}%
\bibitem [{\citenamefont {Ward}\ \emph {et~al.}(2017)\citenamefont {Ward}, \citenamefont {Liu}, \citenamefont {Krishna}, \citenamefont {Hegde}, \citenamefont {Agrawal}, \citenamefont {Choudhary},\ and\ \citenamefont {Wolverton}}]{Voronoi_RF}%
  \BibitemOpen
  \bibfield  {author} {\bibinfo {author} {\bibfnamefont {L.}~\bibnamefont {Ward}}, \bibinfo {author} {\bibfnamefont {R.}~\bibnamefont {Liu}}, \bibinfo {author} {\bibfnamefont {A.}~\bibnamefont {Krishna}}, \bibinfo {author} {\bibfnamefont {V.~I.}\ \bibnamefont {Hegde}}, \bibinfo {author} {\bibfnamefont {A.}~\bibnamefont {Agrawal}}, \bibinfo {author} {\bibfnamefont {A.}~\bibnamefont {Choudhary}},\ and\ \bibinfo {author} {\bibfnamefont {C.}~\bibnamefont {Wolverton}},\ }\href {https://doi.org/10.1103/PhysRevB.96.024104} {\bibfield  {journal} {\bibinfo  {journal} {Physical Review B}\ }\textbf {\bibinfo {volume} {96}},\ \bibinfo {pages} {024104} (\bibinfo {year} {2017})}\BibitemShut {NoStop}%
\bibitem [{\citenamefont {Zuo}\ \emph {et~al.}(2021)\citenamefont {Zuo}, \citenamefont {Qin}, \citenamefont {Chen}, \citenamefont {Ye}, \citenamefont {Li}, \citenamefont {Luo},\ and\ \citenamefont {Ong}}]{BOWSR}%
  \BibitemOpen
  \bibfield  {author} {\bibinfo {author} {\bibfnamefont {Y.}~\bibnamefont {Zuo}}, \bibinfo {author} {\bibfnamefont {M.}~\bibnamefont {Qin}}, \bibinfo {author} {\bibfnamefont {C.}~\bibnamefont {Chen}}, \bibinfo {author} {\bibfnamefont {W.}~\bibnamefont {Ye}}, \bibinfo {author} {\bibfnamefont {X.}~\bibnamefont {Li}}, \bibinfo {author} {\bibfnamefont {J.}~\bibnamefont {Luo}},\ and\ \bibinfo {author} {\bibfnamefont {S.~P.}\ \bibnamefont {Ong}},\ }\href {https://doi.org/10.1016/j.mattod.2021.08.012} {\bibfield  {journal} {\bibinfo  {journal} {Materials Today}\ }\textbf {\bibinfo {volume} {51}},\ \bibinfo {pages} {126} (\bibinfo {year} {2021})}\BibitemShut {NoStop}%
\bibitem [{\citenamefont {Xie}\ and\ \citenamefont {Grossman}(2018)}]{CGCNN}%
  \BibitemOpen
  \bibfield  {author} {\bibinfo {author} {\bibfnamefont {T.}~\bibnamefont {Xie}}\ and\ \bibinfo {author} {\bibfnamefont {J.~C.}\ \bibnamefont {Grossman}},\ }\href {https://doi.org/10.1103/PhysRevLett.120.145301} {\bibfield  {journal} {\bibinfo  {journal} {Physical Review Letters}\ }\textbf {\bibinfo {volume} {120}},\ \bibinfo {pages} {145301} (\bibinfo {year} {2018})}\BibitemShut {NoStop}%
\bibitem [{\citenamefont {Gibson}\ \emph {et~al.}(2022)\citenamefont {Gibson}, \citenamefont {Hire},\ and\ \citenamefont {Hennig}}]{CGCNN+P}%
  \BibitemOpen
  \bibfield  {author} {\bibinfo {author} {\bibfnamefont {J.}~\bibnamefont {Gibson}}, \bibinfo {author} {\bibfnamefont {A.}~\bibnamefont {Hire}},\ and\ \bibinfo {author} {\bibfnamefont {R.~G.}\ \bibnamefont {Hennig}},\ }\href {https://doi.org/10.1038/s41524-022-00891-8} {\bibfield  {journal} {\bibinfo  {journal} {npj Computational Materials}\ }\textbf {\bibinfo {volume} {8}},\ \bibinfo {pages} {211} (\bibinfo {year} {2022})}\BibitemShut {NoStop}%
\bibitem [{\citenamefont {Goodall}\ \emph {et~al.}(2022)\citenamefont {Goodall}, \citenamefont {Parackal}, \citenamefont {Faber}, \citenamefont {Armiento},\ and\ \citenamefont {Lee}}]{Wrenformer}%
  \BibitemOpen
  \bibfield  {author} {\bibinfo {author} {\bibfnamefont {R.~E.~A.}\ \bibnamefont {Goodall}}, \bibinfo {author} {\bibfnamefont {A.~S.}\ \bibnamefont {Parackal}}, \bibinfo {author} {\bibfnamefont {F.~A.}\ \bibnamefont {Faber}}, \bibinfo {author} {\bibfnamefont {R.}~\bibnamefont {Armiento}},\ and\ \bibinfo {author} {\bibfnamefont {A.~A.}\ \bibnamefont {Lee}},\ }\href {https://doi.org/10.1126/sciadv.abn4117} {\bibfield  {journal} {\bibinfo  {journal} {Science Advances}\ }\textbf {\bibinfo {volume} {8}},\ \bibinfo {pages} {eabn4117} (\bibinfo {year} {2022})}\BibitemShut {NoStop}%
\bibitem [{\citenamefont {Riebesell}\ \emph {et~al.}(2025)\citenamefont {Riebesell}, \citenamefont {Goodall}, \citenamefont {Benner}, \citenamefont {Chiang}, \citenamefont {Deng}, \citenamefont {Ceder}, \citenamefont {Asta}, \citenamefont {Lee}, \citenamefont {Jain},\ and\ \citenamefont {Persson}}]{riebesell_matbench_2024}%
  \BibitemOpen
  \bibfield  {author} {\bibinfo {author} {\bibfnamefont {J.}~\bibnamefont {Riebesell}}, \bibinfo {author} {\bibfnamefont {R.~E.~A.}\ \bibnamefont {Goodall}}, \bibinfo {author} {\bibfnamefont {P.}~\bibnamefont {Benner}}, \bibinfo {author} {\bibfnamefont {Y.}~\bibnamefont {Chiang}}, \bibinfo {author} {\bibfnamefont {B.}~\bibnamefont {Deng}}, \bibinfo {author} {\bibfnamefont {G.}~\bibnamefont {Ceder}}, \bibinfo {author} {\bibfnamefont {M.}~\bibnamefont {Asta}}, \bibinfo {author} {\bibfnamefont {A.~A.}\ \bibnamefont {Lee}}, \bibinfo {author} {\bibfnamefont {A.}~\bibnamefont {Jain}},\ and\ \bibinfo {author} {\bibfnamefont {K.~A.}\ \bibnamefont {Persson}},\ }\href {https://doi.org/10.1038/s42256-025-01055-1} {\bibfield  {journal} {\bibinfo  {journal} {Nature Machine Intelligence}\ }\textbf {\bibinfo {volume} {7}},\ \bibinfo {pages} {836} (\bibinfo {year} {2025})},\ \bibinfo {note} {publisher: Nature Publishing Group}\BibitemShut {NoStop}%
\bibitem [{\citenamefont {Yu}\ \emph {et~al.}(2024)\citenamefont {Yu}, \citenamefont {Giantomassi}, \citenamefont {Materzanini}, \citenamefont {Wang},\ and\ \citenamefont {Rignanese}}]{yu_systematic_2024}%
  \BibitemOpen
  \bibfield  {author} {\bibinfo {author} {\bibfnamefont {H.}~\bibnamefont {Yu}}, \bibinfo {author} {\bibfnamefont {M.}~\bibnamefont {Giantomassi}}, \bibinfo {author} {\bibfnamefont {G.}~\bibnamefont {Materzanini}}, \bibinfo {author} {\bibfnamefont {J.}~\bibnamefont {Wang}},\ and\ \bibinfo {author} {\bibfnamefont {G.-M.}\ \bibnamefont {Rignanese}},\ }\href {https://doi.org/10.1002/mgea.58} {\bibfield  {journal} {\bibinfo  {journal} {Materials Genome Engineering Advances}\ }\textbf {\bibinfo {volume} {2}},\ \bibinfo {pages} {e58} (\bibinfo {year} {2024})},\ \bibinfo {note} {\_eprint: https://onlinelibrary.wiley.com/doi/pdf/10.1002/mgea.58}\BibitemShut {NoStop}%
\bibitem [{\citenamefont {Deng}\ \emph {et~al.}(2025)\citenamefont {Deng}, \citenamefont {Choi}, \citenamefont {Zhong}, \citenamefont {Riebesell}, \citenamefont {Anand}, \citenamefont {Li}, \citenamefont {Jun}, \citenamefont {Persson},\ and\ \citenamefont {Ceder}}]{deng_overcoming_nodate}%
  \BibitemOpen
  \bibfield  {author} {\bibinfo {author} {\bibfnamefont {B.}~\bibnamefont {Deng}}, \bibinfo {author} {\bibfnamefont {Y.}~\bibnamefont {Choi}}, \bibinfo {author} {\bibfnamefont {P.}~\bibnamefont {Zhong}}, \bibinfo {author} {\bibfnamefont {J.}~\bibnamefont {Riebesell}}, \bibinfo {author} {\bibfnamefont {S.}~\bibnamefont {Anand}}, \bibinfo {author} {\bibfnamefont {Z.}~\bibnamefont {Li}}, \bibinfo {author} {\bibfnamefont {K.}~\bibnamefont {Jun}}, \bibinfo {author} {\bibfnamefont {K.~A.}\ \bibnamefont {Persson}},\ and\ \bibinfo {author} {\bibfnamefont {G.}~\bibnamefont {Ceder}},\ }\href {https://doi.org/10.1038/s41524-024-01500-6} {\bibfield  {journal} {\bibinfo  {journal} {npj Computational Materials}\ }\textbf {\bibinfo {volume} {11}},\ \bibinfo {pages} {9} (\bibinfo {year} {2025})},\ \bibinfo {note} {publisher: Nature Publishing Group}\BibitemShut {NoStop}%
\bibitem [{\citenamefont {Lee}\ \emph {et~al.}(2025)\citenamefont {Lee}, \citenamefont {Hegde}, \citenamefont {Wolverton},\ and\ \citenamefont {Xia}}]{lee_accelerating_2024}%
  \BibitemOpen
  \bibfield  {author} {\bibinfo {author} {\bibfnamefont {H.}~\bibnamefont {Lee}}, \bibinfo {author} {\bibfnamefont {V.~I.}\ \bibnamefont {Hegde}}, \bibinfo {author} {\bibfnamefont {C.}~\bibnamefont {Wolverton}},\ and\ \bibinfo {author} {\bibfnamefont {Y.}~\bibnamefont {Xia}},\ }\href {https://doi.org/10.1016/j.mtphys.2025.101688} {\bibfield  {journal} {\bibinfo  {journal} {Materials Today Physics}\ }\textbf {\bibinfo {volume} {53}},\ \bibinfo {pages} {101688} (\bibinfo {year} {2025})}\BibitemShut {NoStop}%
\bibitem [{\citenamefont {Simoncelli}\ \emph {et~al.}(2019)\citenamefont {Simoncelli}, \citenamefont {Marzari},\ and\ \citenamefont {Mauri}}]{simoncelli_unified_2019}%
  \BibitemOpen
  \bibfield  {author} {\bibinfo {author} {\bibfnamefont {M.}~\bibnamefont {Simoncelli}}, \bibinfo {author} {\bibfnamefont {N.}~\bibnamefont {Marzari}},\ and\ \bibinfo {author} {\bibfnamefont {F.}~\bibnamefont {Mauri}},\ }\href {https://doi.org/10.1038/s41567-019-0520-x} {\bibfield  {journal} {\bibinfo  {journal} {Nature Physics}\ }\textbf {\bibinfo {volume} {15}},\ \bibinfo {pages} {809} (\bibinfo {year} {2019})}\BibitemShut {NoStop}%
\bibitem [{\citenamefont {Simoncelli}\ \emph {et~al.}(2022)\citenamefont {Simoncelli}, \citenamefont {Marzari},\ and\ \citenamefont {Mauri}}]{simoncelli_wigner_2022}%
  \BibitemOpen
  \bibfield  {author} {\bibinfo {author} {\bibfnamefont {M.}~\bibnamefont {Simoncelli}}, \bibinfo {author} {\bibfnamefont {N.}~\bibnamefont {Marzari}},\ and\ \bibinfo {author} {\bibfnamefont {F.}~\bibnamefont {Mauri}},\ }\href {https://doi.org/10.1103/PhysRevX.12.041011} {\bibfield  {journal} {\bibinfo  {journal} {Physical Review X}\ }\textbf {\bibinfo {volume} {12}},\ \bibinfo {pages} {041011} (\bibinfo {year} {2022})}\BibitemShut {NoStop}%
\bibitem [{\citenamefont {Ritz}\ \emph {et~al.}(2019)\citenamefont {Ritz}, \citenamefont {Li},\ and\ \citenamefont {Benedek}}]{ritz_thermal_2019}%
  \BibitemOpen
  \bibfield  {author} {\bibinfo {author} {\bibfnamefont {E.~T.}\ \bibnamefont {Ritz}}, \bibinfo {author} {\bibfnamefont {S.~J.}\ \bibnamefont {Li}},\ and\ \bibinfo {author} {\bibfnamefont {N.~A.}\ \bibnamefont {Benedek}},\ }\href {https://doi.org/10.1063/1.5125779} {\bibfield  {journal} {\bibinfo  {journal} {Journal of Applied Physics}\ }\textbf {\bibinfo {volume} {126}},\ \bibinfo {pages} {171102} (\bibinfo {year} {2019})}\BibitemShut {NoStop}%
\bibitem [{\citenamefont {Barron}\ \emph {et~al.}(1980)\citenamefont {Barron}, \citenamefont {Collins},\ and\ \citenamefont {White}}]{barron_thermal_1980}%
  \BibitemOpen
  \bibfield  {author} {\bibinfo {author} {\bibfnamefont {T.}~\bibnamefont {Barron}}, \bibinfo {author} {\bibfnamefont {J.}~\bibnamefont {Collins}},\ and\ \bibinfo {author} {\bibfnamefont {G.}~\bibnamefont {White}},\ }\href {https://doi.org/10.1080/00018738000101426} {\bibfield  {journal} {\bibinfo  {journal} {Advances in Physics}\ }\textbf {\bibinfo {volume} {29}},\ \bibinfo {pages} {609} (\bibinfo {year} {1980})}\BibitemShut {NoStop}%
\bibitem [{\citenamefont {Born}\ and\ \citenamefont {Huang}(1998)}]{born_dynamical_1998}%
  \BibitemOpen
  \bibfield  {author} {\bibinfo {author} {\bibfnamefont {M.}~\bibnamefont {Born}}\ and\ \bibinfo {author} {\bibfnamefont {K.}~\bibnamefont {Huang}},\ }\href@noop {} {\emph {\bibinfo {title} {Dynamical Theory of Crystal Lattices}}},\ Oxford Classic Texts in the Physical Sciences\ (\bibinfo  {publisher} {Oxford University Press},\ \bibinfo {address} {Oxford, New York},\ \bibinfo {year} {1998})\BibitemShut {NoStop}%
\bibitem [{\citenamefont {Wallace}(1972)}]{wallace_thermodynamics_1972}%
  \BibitemOpen
  \bibfield  {author} {\bibinfo {author} {\bibfnamefont {D.~C.}\ \bibnamefont {Wallace}},\ }\href@noop {} {\emph {\bibinfo {title} {Thermodynamics of Crystals}}}\ (\bibinfo  {publisher} {Dover Publications},\ \bibinfo {year} {1972})\BibitemShut {NoStop}%
\bibitem [{\citenamefont {Riebesell}\ \emph {et~al.}(2024)\citenamefont {Riebesell}, \citenamefont {Goodall}, \citenamefont {Benner}, \citenamefont {Chiang}, \citenamefont {Deng}, \citenamefont {Lee}, \citenamefont {Jain},\ and\ \citenamefont {Persson}}]{matbench_discovery_url}%
  \BibitemOpen
  \bibfield  {author} {\bibinfo {author} {\bibfnamefont {J.}~\bibnamefont {Riebesell}}, \bibinfo {author} {\bibfnamefont {R.~E.~A.}\ \bibnamefont {Goodall}}, \bibinfo {author} {\bibfnamefont {P.}~\bibnamefont {Benner}}, \bibinfo {author} {\bibfnamefont {Y.}~\bibnamefont {Chiang}}, \bibinfo {author} {\bibfnamefont {B.}~\bibnamefont {Deng}}, \bibinfo {author} {\bibfnamefont {A.~A.}\ \bibnamefont {Lee}}, \bibinfo {author} {\bibfnamefont {A.}~\bibnamefont {Jain}},\ and\ \bibinfo {author} {\bibfnamefont {K.~A.}\ \bibnamefont {Persson}},\ }\href {https://matbench-discovery.materialsproject.org/} {\bibinfo {title} {Matbench-discovery website}} (\bibinfo {year} {2024})\BibitemShut {NoStop}%
\bibitem [{\citenamefont {Jain}\ \emph {et~al.}(2013{\natexlab{a}})\citenamefont {Jain}, \citenamefont {Ong}, \citenamefont {Hautier}, \citenamefont {Chen}, \citenamefont {Richards}, \citenamefont {Dacek}, \citenamefont {Cholia}, \citenamefont {Gunter}, \citenamefont {Skinner}, \citenamefont {Ceder},\ and\ \citenamefont {Persson}}]{materialsproject}%
  \BibitemOpen
  \bibfield  {author} {\bibinfo {author} {\bibfnamefont {A.}~\bibnamefont {Jain}}, \bibinfo {author} {\bibfnamefont {S.~P.}\ \bibnamefont {Ong}}, \bibinfo {author} {\bibfnamefont {G.}~\bibnamefont {Hautier}}, \bibinfo {author} {\bibfnamefont {W.}~\bibnamefont {Chen}}, \bibinfo {author} {\bibfnamefont {W.~D.}\ \bibnamefont {Richards}}, \bibinfo {author} {\bibfnamefont {S.}~\bibnamefont {Dacek}}, \bibinfo {author} {\bibfnamefont {S.}~\bibnamefont {Cholia}}, \bibinfo {author} {\bibfnamefont {D.}~\bibnamefont {Gunter}}, \bibinfo {author} {\bibfnamefont {D.}~\bibnamefont {Skinner}}, \bibinfo {author} {\bibfnamefont {G.}~\bibnamefont {Ceder}},\ and\ \bibinfo {author} {\bibfnamefont {K.~A.}\ \bibnamefont {Persson}},\ }\href {https://doi.org/10.1063/1.4812323} {\bibfield  {journal} {\bibinfo  {journal} {APL Materials}\ }\textbf {\bibinfo {volume} {1}},\ \bibinfo {pages} {011002} (\bibinfo {year} {2013}{\natexlab{a}})}\BibitemShut {NoStop}%
\bibitem [{\citenamefont {Ltd.}(2024)}]{orb_github}%
  \BibitemOpen
  \bibfield  {author} {\bibinfo {author} {\bibfnamefont {O.~M.}\ \bibnamefont {Ltd.}},\ }\href@noop {} {\bibinfo {title} {orb-models}},\ \bibinfo {howpublished} {\url{https://github.com/orbital-materials/orb-models}} (\bibinfo {year} {2024})\BibitemShut {NoStop}%
\bibitem [{\citenamefont {Togo}\ \emph {et~al.}(2015{\natexlab{a}})\citenamefont {Togo}, \citenamefont {Chaput},\ and\ \citenamefont {Tanaka}}]{phono3py}%
  \BibitemOpen
  \bibfield  {author} {\bibinfo {author} {\bibfnamefont {A.}~\bibnamefont {Togo}}, \bibinfo {author} {\bibfnamefont {L.}~\bibnamefont {Chaput}},\ and\ \bibinfo {author} {\bibfnamefont {I.}~\bibnamefont {Tanaka}},\ }\href {https://doi.org/10.1103/PhysRevB.91.094306} {\bibfield  {journal} {\bibinfo  {journal} {Phys. Rev. B}\ }\textbf {\bibinfo {volume} {91}},\ \bibinfo {pages} {094306} (\bibinfo {year} {2015}{\natexlab{a}})}\BibitemShut {NoStop}%
\bibitem [{\citenamefont {Seko}\ \emph {et~al.}(2015)\citenamefont {Seko}, \citenamefont {Togo}, \citenamefont {Hayashi}, \citenamefont {Tsuda}, \citenamefont {Chaput},\ and\ \citenamefont {Tanaka}}]{seko_prediction_2015}%
  \BibitemOpen
  \bibfield  {author} {\bibinfo {author} {\bibfnamefont {A.}~\bibnamefont {Seko}}, \bibinfo {author} {\bibfnamefont {A.}~\bibnamefont {Togo}}, \bibinfo {author} {\bibfnamefont {H.}~\bibnamefont {Hayashi}}, \bibinfo {author} {\bibfnamefont {K.}~\bibnamefont {Tsuda}}, \bibinfo {author} {\bibfnamefont {L.}~\bibnamefont {Chaput}},\ and\ \bibinfo {author} {\bibfnamefont {I.}~\bibnamefont {Tanaka}},\ }\href {https://doi.org/10.1103/PhysRevLett.115.205901} {\bibfield  {journal} {\bibinfo  {journal} {Physical Review Letters}\ }\textbf {\bibinfo {volume} {115}},\ \bibinfo {pages} {205901} (\bibinfo {year} {2015})}\BibitemShut {NoStop}%
\bibitem [{\citenamefont {Simoncelli}\ \emph {et~al.}(2020)\citenamefont {Simoncelli}, \citenamefont {Marzari},\ and\ \citenamefont {Cepellotti}}]{PhysRevX.10.011019}%
  \BibitemOpen
  \bibfield  {author} {\bibinfo {author} {\bibfnamefont {M.}~\bibnamefont {Simoncelli}}, \bibinfo {author} {\bibfnamefont {N.}~\bibnamefont {Marzari}},\ and\ \bibinfo {author} {\bibfnamefont {A.}~\bibnamefont {Cepellotti}},\ }\href {https://doi.org/10.1103/PhysRevX.10.011019} {\bibfield  {journal} {\bibinfo  {journal} {Phys. Rev. X}\ }\textbf {\bibinfo {volume} {10}},\ \bibinfo {pages} {011019} (\bibinfo {year} {2020})}\BibitemShut {NoStop}%
\bibitem [{\citenamefont {Simoncelli}\ \emph {et~al.}(2023)\citenamefont {Simoncelli}, \citenamefont {Mauri},\ and\ \citenamefont {Marzari}}]{simoncelli_thermal_2023}%
  \BibitemOpen
  \bibfield  {author} {\bibinfo {author} {\bibfnamefont {M.}~\bibnamefont {Simoncelli}}, \bibinfo {author} {\bibfnamefont {F.}~\bibnamefont {Mauri}},\ and\ \bibinfo {author} {\bibfnamefont {N.}~\bibnamefont {Marzari}},\ }\href {https://doi.org/10.1038/s41524-023-01033-4} {\bibfield  {journal} {\bibinfo  {journal} {npj Computational Materials}\ }\textbf {\bibinfo {volume} {9}},\ \bibinfo {pages} {1} (\bibinfo {year} {2023})}\BibitemShut {NoStop}%
\bibitem [{\citenamefont {Pazhedath}\ \emph {et~al.}(2024)\citenamefont {Pazhedath}, \citenamefont {Bastonero}, \citenamefont {Marzari},\ and\ \citenamefont {Simoncelli}}]{pazhedath_first-principles_2023}%
  \BibitemOpen
  \bibfield  {author} {\bibinfo {author} {\bibfnamefont {A.}~\bibnamefont {Pazhedath}}, \bibinfo {author} {\bibfnamefont {L.}~\bibnamefont {Bastonero}}, \bibinfo {author} {\bibfnamefont {N.}~\bibnamefont {Marzari}},\ and\ \bibinfo {author} {\bibfnamefont {M.}~\bibnamefont {Simoncelli}},\ }\href {https://doi.org/10.1103/PhysRevApplied.22.024064} {\bibfield  {journal} {\bibinfo  {journal} {Phys. Rev. Appl.}\ }\textbf {\bibinfo {volume} {22}},\ \bibinfo {pages} {024064} (\bibinfo {year} {2024})}\BibitemShut {NoStop}%
\bibitem [{\citenamefont {Padture}(2016)}]{padture_advanced_2016}%
  \BibitemOpen
  \bibfield  {author} {\bibinfo {author} {\bibfnamefont {N.~P.}\ \bibnamefont {Padture}},\ }\href {https://doi.org/10.1038/nmat4687} {\bibfield  {journal} {\bibinfo  {journal} {Nature Materials}\ }\textbf {\bibinfo {volume} {15}},\ \bibinfo {pages} {804} (\bibinfo {year} {2016})}\BibitemShut {NoStop}%
\bibitem [{\citenamefont {Lehmann}\ \emph {et~al.}(2003)\citenamefont {Lehmann}, \citenamefont {Pitzer}, \citenamefont {Pracht}, \citenamefont {Vassen},\ and\ \citenamefont {Stöver}}]{lehmann_thermal_2003}%
  \BibitemOpen
  \bibfield  {author} {\bibinfo {author} {\bibfnamefont {H.}~\bibnamefont {Lehmann}}, \bibinfo {author} {\bibfnamefont {D.}~\bibnamefont {Pitzer}}, \bibinfo {author} {\bibfnamefont {G.}~\bibnamefont {Pracht}}, \bibinfo {author} {\bibfnamefont {R.}~\bibnamefont {Vassen}},\ and\ \bibinfo {author} {\bibfnamefont {D.}~\bibnamefont {Stöver}},\ }\href {https://doi.org/10.1111/j.1151-2916.2003.tb03473.x} {\bibfield  {journal} {\bibinfo  {journal} {Journal of the American Ceramic Society}\ }\textbf {\bibinfo {volume} {86}},\ \bibinfo {pages} {1338} (\bibinfo {year} {2003})},\ \bibinfo {note} {\_eprint: https://onlinelibrary.wiley.com/doi/pdf/10.1111/j.1151-2916.2003.tb03473.x}\BibitemShut {NoStop}%
\bibitem [{\citenamefont {Vassen}\ \emph {et~al.}(2004)\citenamefont {Vassen}, \citenamefont {Cao}, \citenamefont {Tietz}, \citenamefont {Basu},\ and\ \citenamefont {Stöver}}]{vassen_zirconates_2004}%
  \BibitemOpen
  \bibfield  {author} {\bibinfo {author} {\bibfnamefont {R.}~\bibnamefont {Vassen}}, \bibinfo {author} {\bibfnamefont {X.}~\bibnamefont {Cao}}, \bibinfo {author} {\bibfnamefont {F.}~\bibnamefont {Tietz}}, \bibinfo {author} {\bibfnamefont {D.}~\bibnamefont {Basu}},\ and\ \bibinfo {author} {\bibfnamefont {D.}~\bibnamefont {Stöver}},\ }\href {https://doi.org/10.1111/j.1151-2916.2000.tb01506.x} {\bibfield  {journal} {\bibinfo  {journal} {Journal of the American Ceramic Society}\ }\textbf {\bibinfo {volume} {83}},\ \bibinfo {pages} {2023} (\bibinfo {year} {2004})}\BibitemShut {NoStop}%
\bibitem [{\citenamefont {Plata}\ \emph {et~al.}(2017)\citenamefont {Plata}, \citenamefont {Nath}, \citenamefont {Usanmaz}, \citenamefont {Carrete}, \citenamefont {Toher}, \citenamefont {de~Jong}, \citenamefont {Asta}, \citenamefont {Fornari}, \citenamefont {Nardelli},\ and\ \citenamefont {Curtarolo}}]{plata_efficient_2017}%
  \BibitemOpen
  \bibfield  {author} {\bibinfo {author} {\bibfnamefont {J.~J.}\ \bibnamefont {Plata}}, \bibinfo {author} {\bibfnamefont {P.}~\bibnamefont {Nath}}, \bibinfo {author} {\bibfnamefont {D.}~\bibnamefont {Usanmaz}}, \bibinfo {author} {\bibfnamefont {J.}~\bibnamefont {Carrete}}, \bibinfo {author} {\bibfnamefont {C.}~\bibnamefont {Toher}}, \bibinfo {author} {\bibfnamefont {M.}~\bibnamefont {de~Jong}}, \bibinfo {author} {\bibfnamefont {M.}~\bibnamefont {Asta}}, \bibinfo {author} {\bibfnamefont {M.}~\bibnamefont {Fornari}}, \bibinfo {author} {\bibfnamefont {M.~B.}\ \bibnamefont {Nardelli}},\ and\ \bibinfo {author} {\bibfnamefont {S.}~\bibnamefont {Curtarolo}},\ }\href {https://doi.org/10.1038/s41524-017-0046-7} {\bibfield  {journal} {\bibinfo  {journal} {npj Computational Materials}\ }\textbf {\bibinfo {volume} {3}},\ \bibinfo {pages} {1} (\bibinfo {year} {2017})}\BibitemShut {NoStop}%
\bibitem [{\citenamefont {Knoop}\ \emph {et~al.}(2023)\citenamefont {Knoop}, \citenamefont {Purcell}, \citenamefont {Scheffler},\ and\ \citenamefont {Carbogno}}]{knoop_anharmonicity_2023}%
  \BibitemOpen
  \bibfield  {author} {\bibinfo {author} {\bibfnamefont {F.}~\bibnamefont {Knoop}}, \bibinfo {author} {\bibfnamefont {T.~A.}\ \bibnamefont {Purcell}}, \bibinfo {author} {\bibfnamefont {M.}~\bibnamefont {Scheffler}},\ and\ \bibinfo {author} {\bibfnamefont {C.}~\bibnamefont {Carbogno}},\ }\href {https://doi.org/10.1103/PhysRevLett.130.236301} {\bibfield  {journal} {\bibinfo  {journal} {Physical Review Letters}\ }\textbf {\bibinfo {volume} {130}},\ \bibinfo {pages} {236301} (\bibinfo {year} {2023})}\BibitemShut {NoStop}%
\bibitem [{\citenamefont {Bandi}\ \emph {et~al.}(2024)\citenamefont {Bandi}, \citenamefont {Jiang},\ and\ \citenamefont {Marianetti}}]{bandi_benchmarking_2024}%
  \BibitemOpen
  \bibfield  {author} {\bibinfo {author} {\bibfnamefont {S.}~\bibnamefont {Bandi}}, \bibinfo {author} {\bibfnamefont {C.}~\bibnamefont {Jiang}},\ and\ \bibinfo {author} {\bibfnamefont {C.~A.}\ \bibnamefont {Marianetti}},\ }\href {https://iopscience.iop.org/article/10.1088/2632-2153/ad674a} {\bibfield  {journal} {\bibinfo  {journal} {Machine Learning: Science and Technology}\ } (\bibinfo {year} {2024})}\BibitemShut {NoStop}%
\bibitem [{\citenamefont {Perdew}\ \emph {et~al.}(1996)\citenamefont {Perdew}, \citenamefont {Burke},\ and\ \citenamefont {Ernzerhof}}]{perdew_generalized_1996}%
  \BibitemOpen
  \bibfield  {author} {\bibinfo {author} {\bibfnamefont {J.~P.}\ \bibnamefont {Perdew}}, \bibinfo {author} {\bibfnamefont {K.}~\bibnamefont {Burke}},\ and\ \bibinfo {author} {\bibfnamefont {M.}~\bibnamefont {Ernzerhof}},\ }\href {https://doi.org/10.1103/PhysRevLett.77.3865} {\bibfield  {journal} {\bibinfo  {journal} {Physical Review Letters}\ }\textbf {\bibinfo {volume} {77}},\ \bibinfo {pages} {3865} (\bibinfo {year} {1996})}\BibitemShut {NoStop}%
\bibitem [{\citenamefont {Gonze}\ and\ \citenamefont {Lee}(1997{\natexlab{a}})}]{gonze_dynamical_1997}%
  \BibitemOpen
  \bibfield  {author} {\bibinfo {author} {\bibfnamefont {X.}~\bibnamefont {Gonze}}\ and\ \bibinfo {author} {\bibfnamefont {C.}~\bibnamefont {Lee}},\ }\href {https://doi.org/10.1103/PhysRevB.55.10355} {\bibfield  {journal} {\bibinfo  {journal} {Physical Review B}\ }\textbf {\bibinfo {volume} {55}},\ \bibinfo {pages} {10355} (\bibinfo {year} {1997}{\natexlab{a}})}\BibitemShut {NoStop}%
\bibitem [{\citenamefont {Ziman}(1960)}]{ziman1960electrons}%
  \BibitemOpen
  \bibfield  {author} {\bibinfo {author} {\bibfnamefont {J.~M.}\ \bibnamefont {Ziman}},\ }\href {https://global.oup.com/academic/product/electrons-and-phonons-9780198507796?cc=gb&lang=en&#} {\emph {\bibinfo {title} {Electrons and phonons: the theory of transport phenomena in solids}}}\ (\bibinfo  {publisher} {Oxford university press},\ \bibinfo {year} {1960})\BibitemShut {NoStop}%
\bibitem [{\citenamefont {Togo}\ \emph {et~al.}(2015{\natexlab{b}})\citenamefont {Togo}, \citenamefont {Chaput},\ and\ \citenamefont {Tanaka}}]{togo_2015_distribution}%
  \BibitemOpen
  \bibfield  {author} {\bibinfo {author} {\bibfnamefont {A.}~\bibnamefont {Togo}}, \bibinfo {author} {\bibfnamefont {L.}~\bibnamefont {Chaput}},\ and\ \bibinfo {author} {\bibfnamefont {I.}~\bibnamefont {Tanaka}},\ }\href {https://doi.org/10.1103/PhysRevB.91.094306} {\bibfield  {journal} {\bibinfo  {journal} {Phys. Rev. B}\ }\textbf {\bibinfo {volume} {91}},\ \bibinfo {pages} {094306} (\bibinfo {year} {2015}{\natexlab{b}})}\BibitemShut {NoStop}%
\bibitem [{\citenamefont {Kaplan}\ \emph {et~al.}(2025)\citenamefont {Kaplan}, \citenamefont {Liu}, \citenamefont {Qi}, \citenamefont {Ko}, \citenamefont {Deng}, \citenamefont {Riebesell}, \citenamefont {Ceder}, \citenamefont {Persson},\ and\ \citenamefont {Ong}}]{kaplanFoundationalPotentialEnergy2025}%
  \BibitemOpen
  \bibfield  {author} {\bibinfo {author} {\bibfnamefont {A.~D.}\ \bibnamefont {Kaplan}}, \bibinfo {author} {\bibfnamefont {R.}~\bibnamefont {Liu}}, \bibinfo {author} {\bibfnamefont {J.}~\bibnamefont {Qi}}, \bibinfo {author} {\bibfnamefont {T.~W.}\ \bibnamefont {Ko}}, \bibinfo {author} {\bibfnamefont {B.}~\bibnamefont {Deng}}, \bibinfo {author} {\bibfnamefont {J.}~\bibnamefont {Riebesell}}, \bibinfo {author} {\bibfnamefont {G.}~\bibnamefont {Ceder}}, \bibinfo {author} {\bibfnamefont {K.~A.}\ \bibnamefont {Persson}},\ and\ \bibinfo {author} {\bibfnamefont {S.~P.}\ \bibnamefont {Ong}},\ }\href {https://doi.org/10.48550/arXiv.2503.04070} {\bibinfo {title} {A {Foundational} {Potential} {Energy} {Surface} {Dataset} for {Materials}}} (\bibinfo {year} {2025}),\ \bibinfo {note} {arXiv:2503.04070 [cond-mat]}\BibitemShut {NoStop}%
\bibitem [{\citenamefont {Hakansson}\ and\ \citenamefont {Ross}(1989)}]{hakansson_thermal_1989}%
  \BibitemOpen
  \bibfield  {author} {\bibinfo {author} {\bibfnamefont {B.}~\bibnamefont {Hakansson}}\ and\ \bibinfo {author} {\bibfnamefont {R.~G.}\ \bibnamefont {Ross}},\ }\href {https://doi.org/10.1088/0953-8984/1/25/009} {\bibfield  {journal} {\bibinfo  {journal} {Journal of Physics: Condensed Matter}\ }\textbf {\bibinfo {volume} {1}},\ \bibinfo {pages} {3977} (\bibinfo {year} {1989})}\BibitemShut {NoStop}%
\bibitem [{\citenamefont {Eriksson}\ \emph {et~al.}(2019)\citenamefont {Eriksson}, \citenamefont {Fransson},\ and\ \citenamefont {Erhart}}]{eriksson_hiphive_2019}%
  \BibitemOpen
  \bibfield  {author} {\bibinfo {author} {\bibfnamefont {F.}~\bibnamefont {Eriksson}}, \bibinfo {author} {\bibfnamefont {E.}~\bibnamefont {Fransson}},\ and\ \bibinfo {author} {\bibfnamefont {P.}~\bibnamefont {Erhart}},\ }\href {https://doi.org/10.1002/adts.201800184} {\bibfield  {journal} {\bibinfo  {journal} {Advanced Theory and Simulations}\ }\textbf {\bibinfo {volume} {2}},\ \bibinfo {pages} {1800184} (\bibinfo {year} {2019})}\BibitemShut {NoStop}%
\bibitem [{\citenamefont {Togo}(2023)}]{togo_first-principles_2023}%
  \BibitemOpen
  \bibfield  {author} {\bibinfo {author} {\bibfnamefont {A.}~\bibnamefont {Togo}},\ }\href {https://doi.org/10.7566/JPSJ.92.012001} {\bibfield  {journal} {\bibinfo  {journal} {Journal of the Physical Society of Japan}\ }\textbf {\bibinfo {volume} {92}},\ \bibinfo {pages} {012001} (\bibinfo {year} {2023})}\BibitemShut {NoStop}%
\bibitem [{\citenamefont {Mounet}\ \emph {et~al.}(2018)\citenamefont {Mounet}, \citenamefont {Gibertini}, \citenamefont {Schwaller}, \citenamefont {Campi}, \citenamefont {Merkys}, \citenamefont {Marrazzo}, \citenamefont {Sohier}, \citenamefont {Castelli}, \citenamefont {Cepellotti}, \citenamefont {Pizzi},\ and\ \citenamefont {Marzari}}]{mounetTwodimensionalMaterialsHighthroughput2018}%
  \BibitemOpen
  \bibfield  {author} {\bibinfo {author} {\bibfnamefont {N.}~\bibnamefont {Mounet}}, \bibinfo {author} {\bibfnamefont {M.}~\bibnamefont {Gibertini}}, \bibinfo {author} {\bibfnamefont {P.}~\bibnamefont {Schwaller}}, \bibinfo {author} {\bibfnamefont {D.}~\bibnamefont {Campi}}, \bibinfo {author} {\bibfnamefont {A.}~\bibnamefont {Merkys}}, \bibinfo {author} {\bibfnamefont {A.}~\bibnamefont {Marrazzo}}, \bibinfo {author} {\bibfnamefont {T.}~\bibnamefont {Sohier}}, \bibinfo {author} {\bibfnamefont {I.~E.}\ \bibnamefont {Castelli}}, \bibinfo {author} {\bibfnamefont {A.}~\bibnamefont {Cepellotti}}, \bibinfo {author} {\bibfnamefont {G.}~\bibnamefont {Pizzi}},\ and\ \bibinfo {author} {\bibfnamefont {N.}~\bibnamefont {Marzari}},\ }\href {https://doi.org/10.1038/s41565-017-0035-5} {\bibfield  {journal} {\bibinfo  {journal} {Nature Nanotechnology}\ }\textbf {\bibinfo {volume} {13}},\ \bibinfo {pages} {246} (\bibinfo {year} {2018})},\ \bibinfo {note} {publisher: Nature Publishing Group}\BibitemShut {NoStop}%
\bibitem [{\citenamefont {Liang}\ \emph {et~al.}(2019)\citenamefont {Liang}, \citenamefont {Dwaraknath},\ and\ \citenamefont {Persson}}]{liang_high-throughput_2019}%
  \BibitemOpen
  \bibfield  {author} {\bibinfo {author} {\bibfnamefont {Q.}~\bibnamefont {Liang}}, \bibinfo {author} {\bibfnamefont {S.}~\bibnamefont {Dwaraknath}},\ and\ \bibinfo {author} {\bibfnamefont {K.~A.}\ \bibnamefont {Persson}},\ }\href {https://doi.org/10.1038/s41597-019-0138-y} {\bibfield  {journal} {\bibinfo  {journal} {Scientific Data}\ }\textbf {\bibinfo {volume} {6}},\ \bibinfo {pages} {135} (\bibinfo {year} {2019})}\BibitemShut {NoStop}%
\bibitem [{\citenamefont {Bagheri}\ and\ \citenamefont {Komsa}(2023)}]{bagheri_high-throughput_2023}%
  \BibitemOpen
  \bibfield  {author} {\bibinfo {author} {\bibfnamefont {M.}~\bibnamefont {Bagheri}}\ and\ \bibinfo {author} {\bibfnamefont {H.-P.}\ \bibnamefont {Komsa}},\ }\href {https://doi.org/10.1038/s41597-023-01988-5} {\bibfield  {journal} {\bibinfo  {journal} {Scientific Data}\ }\textbf {\bibinfo {volume} {10}},\ \bibinfo {pages} {80} (\bibinfo {year} {2023})}\BibitemShut {NoStop}%
\bibitem [{\citenamefont {Uhrin}\ \emph {et~al.}(2021)\citenamefont {Uhrin}, \citenamefont {Huber}, \citenamefont {Yu}, \citenamefont {Marzari},\ and\ \citenamefont {Pizzi}}]{Uhrin2021}%
  \BibitemOpen
  \bibfield  {author} {\bibinfo {author} {\bibfnamefont {M.}~\bibnamefont {Uhrin}}, \bibinfo {author} {\bibfnamefont {S.~P.}\ \bibnamefont {Huber}}, \bibinfo {author} {\bibfnamefont {J.}~\bibnamefont {Yu}}, \bibinfo {author} {\bibfnamefont {N.}~\bibnamefont {Marzari}},\ and\ \bibinfo {author} {\bibfnamefont {G.}~\bibnamefont {Pizzi}},\ }\href {https://doi.org/10.1016/j.commatsci.2020.110086} {\bibfield  {journal} {\bibinfo  {journal} {Comput. Mater. Sci.}\ }\textbf {\bibinfo {volume} {187}},\ \bibinfo {pages} {110086} (\bibinfo {year} {2021})}\BibitemShut {NoStop}%
\bibitem [{\citenamefont {Roy}\ \emph {et~al.}(2021)\citenamefont {Roy}, \citenamefont {Mandal},\ and\ \citenamefont {Pathak}}]{roy_machine_2021}%
  \BibitemOpen
  \bibfield  {author} {\bibinfo {author} {\bibfnamefont {D.}~\bibnamefont {Roy}}, \bibinfo {author} {\bibfnamefont {S.~C.}\ \bibnamefont {Mandal}},\ and\ \bibinfo {author} {\bibfnamefont {B.}~\bibnamefont {Pathak}},\ }\href {https://doi.org/10.1021/acsami.1c16696} {\bibfield  {journal} {\bibinfo  {journal} {ACS Applied Materials \& Interfaces}\ }\textbf {\bibinfo {volume} {13}},\ \bibinfo {pages} {56151} (\bibinfo {year} {2021})}\BibitemShut {NoStop}%
\bibitem [{\citenamefont {Jain}\ \emph {et~al.}(2013{\natexlab{b}})\citenamefont {Jain}, \citenamefont {Ong}, \citenamefont {Hautier}, \citenamefont {Chen}, \citenamefont {Richards}, \citenamefont {Dacek}, \citenamefont {Cholia}, \citenamefont {Gunter}, \citenamefont {Skinner}, \citenamefont {Ceder},\ and\ \citenamefont {Persson}}]{jain_commentary_2013}%
  \BibitemOpen
  \bibfield  {author} {\bibinfo {author} {\bibfnamefont {A.}~\bibnamefont {Jain}}, \bibinfo {author} {\bibfnamefont {S.~P.}\ \bibnamefont {Ong}}, \bibinfo {author} {\bibfnamefont {G.}~\bibnamefont {Hautier}}, \bibinfo {author} {\bibfnamefont {W.}~\bibnamefont {Chen}}, \bibinfo {author} {\bibfnamefont {W.~D.}\ \bibnamefont {Richards}}, \bibinfo {author} {\bibfnamefont {S.}~\bibnamefont {Dacek}}, \bibinfo {author} {\bibfnamefont {S.}~\bibnamefont {Cholia}}, \bibinfo {author} {\bibfnamefont {D.}~\bibnamefont {Gunter}}, \bibinfo {author} {\bibfnamefont {D.}~\bibnamefont {Skinner}}, \bibinfo {author} {\bibfnamefont {G.}~\bibnamefont {Ceder}},\ and\ \bibinfo {author} {\bibfnamefont {K.~A.}\ \bibnamefont {Persson}},\ }\href {https://doi.org/10.1063/1.4812323} {\bibfield  {journal} {\bibinfo  {journal} {APL Materials}\ }\textbf {\bibinfo {volume} {1}},\ \bibinfo {pages} {011002} (\bibinfo {year} {2013}{\natexlab{b}})}\BibitemShut {NoStop}%
\bibitem [{\citenamefont {Fu}\ \emph {et~al.}(2025)\citenamefont {Fu}, \citenamefont {Wood}, \citenamefont {Barroso-Luque}, \citenamefont {Levine}, \citenamefont {Gao}, \citenamefont {Dzamba},\ and\ \citenamefont {Zitnick}}]{fu_learning_2025}%
  \BibitemOpen
  \bibfield  {author} {\bibinfo {author} {\bibfnamefont {X.}~\bibnamefont {Fu}}, \bibinfo {author} {\bibfnamefont {B.~M.}\ \bibnamefont {Wood}}, \bibinfo {author} {\bibfnamefont {L.}~\bibnamefont {Barroso-Luque}}, \bibinfo {author} {\bibfnamefont {D.~S.}\ \bibnamefont {Levine}}, \bibinfo {author} {\bibfnamefont {M.}~\bibnamefont {Gao}}, \bibinfo {author} {\bibfnamefont {M.}~\bibnamefont {Dzamba}},\ and\ \bibinfo {author} {\bibfnamefont {C.~L.}\ \bibnamefont {Zitnick}},\ }\href {https://doi.org/10.48550/arXiv.2502.12147} {\bibinfo {title} {Learning smooth and expressive interatomic potentials for physical property prediction}} (\bibinfo {year} {2025}),\ \bibinfo {note} {arXiv:2502.12147 [physics] version: 1}\BibitemShut {NoStop}%
\bibitem [{\citenamefont {Rhodes}\ \emph {et~al.}(2025)\citenamefont {Rhodes}, \citenamefont {Vandenhaute}, \citenamefont {Šimkus}, \citenamefont {Gin}, \citenamefont {Godwin}, \citenamefont {Duignan},\ and\ \citenamefont {Neumann}}]{rhodes_orb-v3_2025}%
  \BibitemOpen
  \bibfield  {author} {\bibinfo {author} {\bibfnamefont {B.}~\bibnamefont {Rhodes}}, \bibinfo {author} {\bibfnamefont {S.}~\bibnamefont {Vandenhaute}}, \bibinfo {author} {\bibfnamefont {V.}~\bibnamefont {Šimkus}}, \bibinfo {author} {\bibfnamefont {J.}~\bibnamefont {Gin}}, \bibinfo {author} {\bibfnamefont {J.}~\bibnamefont {Godwin}}, \bibinfo {author} {\bibfnamefont {T.}~\bibnamefont {Duignan}},\ and\ \bibinfo {author} {\bibfnamefont {M.}~\bibnamefont {Neumann}},\ }\href {https://doi.org/10.48550/arXiv.2504.06231} {\bibinfo {title} {Orb-v3: atomistic simulation at scale}} (\bibinfo {year} {2025}),\ \bibinfo {note} {arXiv:2504.06231 [cond-mat]}\BibitemShut {NoStop}%
\bibitem [{egi(2025)}]{egip_nodate}%
  \BibitemOpen
  \href {https://www.radical-ai.com/news/EGIP} {\bibinfo {title} {{EGIP}: {Another} step forward in {ML}-first materials {\textbar} {Radical} {AI}}} (\bibinfo {year} {2025})\BibitemShut {NoStop}%
\bibitem [{\citenamefont {Baroni}\ \emph {et~al.}(2001)\citenamefont {Baroni}, \citenamefont {de~Gironcoli}, \citenamefont {Dal~Corso},\ and\ \citenamefont {Giannozzi}}]{baroni_phonons_2001}%
  \BibitemOpen
  \bibfield  {author} {\bibinfo {author} {\bibfnamefont {S.}~\bibnamefont {Baroni}}, \bibinfo {author} {\bibfnamefont {S.}~\bibnamefont {de~Gironcoli}}, \bibinfo {author} {\bibfnamefont {A.}~\bibnamefont {Dal~Corso}},\ and\ \bibinfo {author} {\bibfnamefont {P.}~\bibnamefont {Giannozzi}},\ }\href {https://doi.org/10.1103/RevModPhys.73.515} {\bibfield  {journal} {\bibinfo  {journal} {Reviews of Modern Physics}\ }\textbf {\bibinfo {volume} {73}},\ \bibinfo {pages} {515} (\bibinfo {year} {2001})}\BibitemShut {NoStop}%
\bibitem [{\citenamefont {Li}\ \emph {et~al.}(2025)\citenamefont {Li}, \citenamefont {Chen}, \citenamefont {Wang}, \citenamefont {Yang}, \citenamefont {Lu}, \citenamefont {Li}, \citenamefont {Chen}, \citenamefont {Zhu}, \citenamefont {Liu}, \citenamefont {Tan}, \citenamefont {Tang}, \citenamefont {Zhou}, \citenamefont {Zeni}, \citenamefont {Fowler}, \citenamefont {Zügner}, \citenamefont {Pinsler}, \citenamefont {Horton}, \citenamefont {Xie}, \citenamefont {Liu}, \citenamefont {Liu}, \citenamefont {Qin}, \citenamefont {Lv}, \citenamefont {Donadio},\ and\ \citenamefont {Hao}}]{li_probing_2025}%
  \BibitemOpen
  \bibfield  {author} {\bibinfo {author} {\bibfnamefont {J.}~\bibnamefont {Li}}, \bibinfo {author} {\bibfnamefont {Z.}~\bibnamefont {Chen}}, \bibinfo {author} {\bibfnamefont {Q.}~\bibnamefont {Wang}}, \bibinfo {author} {\bibfnamefont {H.}~\bibnamefont {Yang}}, \bibinfo {author} {\bibfnamefont {Z.}~\bibnamefont {Lu}}, \bibinfo {author} {\bibfnamefont {G.}~\bibnamefont {Li}}, \bibinfo {author} {\bibfnamefont {S.}~\bibnamefont {Chen}}, \bibinfo {author} {\bibfnamefont {Y.}~\bibnamefont {Zhu}}, \bibinfo {author} {\bibfnamefont {X.}~\bibnamefont {Liu}}, \bibinfo {author} {\bibfnamefont {J.}~\bibnamefont {Tan}}, \bibinfo {author} {\bibfnamefont {M.}~\bibnamefont {Tang}}, \bibinfo {author} {\bibfnamefont {Y.}~\bibnamefont {Zhou}}, \bibinfo {author} {\bibfnamefont {C.}~\bibnamefont {Zeni}}, \bibinfo {author} {\bibfnamefont {A.}~\bibnamefont {Fowler}}, \bibinfo {author} {\bibfnamefont {D.}~\bibnamefont {Zügner}}, \bibinfo {author} {\bibfnamefont {R.}~\bibnamefont {Pinsler}}, \bibinfo {author} {\bibfnamefont {M.}~\bibnamefont {Horton}}, \bibinfo {author} {\bibfnamefont {T.}~\bibnamefont {Xie}}, \bibinfo {author} {\bibfnamefont {T.-Y.}\ \bibnamefont {Liu}}, \bibinfo {author} {\bibfnamefont {H.}~\bibnamefont {Liu}}, \bibinfo {author} {\bibfnamefont {T.}~\bibnamefont {Qin}}, \bibinfo {author} {\bibfnamefont {B.}~\bibnamefont {Lv}}, \bibinfo {author} {\bibfnamefont {D.}~\bibnamefont {Donadio}},\ and\ \bibinfo {author} {\bibfnamefont {H.}~\bibnamefont {Hao}},\ }\href {https://doi.org/10.48550/arXiv.2503.11568} {\bibinfo {title} {Probing the limit of heat transfer in inorganic crystals with deep learning}} (\bibinfo {year} {2025})\BibitemShut {NoStop}%
\bibitem [{\citenamefont {Carrete}\ \emph {et~al.}(2017)\citenamefont {Carrete}, \citenamefont {Vermeersch}, \citenamefont {Katre}, \citenamefont {van Roekeghem}, \citenamefont {Wang}, \citenamefont {Madsen},\ and\ \citenamefont {Mingo}}]{carrete2017almabte}%
  \BibitemOpen
  \bibfield  {author} {\bibinfo {author} {\bibfnamefont {J.}~\bibnamefont {Carrete}}, \bibinfo {author} {\bibfnamefont {B.}~\bibnamefont {Vermeersch}}, \bibinfo {author} {\bibfnamefont {A.}~\bibnamefont {Katre}}, \bibinfo {author} {\bibfnamefont {A.}~\bibnamefont {van Roekeghem}}, \bibinfo {author} {\bibfnamefont {T.}~\bibnamefont {Wang}}, \bibinfo {author} {\bibfnamefont {G.~K.}\ \bibnamefont {Madsen}},\ and\ \bibinfo {author} {\bibfnamefont {N.}~\bibnamefont {Mingo}},\ }\href {https://doi.org/10.1016/j.cpc.2017.06.023} {\bibfield  {journal} {\bibinfo  {journal} {Computer Physics Communications}\ }\textbf {\bibinfo {volume} {220}},\ \bibinfo {pages} {351} (\bibinfo {year} {2017})}\BibitemShut {NoStop}%
\bibitem [{\citenamefont {Tadano}\ \emph {et~al.}(2014{\natexlab{a}})\citenamefont {Tadano}, \citenamefont {Gohda},\ and\ \citenamefont {Tsuneyuki}}]{alamode}%
  \BibitemOpen
  \bibfield  {author} {\bibinfo {author} {\bibfnamefont {T.}~\bibnamefont {Tadano}}, \bibinfo {author} {\bibfnamefont {Y.}~\bibnamefont {Gohda}},\ and\ \bibinfo {author} {\bibfnamefont {S.}~\bibnamefont {Tsuneyuki}},\ }\href {https://doi.org/https://doi.org/10.1088/0953-8984/26/22/225402} {\bibfield  {journal} {\bibinfo  {journal} {J. Phys. Condens. Matter}\ }\textbf {\bibinfo {volume} {26}},\ \bibinfo {pages} {225402} (\bibinfo {year} {2014}{\natexlab{a}})}\BibitemShut {NoStop}%
\bibitem [{\citenamefont {Barbalinardo}\ \emph {et~al.}(2020)\citenamefont {Barbalinardo}, \citenamefont {Chen}, \citenamefont {Lundgren},\ and\ \citenamefont {Donadio}}]{kaldo}%
  \BibitemOpen
  \bibfield  {author} {\bibinfo {author} {\bibfnamefont {G.}~\bibnamefont {Barbalinardo}}, \bibinfo {author} {\bibfnamefont {Z.}~\bibnamefont {Chen}}, \bibinfo {author} {\bibfnamefont {N.~W.}\ \bibnamefont {Lundgren}},\ and\ \bibinfo {author} {\bibfnamefont {D.}~\bibnamefont {Donadio}},\ }\href {https://doi.org/10.1063/5.0020443} {\bibfield  {journal} {\bibinfo  {journal} {Journal of Applied Physics}\ }\textbf {\bibinfo {volume} {128}},\ \bibinfo {pages} {135104} (\bibinfo {year} {2020})}\BibitemShut {NoStop}%
\bibitem [{\citenamefont {Paulatto}\ \emph {et~al.}(2013)\citenamefont {Paulatto}, \citenamefont {Mauri},\ and\ \citenamefont {Lazzeri}}]{paulatto_anharmonic_2013}%
  \BibitemOpen
  \bibfield  {author} {\bibinfo {author} {\bibfnamefont {L.}~\bibnamefont {Paulatto}}, \bibinfo {author} {\bibfnamefont {F.}~\bibnamefont {Mauri}},\ and\ \bibinfo {author} {\bibfnamefont {M.}~\bibnamefont {Lazzeri}},\ }\href {https://doi.org/10.1103/PhysRevB.87.214303} {\bibfield  {journal} {\bibinfo  {journal} {Physical Review B}\ }\textbf {\bibinfo {volume} {87}},\ \bibinfo {pages} {214303} (\bibinfo {year} {2013})}\BibitemShut {NoStop}%
\bibitem [{\citenamefont {Fugallo}\ \emph {et~al.}(2013)\citenamefont {Fugallo}, \citenamefont {Lazzeri}, \citenamefont {Paulatto},\ and\ \citenamefont {Mauri}}]{fugallo2013ab}%
  \BibitemOpen
  \bibfield  {author} {\bibinfo {author} {\bibfnamefont {G.}~\bibnamefont {Fugallo}}, \bibinfo {author} {\bibfnamefont {M.}~\bibnamefont {Lazzeri}}, \bibinfo {author} {\bibfnamefont {L.}~\bibnamefont {Paulatto}},\ and\ \bibinfo {author} {\bibfnamefont {F.}~\bibnamefont {Mauri}},\ }\href {https://doi.org/10.1103/PhysRevB.88.045430} {\bibfield  {journal} {\bibinfo  {journal} {Phys. Rev. B}\ }\textbf {\bibinfo {volume} {88}},\ \bibinfo {pages} {045430} (\bibinfo {year} {2013})}\BibitemShut {NoStop}%
\bibitem [{\citenamefont {Cepellotti}\ \emph {et~al.}(2022)\citenamefont {Cepellotti}, \citenamefont {Coulter}, \citenamefont {Johansson}, \citenamefont {Fedorova},\ and\ \citenamefont {Kozinsky}}]{cepellotti_phoebe_2022}%
  \BibitemOpen
  \bibfield  {author} {\bibinfo {author} {\bibfnamefont {A.}~\bibnamefont {Cepellotti}}, \bibinfo {author} {\bibfnamefont {J.}~\bibnamefont {Coulter}}, \bibinfo {author} {\bibfnamefont {A.}~\bibnamefont {Johansson}}, \bibinfo {author} {\bibfnamefont {N.~S.}\ \bibnamefont {Fedorova}},\ and\ \bibinfo {author} {\bibfnamefont {B.}~\bibnamefont {Kozinsky}},\ }\href {https://doi.org/10.1088/2515-7639/ac86f6} {\bibfield  {journal} {\bibinfo  {journal} {Journal of Physics: Materials}\ }\textbf {\bibinfo {volume} {5}},\ \bibinfo {pages} {035003} (\bibinfo {year} {2022})}\BibitemShut {NoStop}%
\bibitem [{\citenamefont {Savić}\ \emph {et~al.}(2013)\citenamefont {Savić}, \citenamefont {Donadio}, \citenamefont {Gygi},\ and\ \citenamefont {Galli}}]{savic_dimensionality_2013}%
  \BibitemOpen
  \bibfield  {author} {\bibinfo {author} {\bibfnamefont {I.}~\bibnamefont {Savić}}, \bibinfo {author} {\bibfnamefont {D.}~\bibnamefont {Donadio}}, \bibinfo {author} {\bibfnamefont {F.}~\bibnamefont {Gygi}},\ and\ \bibinfo {author} {\bibfnamefont {G.}~\bibnamefont {Galli}},\ }\href {https://pubs.aip.org/aip/apl/article/102/7/073113/24193} {\bibfield  {journal} {\bibinfo  {journal} {Applied Physics Letters}\ }\textbf {\bibinfo {volume} {102}} (\bibinfo {year} {2013})}\BibitemShut {NoStop}%
\bibitem [{\citenamefont {Tamura}(1983)}]{tamura_isotope}%
  \BibitemOpen
  \bibfield  {author} {\bibinfo {author} {\bibfnamefont {S.-i.}\ \bibnamefont {Tamura}},\ }\href {https://doi.org/10.1103/PhysRevB.27.858} {\bibfield  {journal} {\bibinfo  {journal} {Phys. Rev. B}\ }\textbf {\bibinfo {volume} {27}},\ \bibinfo {pages} {858} (\bibinfo {year} {1983})}\BibitemShut {NoStop}%
\bibitem [{\citenamefont {Foss}\ and\ \citenamefont {Aksamija}(2020)}]{foss_effects_2020}%
  \BibitemOpen
  \bibfield  {author} {\bibinfo {author} {\bibfnamefont {C.~J.}\ \bibnamefont {Foss}}\ and\ \bibinfo {author} {\bibfnamefont {Z.}~\bibnamefont {Aksamija}},\ }\href {https://doi.org/10.1103/PhysRevMaterials.4.124006} {\bibfield  {journal} {\bibinfo  {journal} {Physical Review Materials}\ }\textbf {\bibinfo {volume} {4}},\ \bibinfo {pages} {124006} (\bibinfo {year} {2020})}\BibitemShut {NoStop}%
\bibitem [{\citenamefont {Hahn}\ \emph {et~al.}(2021)\citenamefont {Hahn}, \citenamefont {Melis}, \citenamefont {Bernardini},\ and\ \citenamefont {Colombo}}]{hahn_engineering_2021}%
  \BibitemOpen
  \bibfield  {author} {\bibinfo {author} {\bibfnamefont {K.~R.}\ \bibnamefont {Hahn}}, \bibinfo {author} {\bibfnamefont {C.}~\bibnamefont {Melis}}, \bibinfo {author} {\bibfnamefont {F.}~\bibnamefont {Bernardini}},\ and\ \bibinfo {author} {\bibfnamefont {L.}~\bibnamefont {Colombo}},\ }\href {https://doi.org/10.3389/fmech.2021.712989} {\bibfield  {journal} {\bibinfo  {journal} {Frontiers in Mechanical Engineering}\ }\textbf {\bibinfo {volume} {7}},\ \bibinfo {pages} {712989} (\bibinfo {year} {2021})}\BibitemShut {NoStop}%
\bibitem [{\citenamefont {Abou El~Kheir}\ \emph {et~al.}(2024)\citenamefont {Abou El~Kheir}, \citenamefont {Bonati}, \citenamefont {Parrinello},\ and\ \citenamefont {Bernasconi}}]{abou_el_kheir_unraveling_2024}%
  \BibitemOpen
  \bibfield  {author} {\bibinfo {author} {\bibfnamefont {O.}~\bibnamefont {Abou El~Kheir}}, \bibinfo {author} {\bibfnamefont {L.}~\bibnamefont {Bonati}}, \bibinfo {author} {\bibfnamefont {M.}~\bibnamefont {Parrinello}},\ and\ \bibinfo {author} {\bibfnamefont {M.}~\bibnamefont {Bernasconi}},\ }\href {https://www.nature.com/articles/s41524-024-01217-6} {\bibfield  {journal} {\bibinfo  {journal} {npj Computational Materials}\ }\textbf {\bibinfo {volume} {10}},\ \bibinfo {pages} {33} (\bibinfo {year} {2024})}\BibitemShut {NoStop}%
\bibitem [{\citenamefont {Stewart}\ \emph {et~al.}(2009)\citenamefont {Stewart}, \citenamefont {Savić},\ and\ \citenamefont {Mingo}}]{stewart_first-principles_2009}%
  \BibitemOpen
  \bibfield  {author} {\bibinfo {author} {\bibfnamefont {D.~A.}\ \bibnamefont {Stewart}}, \bibinfo {author} {\bibfnamefont {I.}~\bibnamefont {Savić}},\ and\ \bibinfo {author} {\bibfnamefont {N.}~\bibnamefont {Mingo}},\ }\href {https://doi.org/10.1021/nl802503q} {\bibfield  {journal} {\bibinfo  {journal} {Nano Letters}\ }\textbf {\bibinfo {volume} {9}},\ \bibinfo {pages} {81} (\bibinfo {year} {2009})}\BibitemShut {NoStop}%
\bibitem [{\citenamefont {Togo}\ and\ \citenamefont {Seko}(2024)}]{togo_--fly_2024}%
  \BibitemOpen
  \bibfield  {author} {\bibinfo {author} {\bibfnamefont {A.}~\bibnamefont {Togo}}\ and\ \bibinfo {author} {\bibfnamefont {A.}~\bibnamefont {Seko}},\ }\href {https://doi.org/10.1063/5.0211296} {\bibfield  {journal} {\bibinfo  {journal} {The Journal of Chemical Physics}\ }\textbf {\bibinfo {volume} {160}},\ \bibinfo {pages} {211001} (\bibinfo {year} {2024})}\BibitemShut {NoStop}%
\bibitem [{\citenamefont {Tadano}\ \emph {et~al.}(2014{\natexlab{b}})\citenamefont {Tadano}, \citenamefont {Gohda},\ and\ \citenamefont {Tsuneyuki}}]{tadano_anharmonic_2014}%
  \BibitemOpen
  \bibfield  {author} {\bibinfo {author} {\bibfnamefont {T.}~\bibnamefont {Tadano}}, \bibinfo {author} {\bibfnamefont {Y.}~\bibnamefont {Gohda}},\ and\ \bibinfo {author} {\bibfnamefont {S.}~\bibnamefont {Tsuneyuki}},\ }\href {https://doi.org/10.1088/0953-8984/26/22/225402} {\bibfield  {journal} {\bibinfo  {journal} {Journal of Physics: Condensed Matter}\ }\textbf {\bibinfo {volume} {26}},\ \bibinfo {pages} {225402} (\bibinfo {year} {2014}{\natexlab{b}})}\BibitemShut {NoStop}%
\bibitem [{\citenamefont {Li}\ \emph {et~al.}(2014)\citenamefont {Li}, \citenamefont {Carrete}, \citenamefont {A.~Katcho},\ and\ \citenamefont {Mingo}}]{li_shengbte_2014}%
  \BibitemOpen
  \bibfield  {author} {\bibinfo {author} {\bibfnamefont {W.}~\bibnamefont {Li}}, \bibinfo {author} {\bibfnamefont {J.}~\bibnamefont {Carrete}}, \bibinfo {author} {\bibfnamefont {N.}~\bibnamefont {A.~Katcho}},\ and\ \bibinfo {author} {\bibfnamefont {N.}~\bibnamefont {Mingo}},\ }\href {https://doi.org/10.1016/j.cpc.2014.02.015} {\bibfield  {journal} {\bibinfo  {journal} {Computer Physics Communications}\ }\textbf {\bibinfo {volume} {185}},\ \bibinfo {pages} {1747} (\bibinfo {year} {2014})}\BibitemShut {NoStop}%
\bibitem [{\citenamefont {Hellman}\ and\ \citenamefont {Abrikosov}(2013)}]{hellman_temperature-dependent_2013}%
  \BibitemOpen
  \bibfield  {author} {\bibinfo {author} {\bibfnamefont {O.}~\bibnamefont {Hellman}}\ and\ \bibinfo {author} {\bibfnamefont {I.~A.}\ \bibnamefont {Abrikosov}},\ }\bibfield  {journal} {\bibinfo  {journal} {Physical Review B}\ }\textbf {\bibinfo {volume} {88}},\ \href {https://doi.org/10.1103/PhysRevB.88.144301} {10.1103/PhysRevB.88.144301} (\bibinfo {year} {2013})\BibitemShut {NoStop}%
\bibitem [{\citenamefont {Dragašević}\ and\ \citenamefont {Simoncelli}(2023)}]{dragasevic_viscous_2023}%
  \BibitemOpen
  \bibfield  {author} {\bibinfo {author} {\bibfnamefont {J.}~\bibnamefont {Dragašević}}\ and\ \bibinfo {author} {\bibfnamefont {M.}~\bibnamefont {Simoncelli}},\ }\href {http://arxiv.org/abs/2303.12777} {\bibfield  {journal} {\bibinfo  {journal} {arXiv:2303.12777}\ } (\bibinfo {year} {2023})}\BibitemShut {NoStop}%
\bibitem [{\citenamefont {Di~Lucente}\ \emph {et~al.}(2023)\citenamefont {Di~Lucente}, \citenamefont {Simoncelli},\ and\ \citenamefont {Marzari}}]{di_lucente_crossover_2023}%
  \BibitemOpen
  \bibfield  {author} {\bibinfo {author} {\bibfnamefont {E.}~\bibnamefont {Di~Lucente}}, \bibinfo {author} {\bibfnamefont {M.}~\bibnamefont {Simoncelli}},\ and\ \bibinfo {author} {\bibfnamefont {N.}~\bibnamefont {Marzari}},\ }\href {https://doi.org/10.1103/PhysRevResearch.5.033125} {\bibfield  {journal} {\bibinfo  {journal} {Physical Review Research}\ }\textbf {\bibinfo {volume} {5}},\ \bibinfo {pages} {033125} (\bibinfo {year} {2023})}\BibitemShut {NoStop}%
\bibitem [{\citenamefont {Jain}(2020)}]{jain_multichannel_2020}%
  \BibitemOpen
  \bibfield  {author} {\bibinfo {author} {\bibfnamefont {A.}~\bibnamefont {Jain}},\ }\href {https://doi.org/10.1103/PhysRevB.102.201201} {\bibfield  {journal} {\bibinfo  {journal} {Phys. Rev. B}\ }\textbf {\bibinfo {volume} {102}},\ \bibinfo {pages} {201201} (\bibinfo {year} {2020})}\BibitemShut {NoStop}%
\bibitem [{\citenamefont {Xia}\ \emph {et~al.}(2020)\citenamefont {Xia}, \citenamefont {Hegde}, \citenamefont {Pal}, \citenamefont {Hua}, \citenamefont {Gaines}, \citenamefont {Patel}, \citenamefont {He}, \citenamefont {Aykol},\ and\ \citenamefont {Wolverton}}]{xia_microscopic_2020}%
  \BibitemOpen
  \bibfield  {author} {\bibinfo {author} {\bibfnamefont {Y.}~\bibnamefont {Xia}}, \bibinfo {author} {\bibfnamefont {V.~I.}\ \bibnamefont {Hegde}}, \bibinfo {author} {\bibfnamefont {K.}~\bibnamefont {Pal}}, \bibinfo {author} {\bibfnamefont {X.}~\bibnamefont {Hua}}, \bibinfo {author} {\bibfnamefont {D.}~\bibnamefont {Gaines}}, \bibinfo {author} {\bibfnamefont {S.}~\bibnamefont {Patel}}, \bibinfo {author} {\bibfnamefont {J.}~\bibnamefont {He}}, \bibinfo {author} {\bibfnamefont {M.}~\bibnamefont {Aykol}},\ and\ \bibinfo {author} {\bibfnamefont {C.}~\bibnamefont {Wolverton}},\ }\href {https://doi.org/10.1103/PhysRevX.10.041029} {\bibfield  {journal} {\bibinfo  {journal} {Phys. Rev. X}\ }\textbf {\bibinfo {volume} {10}},\ \bibinfo {pages} {041029} (\bibinfo {year} {2020})}\BibitemShut {NoStop}%
\bibitem [{\citenamefont {Shen}\ \emph {et~al.}(2024)\citenamefont {Shen}, \citenamefont {Ouyang}, \citenamefont {Huang}, \citenamefont {Tung}, \citenamefont {Yang}, \citenamefont {Faizan}, \citenamefont {Perez}, \citenamefont {He}, \citenamefont {Sotnikov}, \citenamefont {Willa}, \citenamefont {Wang}, \citenamefont {Chen},\ and\ \citenamefont {Guilmeau}}]{shen_amorphous-like_2024}%
  \BibitemOpen
  \bibfield  {author} {\bibinfo {author} {\bibfnamefont {X.}~\bibnamefont {Shen}}, \bibinfo {author} {\bibfnamefont {N.}~\bibnamefont {Ouyang}}, \bibinfo {author} {\bibfnamefont {Y.}~\bibnamefont {Huang}}, \bibinfo {author} {\bibfnamefont {Y.-H.}\ \bibnamefont {Tung}}, \bibinfo {author} {\bibfnamefont {C.-C.}\ \bibnamefont {Yang}}, \bibinfo {author} {\bibfnamefont {M.}~\bibnamefont {Faizan}}, \bibinfo {author} {\bibfnamefont {N.}~\bibnamefont {Perez}}, \bibinfo {author} {\bibfnamefont {R.}~\bibnamefont {He}}, \bibinfo {author} {\bibfnamefont {A.}~\bibnamefont {Sotnikov}}, \bibinfo {author} {\bibfnamefont {K.}~\bibnamefont {Willa}}, \bibinfo {author} {\bibfnamefont {C.}~\bibnamefont {Wang}}, \bibinfo {author} {\bibfnamefont {Y.}~\bibnamefont {Chen}},\ and\ \bibinfo {author} {\bibfnamefont {E.}~\bibnamefont {Guilmeau}},\ }\bibfield  {journal} {\bibinfo  {journal} {Advanced Science}\ }\textbf {\bibinfo {volume} {11}},\ \href {https://doi.org/10.1002/advs.202400258} {10.1002/advs.202400258} (\bibinfo {year} {2024})\BibitemShut {NoStop}%
\bibitem [{\citenamefont {Pandey}\ \emph {et~al.}(2022)\citenamefont {Pandey}, \citenamefont {Du}, \citenamefont {Parker},\ and\ \citenamefont {Lindsay}}]{pandey_origin_2022}%
  \BibitemOpen
  \bibfield  {author} {\bibinfo {author} {\bibfnamefont {T.}~\bibnamefont {Pandey}}, \bibinfo {author} {\bibfnamefont {M.-H.}\ \bibnamefont {Du}}, \bibinfo {author} {\bibfnamefont {D.~S.}\ \bibnamefont {Parker}},\ and\ \bibinfo {author} {\bibfnamefont {L.}~\bibnamefont {Lindsay}},\ }\href {https://doi.org/10.1016/j.mtphys.2022.100881} {\bibfield  {journal} {\bibinfo  {journal} {Materials Today Physics}\ }\textbf {\bibinfo {volume} {28}},\ \bibinfo {pages} {100881} (\bibinfo {year} {2022})}\BibitemShut {NoStop}%
\bibitem [{\citenamefont {Fiorentino}\ and\ \citenamefont {Baroni}(2023)}]{fiorentino_green-kubo_2023}%
  \BibitemOpen
  \bibfield  {author} {\bibinfo {author} {\bibfnamefont {A.}~\bibnamefont {Fiorentino}}\ and\ \bibinfo {author} {\bibfnamefont {S.}~\bibnamefont {Baroni}},\ }\href {https://doi.org/10.1103/PhysRevB.107.054311} {\bibfield  {journal} {\bibinfo  {journal} {Physical Review B}\ }\textbf {\bibinfo {volume} {107}},\ \bibinfo {pages} {054311} (\bibinfo {year} {2023})}\BibitemShut {NoStop}%
\bibitem [{\citenamefont {Zhou}\ \emph {et~al.}(2024)\citenamefont {Zhou}, \citenamefont {Xiao}, \citenamefont {Ma}, \citenamefont {Gofryk}, \citenamefont {Jiang}, \citenamefont {Manley}, \citenamefont {Hurley},\ and\ \citenamefont {Marianetti}}]{zhou_phonon_2024}%
  \BibitemOpen
  \bibfield  {author} {\bibinfo {author} {\bibfnamefont {S.}~\bibnamefont {Zhou}}, \bibinfo {author} {\bibfnamefont {E.}~\bibnamefont {Xiao}}, \bibinfo {author} {\bibfnamefont {H.}~\bibnamefont {Ma}}, \bibinfo {author} {\bibfnamefont {K.}~\bibnamefont {Gofryk}}, \bibinfo {author} {\bibfnamefont {C.}~\bibnamefont {Jiang}}, \bibinfo {author} {\bibfnamefont {M.~E.}\ \bibnamefont {Manley}}, \bibinfo {author} {\bibfnamefont {D.~H.}\ \bibnamefont {Hurley}},\ and\ \bibinfo {author} {\bibfnamefont {C.~A.}\ \bibnamefont {Marianetti}},\ }\href {https://doi.org/10.1103/PhysRevLett.132.106502} {\bibfield  {journal} {\bibinfo  {journal} {Physical Review Letters}\ }\textbf {\bibinfo {volume} {132}},\ \bibinfo {pages} {106502} (\bibinfo {year} {2024})}\BibitemShut {NoStop}%
\bibitem [{\citenamefont {Zheng}\ \emph {et~al.}(2024)\citenamefont {Zheng}, \citenamefont {Lin}, \citenamefont {Lin}, \citenamefont {Hautier}, \citenamefont {Guo},\ and\ \citenamefont {Huang}}]{zheng_unravelling_2024}%
  \BibitemOpen
  \bibfield  {author} {\bibinfo {author} {\bibfnamefont {J.}~\bibnamefont {Zheng}}, \bibinfo {author} {\bibfnamefont {C.}~\bibnamefont {Lin}}, \bibinfo {author} {\bibfnamefont {C.}~\bibnamefont {Lin}}, \bibinfo {author} {\bibfnamefont {G.}~\bibnamefont {Hautier}}, \bibinfo {author} {\bibfnamefont {R.}~\bibnamefont {Guo}},\ and\ \bibinfo {author} {\bibfnamefont {B.}~\bibnamefont {Huang}},\ }\href {https://www.nature.com/articles/s41524-024-01211-y} {\bibfield  {journal} {\bibinfo  {journal} {npj Computational Materials}\ }\textbf {\bibinfo {volume} {10}} (\bibinfo {year} {2024})}\BibitemShut {NoStop}%
\bibitem [{\citenamefont {Jia}\ \emph {et~al.}(2023)\citenamefont {Jia}, \citenamefont {Zhao}, \citenamefont {Wu}, \citenamefont {Chen}, \citenamefont {Liu},\ and\ \citenamefont {Wu}}]{jia_cu3bis3_2023}%
  \BibitemOpen
  \bibfield  {author} {\bibinfo {author} {\bibfnamefont {F.}~\bibnamefont {Jia}}, \bibinfo {author} {\bibfnamefont {S.}~\bibnamefont {Zhao}}, \bibinfo {author} {\bibfnamefont {J.}~\bibnamefont {Wu}}, \bibinfo {author} {\bibfnamefont {L.}~\bibnamefont {Chen}}, \bibinfo {author} {\bibfnamefont {T.-H.}\ \bibnamefont {Liu}},\ and\ \bibinfo {author} {\bibfnamefont {L.-M.}\ \bibnamefont {Wu}},\ }\href {https://doi.org/10.1002/ange.202315642} {\bibfield  {journal} {\bibinfo  {journal} {Angewandte Chemie}\ }\textbf {\bibinfo {volume} {135}},\ \bibinfo {pages} {e202315642} (\bibinfo {year} {2023})}\BibitemShut {NoStop}%
\bibitem [{\citenamefont {Tadano}\ and\ \citenamefont {Saidi}(2022)}]{tadano_first-principles_2022}%
  \BibitemOpen
  \bibfield  {author} {\bibinfo {author} {\bibfnamefont {T.}~\bibnamefont {Tadano}}\ and\ \bibinfo {author} {\bibfnamefont {W.~A.}\ \bibnamefont {Saidi}},\ }\href {https://doi.org/10.1103/PhysRevLett.129.185901} {\bibfield  {journal} {\bibinfo  {journal} {Physical Review Letters}\ }\textbf {\bibinfo {volume} {129}},\ \bibinfo {pages} {185901} (\bibinfo {year} {2022})}\BibitemShut {NoStop}%
\bibitem [{\citenamefont {Bernges}\ \emph {et~al.}(2022)\citenamefont {Bernges}, \citenamefont {Hanus}, \citenamefont {Wankmiller}, \citenamefont {Imasato}, \citenamefont {Lin}, \citenamefont {Ghidiu}, \citenamefont {Gerlitz}, \citenamefont {Peterlechner}, \citenamefont {Graham}, \citenamefont {Hautier}, \citenamefont {Pei}, \citenamefont {Hansen}, \citenamefont {Wilde}, \citenamefont {Snyder}, \citenamefont {George}, \citenamefont {Agne},\ and\ \citenamefont {Zeier}}]{bernges_considering_2022}%
  \BibitemOpen
  \bibfield  {author} {\bibinfo {author} {\bibfnamefont {T.}~\bibnamefont {Bernges}}, \bibinfo {author} {\bibfnamefont {R.}~\bibnamefont {Hanus}}, \bibinfo {author} {\bibfnamefont {B.}~\bibnamefont {Wankmiller}}, \bibinfo {author} {\bibfnamefont {K.}~\bibnamefont {Imasato}}, \bibinfo {author} {\bibfnamefont {S.}~\bibnamefont {Lin}}, \bibinfo {author} {\bibfnamefont {M.}~\bibnamefont {Ghidiu}}, \bibinfo {author} {\bibfnamefont {M.}~\bibnamefont {Gerlitz}}, \bibinfo {author} {\bibfnamefont {M.}~\bibnamefont {Peterlechner}}, \bibinfo {author} {\bibfnamefont {S.}~\bibnamefont {Graham}}, \bibinfo {author} {\bibfnamefont {G.}~\bibnamefont {Hautier}}, \bibinfo {author} {\bibfnamefont {Y.}~\bibnamefont {Pei}}, \bibinfo {author} {\bibfnamefont {M.~R.}\ \bibnamefont {Hansen}}, \bibinfo {author} {\bibfnamefont {G.}~\bibnamefont {Wilde}}, \bibinfo {author} {\bibfnamefont {G.~J.}\ \bibnamefont {Snyder}}, \bibinfo {author} {\bibfnamefont {J.}~\bibnamefont {George}}, \bibinfo {author} {\bibfnamefont {M.~T.}\ \bibnamefont {Agne}},\ and\ \bibinfo {author} {\bibfnamefont {W.~G.}\ \bibnamefont {Zeier}},\ }\href {https://doi.org/10.1002/aenm.202200717} {\bibfield  {journal} {\bibinfo  {journal} {Advanced Energy Materials}\ }\textbf {\bibinfo {volume} {12}},\ \bibinfo {pages} {2200717} (\bibinfo {year} {2022})}\BibitemShut {NoStop}%
\bibitem [{\citenamefont {Yang}\ \emph {et~al.}(2022)\citenamefont {Yang}, \citenamefont {Tiwari},\ and\ \citenamefont {Feng}}]{yang_reduced_2022}%
  \BibitemOpen
  \bibfield  {author} {\bibinfo {author} {\bibfnamefont {X.}~\bibnamefont {Yang}}, \bibinfo {author} {\bibfnamefont {J.}~\bibnamefont {Tiwari}},\ and\ \bibinfo {author} {\bibfnamefont {T.}~\bibnamefont {Feng}},\ }\href {https://doi.org/10.1016/j.mtphys.2022.100689} {\bibfield  {journal} {\bibinfo  {journal} {Materials Today Physics}\ }\textbf {\bibinfo {volume} {24}},\ \bibinfo {pages} {100689} (\bibinfo {year} {2022})}\BibitemShut {NoStop}%
\bibitem [{\citenamefont {Zeng}\ \emph {et~al.}(2024)\citenamefont {Zeng}, \citenamefont {Shen}, \citenamefont {Cheng}, \citenamefont {Perez}, \citenamefont {Ouyang}, \citenamefont {Fan}, \citenamefont {Lemoine}, \citenamefont {Raveau}, \citenamefont {Guilmeau},\ and\ \citenamefont {Chen}}]{zeng_pushing_2024}%
  \BibitemOpen
  \bibfield  {author} {\bibinfo {author} {\bibfnamefont {Z.}~\bibnamefont {Zeng}}, \bibinfo {author} {\bibfnamefont {X.}~\bibnamefont {Shen}}, \bibinfo {author} {\bibfnamefont {R.}~\bibnamefont {Cheng}}, \bibinfo {author} {\bibfnamefont {O.}~\bibnamefont {Perez}}, \bibinfo {author} {\bibfnamefont {N.}~\bibnamefont {Ouyang}}, \bibinfo {author} {\bibfnamefont {Z.}~\bibnamefont {Fan}}, \bibinfo {author} {\bibfnamefont {P.}~\bibnamefont {Lemoine}}, \bibinfo {author} {\bibfnamefont {B.}~\bibnamefont {Raveau}}, \bibinfo {author} {\bibfnamefont {E.}~\bibnamefont {Guilmeau}},\ and\ \bibinfo {author} {\bibfnamefont {Y.}~\bibnamefont {Chen}},\ }\href {https://www.nature.com/articles/s41467-024-46799-3} {\bibfield  {journal} {\bibinfo  {journal} {Nature Communications}\ }\textbf {\bibinfo {volume} {15}} (\bibinfo {year} {2024})}\BibitemShut {NoStop}%
\bibitem [{\citenamefont {Caldarelli}\ \emph {et~al.}(2022)\citenamefont {Caldarelli}, \citenamefont {Simoncelli}, \citenamefont {Marzari}, \citenamefont {Mauri},\ and\ \citenamefont {Benfatto}}]{caldarelli_many-body_2022}%
  \BibitemOpen
  \bibfield  {author} {\bibinfo {author} {\bibfnamefont {G.}~\bibnamefont {Caldarelli}}, \bibinfo {author} {\bibfnamefont {M.}~\bibnamefont {Simoncelli}}, \bibinfo {author} {\bibfnamefont {N.}~\bibnamefont {Marzari}}, \bibinfo {author} {\bibfnamefont {F.}~\bibnamefont {Mauri}},\ and\ \bibinfo {author} {\bibfnamefont {L.}~\bibnamefont {Benfatto}},\ }\href {https://doi.org/10.1103/PhysRevB.106.024312} {\bibfield  {journal} {\bibinfo  {journal} {Physical Review B}\ }\textbf {\bibinfo {volume} {106}},\ \bibinfo {pages} {024312} (\bibinfo {year} {2022})}\BibitemShut {NoStop}%
\bibitem [{\citenamefont {Xia}\ \emph {et~al.}(2023)\citenamefont {Xia}, \citenamefont {Gaines}, \citenamefont {He}, \citenamefont {Pal}, \citenamefont {Li}, \citenamefont {Kanatzidis}, \citenamefont {Ozoliņš},\ and\ \citenamefont {Wolverton}}]{xia_unified_2023}%
  \BibitemOpen
  \bibfield  {author} {\bibinfo {author} {\bibfnamefont {Y.}~\bibnamefont {Xia}}, \bibinfo {author} {\bibfnamefont {D.}~\bibnamefont {Gaines}}, \bibinfo {author} {\bibfnamefont {J.}~\bibnamefont {He}}, \bibinfo {author} {\bibfnamefont {K.}~\bibnamefont {Pal}}, \bibinfo {author} {\bibfnamefont {Z.}~\bibnamefont {Li}}, \bibinfo {author} {\bibfnamefont {M.~G.}\ \bibnamefont {Kanatzidis}}, \bibinfo {author} {\bibfnamefont {V.}~\bibnamefont {Ozoliņš}},\ and\ \bibinfo {author} {\bibfnamefont {C.}~\bibnamefont {Wolverton}},\ }\href {https://doi.org/10.1073/pnas.2302541120} {\bibfield  {journal} {\bibinfo  {journal} {Proceedings of the National Academy of Sciences}\ }\textbf {\bibinfo {volume} {120}},\ \bibinfo {pages} {e2302541120} (\bibinfo {year} {2023})}\BibitemShut {NoStop}%
\bibitem [{\citenamefont {Luo}\ \emph {et~al.}(2020)\citenamefont {Luo}, \citenamefont {Yang}, \citenamefont {Feng}, \citenamefont {Wang},\ and\ \citenamefont {Ruan}}]{luo_vibrational_2020}%
  \BibitemOpen
  \bibfield  {author} {\bibinfo {author} {\bibfnamefont {Y.}~\bibnamefont {Luo}}, \bibinfo {author} {\bibfnamefont {X.}~\bibnamefont {Yang}}, \bibinfo {author} {\bibfnamefont {T.}~\bibnamefont {Feng}}, \bibinfo {author} {\bibfnamefont {J.}~\bibnamefont {Wang}},\ and\ \bibinfo {author} {\bibfnamefont {X.}~\bibnamefont {Ruan}},\ }\href {https://doi.org/10.1038/s41467-020-16371-w} {\bibfield  {journal} {\bibinfo  {journal} {Nature Communications}\ }\textbf {\bibinfo {volume} {11}},\ \bibinfo {pages} {2554} (\bibinfo {year} {2020})}\BibitemShut {NoStop}%
\bibitem [{\citenamefont {Harper}\ \emph {et~al.}(2024)\citenamefont {Harper}, \citenamefont {Iwanowski}, \citenamefont {Witt}, \citenamefont {Payne},\ and\ \citenamefont {Simoncelli}}]{harper_vibrational_2023}%
  \BibitemOpen
  \bibfield  {author} {\bibinfo {author} {\bibfnamefont {A.~F.}\ \bibnamefont {Harper}}, \bibinfo {author} {\bibfnamefont {K.}~\bibnamefont {Iwanowski}}, \bibinfo {author} {\bibfnamefont {W.~C.}\ \bibnamefont {Witt}}, \bibinfo {author} {\bibfnamefont {M.~C.}\ \bibnamefont {Payne}},\ and\ \bibinfo {author} {\bibfnamefont {M.}~\bibnamefont {Simoncelli}},\ }\href {https://doi.org/10.1103/PhysRevMaterials.8.043601} {\bibfield  {journal} {\bibinfo  {journal} {Physical Review Materials}\ }\textbf {\bibinfo {volume} {8}},\ \bibinfo {pages} {043601} (\bibinfo {year} {2024})}\BibitemShut {NoStop}%
\bibitem [{\citenamefont {Liu}\ \emph {et~al.}(2023)\citenamefont {Liu}, \citenamefont {Liang}, \citenamefont {Yang}, \citenamefont {Yang}, \citenamefont {Yang}, \citenamefont {Song}, \citenamefont {Mei}, \citenamefont {Cs{\'a}nyi},\ and\ \citenamefont {Cao}}]{liu2023unraveling}%
  \BibitemOpen
  \bibfield  {author} {\bibinfo {author} {\bibfnamefont {Y.}~\bibnamefont {Liu}}, \bibinfo {author} {\bibfnamefont {H.}~\bibnamefont {Liang}}, \bibinfo {author} {\bibfnamefont {L.}~\bibnamefont {Yang}}, \bibinfo {author} {\bibfnamefont {G.}~\bibnamefont {Yang}}, \bibinfo {author} {\bibfnamefont {H.}~\bibnamefont {Yang}}, \bibinfo {author} {\bibfnamefont {S.}~\bibnamefont {Song}}, \bibinfo {author} {\bibfnamefont {Z.}~\bibnamefont {Mei}}, \bibinfo {author} {\bibfnamefont {G.}~\bibnamefont {Cs{\'a}nyi}},\ and\ \bibinfo {author} {\bibfnamefont {B.}~\bibnamefont {Cao}},\ }\href {https://doi.org/https://doi.org/10.1002/adma.202210873} {\bibfield  {journal} {\bibinfo  {journal} {Advanced Materials}\ }\textbf {\bibinfo {volume} {35}},\ \bibinfo {pages} {2210873} (\bibinfo {year} {2023})}\BibitemShut {NoStop}%
\bibitem [{\citenamefont {Ndour}\ \emph {et~al.}(2023)\citenamefont {Ndour}, \citenamefont {Jund},\ and\ \citenamefont {Chaput}}]{ndour_practical_2023}%
  \BibitemOpen
  \bibfield  {author} {\bibinfo {author} {\bibfnamefont {M.}~\bibnamefont {Ndour}}, \bibinfo {author} {\bibfnamefont {P.}~\bibnamefont {Jund}},\ and\ \bibinfo {author} {\bibfnamefont {L.}~\bibnamefont {Chaput}},\ }\href {https://doi.org/10.1016/j.jnoncrysol.2023.122618} {\bibfield  {journal} {\bibinfo  {journal} {Journal of Non-Crystalline Solids}\ }\textbf {\bibinfo {volume} {621}},\ \bibinfo {pages} {122618} (\bibinfo {year} {2023})}\BibitemShut {NoStop}%
\bibitem [{\citenamefont {Lundgren}\ \emph {et~al.}(2021)\citenamefont {Lundgren}, \citenamefont {Barbalinardo},\ and\ \citenamefont {Donadio}}]{lundgren_mode_2021}%
  \BibitemOpen
  \bibfield  {author} {\bibinfo {author} {\bibfnamefont {N.~W.}\ \bibnamefont {Lundgren}}, \bibinfo {author} {\bibfnamefont {G.}~\bibnamefont {Barbalinardo}},\ and\ \bibinfo {author} {\bibfnamefont {D.}~\bibnamefont {Donadio}},\ }\href {https://doi.org/10.1103/PhysRevB.103.024204} {\bibfield  {journal} {\bibinfo  {journal} {Physical Review B}\ }\textbf {\bibinfo {volume} {103}},\ \bibinfo {pages} {024204} (\bibinfo {year} {2021})}\BibitemShut {NoStop}%
\bibitem [{\citenamefont {Kielar}\ \emph {et~al.}(2024)\citenamefont {Kielar}, \citenamefont {Li}, \citenamefont {Huang}, \citenamefont {Hu}, \citenamefont {Slebodnick}, \citenamefont {Alatas},\ and\ \citenamefont {Tian}}]{kielar_anomalous_2024}%
  \BibitemOpen
  \bibfield  {author} {\bibinfo {author} {\bibfnamefont {S.}~\bibnamefont {Kielar}}, \bibinfo {author} {\bibfnamefont {C.}~\bibnamefont {Li}}, \bibinfo {author} {\bibfnamefont {H.}~\bibnamefont {Huang}}, \bibinfo {author} {\bibfnamefont {R.}~\bibnamefont {Hu}}, \bibinfo {author} {\bibfnamefont {C.}~\bibnamefont {Slebodnick}}, \bibinfo {author} {\bibfnamefont {A.}~\bibnamefont {Alatas}},\ and\ \bibinfo {author} {\bibfnamefont {Z.}~\bibnamefont {Tian}},\ }\href {https://doi.org/10.1038/s41467-024-51377-8} {\bibfield  {journal} {\bibinfo  {journal} {Nature Communications}\ }\textbf {\bibinfo {volume} {15}},\ \bibinfo {pages} {6981} (\bibinfo {year} {2024})}\BibitemShut {NoStop}%
\bibitem [{\citenamefont {Simoncelli}\ \emph {et~al.}(2024)\citenamefont {Simoncelli}, \citenamefont {Fournier}, \citenamefont {Marangolo}, \citenamefont {Balan}, \citenamefont {Béneut}, \citenamefont {Baptiste}, \citenamefont {Doisneau}, \citenamefont {Marzari},\ and\ \citenamefont {Mauri}}]{simoncelli_temperature-invariant_2024}%
  \BibitemOpen
  \bibfield  {author} {\bibinfo {author} {\bibfnamefont {M.}~\bibnamefont {Simoncelli}}, \bibinfo {author} {\bibfnamefont {D.}~\bibnamefont {Fournier}}, \bibinfo {author} {\bibfnamefont {M.}~\bibnamefont {Marangolo}}, \bibinfo {author} {\bibfnamefont {E.}~\bibnamefont {Balan}}, \bibinfo {author} {\bibfnamefont {K.}~\bibnamefont {Béneut}}, \bibinfo {author} {\bibfnamefont {B.}~\bibnamefont {Baptiste}}, \bibinfo {author} {\bibfnamefont {B.}~\bibnamefont {Doisneau}}, \bibinfo {author} {\bibfnamefont {N.}~\bibnamefont {Marzari}},\ and\ \bibinfo {author} {\bibfnamefont {F.}~\bibnamefont {Mauri}},\ }\href {https://doi.org/10.48550/arXiv.2405.13161} {\bibinfo {title} {Temperature-invariant heat conductivity from compensating crystalline and glassy transport: from the {Steinbach} meteorite to furnace bricks}} (\bibinfo {year} {2024})\BibitemShut {NoStop}%
\bibitem [{\citenamefont {Larsen}\ \emph {et~al.}(2017)\citenamefont {Larsen}, \citenamefont {Mortensen}, \citenamefont {Blomqvist}, \citenamefont {Castelli}, \citenamefont {Christensen}, \citenamefont {Dułak}, \citenamefont {Friis}, \citenamefont {Groves}, \citenamefont {Hammer}, \citenamefont {Hargus}, \citenamefont {Hermes}, \citenamefont {Jennings}, \citenamefont {Jensen}, \citenamefont {Kermode}, \citenamefont {Kitchin}, \citenamefont {Kolsbjerg}, \citenamefont {Kubal}, \citenamefont {Kaasbjerg}, \citenamefont {Lysgaard}, \citenamefont {Maronsson}, \citenamefont {Maxson}, \citenamefont {Olsen}, \citenamefont {Pastewka}, \citenamefont {Peterson}, \citenamefont {Rostgaard}, \citenamefont {Schiøtz}, \citenamefont {Schütt}, \citenamefont {Strange}, \citenamefont {Thygesen}, \citenamefont {Vegge}, \citenamefont {Vilhelmsen}, \citenamefont {Walter}, \citenamefont {Zeng},\ and\ \citenamefont {Jacobsen}}]{ase-paper}%
  \BibitemOpen
  \bibfield  {author} {\bibinfo {author} {\bibfnamefont {A.~H.}\ \bibnamefont {Larsen}}, \bibinfo {author} {\bibfnamefont {J.~J.}\ \bibnamefont {Mortensen}}, \bibinfo {author} {\bibfnamefont {J.}~\bibnamefont {Blomqvist}}, \bibinfo {author} {\bibfnamefont {I.~E.}\ \bibnamefont {Castelli}}, \bibinfo {author} {\bibfnamefont {R.}~\bibnamefont {Christensen}}, \bibinfo {author} {\bibfnamefont {M.}~\bibnamefont {Dułak}}, \bibinfo {author} {\bibfnamefont {J.}~\bibnamefont {Friis}}, \bibinfo {author} {\bibfnamefont {M.~N.}\ \bibnamefont {Groves}}, \bibinfo {author} {\bibfnamefont {B.}~\bibnamefont {Hammer}}, \bibinfo {author} {\bibfnamefont {C.}~\bibnamefont {Hargus}}, \bibinfo {author} {\bibfnamefont {E.~D.}\ \bibnamefont {Hermes}}, \bibinfo {author} {\bibfnamefont {P.~C.}\ \bibnamefont {Jennings}}, \bibinfo {author} {\bibfnamefont {P.~B.}\ \bibnamefont {Jensen}}, \bibinfo {author} {\bibfnamefont {J.}~\bibnamefont {Kermode}}, \bibinfo {author} {\bibfnamefont {J.~R.}\ \bibnamefont {Kitchin}}, \bibinfo {author} {\bibfnamefont {E.~L.}\ \bibnamefont {Kolsbjerg}}, \bibinfo {author} {\bibfnamefont {J.}~\bibnamefont {Kubal}}, \bibinfo {author} {\bibfnamefont {K.}~\bibnamefont {Kaasbjerg}}, \bibinfo {author} {\bibfnamefont {S.}~\bibnamefont {Lysgaard}}, \bibinfo {author} {\bibfnamefont {J.~B.}\ \bibnamefont {Maronsson}}, \bibinfo {author} {\bibfnamefont {T.}~\bibnamefont {Maxson}}, \bibinfo {author} {\bibfnamefont {T.}~\bibnamefont {Olsen}}, \bibinfo {author} {\bibfnamefont {L.}~\bibnamefont {Pastewka}}, \bibinfo {author} {\bibfnamefont {A.}~\bibnamefont {Peterson}}, \bibinfo {author} {\bibfnamefont {C.}~\bibnamefont {Rostgaard}}, \bibinfo {author} {\bibfnamefont {J.}~\bibnamefont {Schiøtz}}, \bibinfo {author} {\bibfnamefont {O.}~\bibnamefont {Schütt}}, \bibinfo {author} {\bibfnamefont {M.}~\bibnamefont {Strange}}, \bibinfo {author} {\bibfnamefont {K.~S.}\ \bibnamefont {Thygesen}}, \bibinfo {author} {\bibfnamefont {T.}~\bibnamefont {Vegge}}, \bibinfo {author} {\bibfnamefont {L.}~\bibnamefont {Vilhelmsen}}, \bibinfo {author} {\bibfnamefont {M.}~\bibnamefont {Walter}}, \bibinfo {author} {\bibfnamefont {Z.}~\bibnamefont {Zeng}},\ and\ \bibinfo {author} {\bibfnamefont {K.~W.}\ \bibnamefont {Jacobsen}},\ }\href {http://stacks.iop.org/0953-8984/29/i=27/a=273002} {\bibfield  {journal} {\bibinfo  {journal} {Journal of Physics: Condensed Matter}\ }\textbf {\bibinfo {volume} {29}},\ \bibinfo {pages} {273002} (\bibinfo {year} {2017})}\BibitemShut {NoStop}%
\bibitem [{\citenamefont {Togo}\ \emph {et~al.}(2023)\citenamefont {Togo}, \citenamefont {Chaput}, \citenamefont {Tadano},\ and\ \citenamefont {Tanaka}}]{togo_implementation_2023}%
  \BibitemOpen
  \bibfield  {author} {\bibinfo {author} {\bibfnamefont {A.}~\bibnamefont {Togo}}, \bibinfo {author} {\bibfnamefont {L.}~\bibnamefont {Chaput}}, \bibinfo {author} {\bibfnamefont {T.}~\bibnamefont {Tadano}},\ and\ \bibinfo {author} {\bibfnamefont {I.}~\bibnamefont {Tanaka}},\ }\href {https://doi.org/10.1088/1361-648X/acd831} {\bibfield  {journal} {\bibinfo  {journal} {Journal of Physics: Condensed Matter}\ }\textbf {\bibinfo {volume} {35}},\ \bibinfo {pages} {353001} (\bibinfo {year} {2023})}\BibitemShut {NoStop}%
\bibitem [{\citenamefont {Togo}\ and\ \citenamefont {Tanaka}(2018)}]{spglib}%
  \BibitemOpen
  \bibfield  {author} {\bibinfo {author} {\bibfnamefont {A.}~\bibnamefont {Togo}}\ and\ \bibinfo {author} {\bibfnamefont {I.}~\bibnamefont {Tanaka}},\ }\href {https://doi.org/10.48550/arXiv.1808.01590} {\bibinfo {title} {{\$\texttt\{Spglib\}\$: a software library for crystal symmetry search}}} (\bibinfo {year} {2018})\BibitemShut {NoStop}%
\bibitem [{\citenamefont {Gonze}\ and\ \citenamefont {Lee}(1997{\natexlab{b}})}]{Gonze1997}%
  \BibitemOpen
  \bibfield  {author} {\bibinfo {author} {\bibfnamefont {X.}~\bibnamefont {Gonze}}\ and\ \bibinfo {author} {\bibfnamefont {C.}~\bibnamefont {Lee}},\ }\href {https://doi.org/10.1103/physrevb.55.10355} {\bibfield  {journal} {\bibinfo  {journal} {Physical Review B}\ }\textbf {\bibinfo {volume} {55}},\ \bibinfo {pages} {10355} (\bibinfo {year} {1997}{\natexlab{b}})}\BibitemShut {NoStop}%
\bibitem [{\citenamefont {Wang}\ \emph {et~al.}(2010)\citenamefont {Wang}, \citenamefont {Wang}, \citenamefont {Wang}, \citenamefont {Mei}, \citenamefont {Shang}, \citenamefont {Chen},\ and\ \citenamefont {Liu}}]{Wang_2010}%
  \BibitemOpen
  \bibfield  {author} {\bibinfo {author} {\bibfnamefont {Y.}~\bibnamefont {Wang}}, \bibinfo {author} {\bibfnamefont {J.~J.}\ \bibnamefont {Wang}}, \bibinfo {author} {\bibfnamefont {W.~Y.}\ \bibnamefont {Wang}}, \bibinfo {author} {\bibfnamefont {Z.~G.}\ \bibnamefont {Mei}}, \bibinfo {author} {\bibfnamefont {S.~L.}\ \bibnamefont {Shang}}, \bibinfo {author} {\bibfnamefont {L.~Q.}\ \bibnamefont {Chen}},\ and\ \bibinfo {author} {\bibfnamefont {Z.~K.}\ \bibnamefont {Liu}},\ }\href {https://doi.org/10.1088/0953-8984/22/20/202201} {\bibfield  {journal} {\bibinfo  {journal} {Journal of Physics: Condensed Matter}\ }\textbf {\bibinfo {volume} {22}},\ \bibinfo {pages} {202201} (\bibinfo {year} {2010})}\BibitemShut {NoStop}%
\bibitem [{noa()}]{noauthor_csd3_nodate}%
  \BibitemOpen
  \href {https://docs.hpc.cam.ac.uk/hpc/user-guide/upgrade2021.html} {\bibinfo {title} {{The Cambridge Service for Data Driven Discovery (CSD3), cost of CPU and GPU hours (2024-09-02).}}}\BibitemShut {Stop}%
\bibitem [{\citenamefont {Togo}\ \emph {et~al.}(2024)\citenamefont {Togo}, \citenamefont {Shinohara},\ and\ \citenamefont {Tanaka}}]{togo_spglib_nodate}%
  \BibitemOpen
  \bibfield  {author} {\bibinfo {author} {\bibfnamefont {A.}~\bibnamefont {Togo}}, \bibinfo {author} {\bibfnamefont {K.}~\bibnamefont {Shinohara}},\ and\ \bibinfo {author} {\bibfnamefont {I.}~\bibnamefont {Tanaka}},\ }\href {https://doi.org/10.1080/27660400.2024.2384822} {\bibfield  {journal} {\bibinfo  {journal} {Science and Technology of Advanced Materials: Methods}\ }\textbf {\bibinfo {volume} {0}},\ \bibinfo {pages} {2384822} (\bibinfo {year} {2024})}\BibitemShut {NoStop}%
\bibitem [{\citenamefont {Reicht}\ \emph {et~al.}(2024)\citenamefont {Reicht}, \citenamefont {Legenstein}, \citenamefont {Wieser},\ and\ \citenamefont {Zojer}}]{reicht_designing_2024}%
  \BibitemOpen
  \bibfield  {author} {\bibinfo {author} {\bibfnamefont {L.}~\bibnamefont {Reicht}}, \bibinfo {author} {\bibfnamefont {L.}~\bibnamefont {Legenstein}}, \bibinfo {author} {\bibfnamefont {S.}~\bibnamefont {Wieser}},\ and\ \bibinfo {author} {\bibfnamefont {E.}~\bibnamefont {Zojer}},\ }\href {https://doi.org/10.3390/molecules29163724} {\bibfield  {journal} {\bibinfo  {journal} {Molecules}\ }\textbf {\bibinfo {volume} {29}},\ \bibinfo {pages} {3724} (\bibinfo {year} {2024})}\BibitemShut {NoStop}%
\bibitem [{\citenamefont {Kriuchevskyi}\ \emph {et~al.}(2020)\citenamefont {Kriuchevskyi}, \citenamefont {Palyulin}, \citenamefont {Milkus}, \citenamefont {Elder}, \citenamefont {Sirk},\ and\ \citenamefont {Zaccone}}]{PhysRevB.102.024108}%
  \BibitemOpen
  \bibfield  {author} {\bibinfo {author} {\bibfnamefont {I.}~\bibnamefont {Kriuchevskyi}}, \bibinfo {author} {\bibfnamefont {V.~V.}\ \bibnamefont {Palyulin}}, \bibinfo {author} {\bibfnamefont {R.}~\bibnamefont {Milkus}}, \bibinfo {author} {\bibfnamefont {R.~M.}\ \bibnamefont {Elder}}, \bibinfo {author} {\bibfnamefont {T.~W.}\ \bibnamefont {Sirk}},\ and\ \bibinfo {author} {\bibfnamefont {A.}~\bibnamefont {Zaccone}},\ }\href {https://doi.org/10.1103/PhysRevB.102.024108} {\bibfield  {journal} {\bibinfo  {journal} {Phys. Rev. B}\ }\textbf {\bibinfo {volume} {102}},\ \bibinfo {pages} {024108} (\bibinfo {year} {2020})}\BibitemShut {NoStop}%
\bibitem [{\citenamefont {de~Araujo~Oliveira}\ \emph {et~al.}(2024)\citenamefont {de~Araujo~Oliveira}, \citenamefont {Fan}, \citenamefont {Harju},\ and\ \citenamefont {Pereira}}]{de_araujo_oliveira_tuning_2024}%
  \BibitemOpen
  \bibfield  {author} {\bibinfo {author} {\bibfnamefont {H.}~\bibnamefont {de~Araujo~Oliveira}}, \bibinfo {author} {\bibfnamefont {Z.}~\bibnamefont {Fan}}, \bibinfo {author} {\bibfnamefont {A.}~\bibnamefont {Harju}},\ and\ \bibinfo {author} {\bibfnamefont {L.~F.~C.}\ \bibnamefont {Pereira}},\ }\href {https://doi.org/10.1021/acsanm.4c01875} {\bibfield  {journal} {\bibinfo  {journal} {ACS Applied Nano Materials}\ } (\bibinfo {year} {2024})}\BibitemShut {NoStop}%
\bibitem [{\citenamefont {Bosoni}\ \emph {et~al.}(2020)\citenamefont {Bosoni}, \citenamefont {Campi}, \citenamefont {Donadio}, \citenamefont {Sosso}, \citenamefont {Behler},\ and\ \citenamefont {Bernasconi}}]{bosoni_atomistic_2020}%
  \BibitemOpen
  \bibfield  {author} {\bibinfo {author} {\bibfnamefont {E.}~\bibnamefont {Bosoni}}, \bibinfo {author} {\bibfnamefont {D.}~\bibnamefont {Campi}}, \bibinfo {author} {\bibfnamefont {D.}~\bibnamefont {Donadio}}, \bibinfo {author} {\bibfnamefont {G.~C.}\ \bibnamefont {Sosso}}, \bibinfo {author} {\bibfnamefont {J.}~\bibnamefont {Behler}},\ and\ \bibinfo {author} {\bibfnamefont {M.}~\bibnamefont {Bernasconi}},\ }\href {https://doi.org/10.1088/1361-6463/ab5478} {\bibfield  {journal} {\bibinfo  {journal} {Journal of Physics D: Applied Physics}\ }\textbf {\bibinfo {volume} {53}},\ \bibinfo {pages} {054001} (\bibinfo {year} {2020})}\BibitemShut {NoStop}%
\bibitem [{\citenamefont {DeAngelis}\ \emph {et~al.}(2019)\citenamefont {DeAngelis}, \citenamefont {Muraleedharan}, \citenamefont {Moon}, \citenamefont {Seyf}, \citenamefont {Minnich}, \citenamefont {McGaughey},\ and\ \citenamefont {Henry}}]{deangelis_thermal_2019}%
  \BibitemOpen
  \bibfield  {author} {\bibinfo {author} {\bibfnamefont {F.}~\bibnamefont {DeAngelis}}, \bibinfo {author} {\bibfnamefont {M.~G.}\ \bibnamefont {Muraleedharan}}, \bibinfo {author} {\bibfnamefont {J.}~\bibnamefont {Moon}}, \bibinfo {author} {\bibfnamefont {H.~R.}\ \bibnamefont {Seyf}}, \bibinfo {author} {\bibfnamefont {A.~J.}\ \bibnamefont {Minnich}}, \bibinfo {author} {\bibfnamefont {A.~J.~H.}\ \bibnamefont {McGaughey}},\ and\ \bibinfo {author} {\bibfnamefont {A.}~\bibnamefont {Henry}},\ }\href {https://doi.org/10.1080/15567265.2018.1519004} {\bibfield  {journal} {\bibinfo  {journal} {Nanoscale and Microscale Thermophysical Engineering}\ }\textbf {\bibinfo {volume} {23}},\ \bibinfo {pages} {81} (\bibinfo {year} {2019})}\BibitemShut {NoStop}%
\bibitem [{\citenamefont {Guerboub}\ \emph {et~al.}(2023)\citenamefont {Guerboub}, \citenamefont {Wansi~Wendji}, \citenamefont {Massobrio}, \citenamefont {Bouzid}, \citenamefont {Boero}, \citenamefont {Ori},\ and\ \citenamefont {Martin}}]{guerboub_impact_2023}%
  \BibitemOpen
  \bibfield  {author} {\bibinfo {author} {\bibfnamefont {M.}~\bibnamefont {Guerboub}}, \bibinfo {author} {\bibfnamefont {S.~D.}\ \bibnamefont {Wansi~Wendji}}, \bibinfo {author} {\bibfnamefont {C.}~\bibnamefont {Massobrio}}, \bibinfo {author} {\bibfnamefont {A.}~\bibnamefont {Bouzid}}, \bibinfo {author} {\bibfnamefont {M.}~\bibnamefont {Boero}}, \bibinfo {author} {\bibfnamefont {G.}~\bibnamefont {Ori}},\ and\ \bibinfo {author} {\bibfnamefont {E.}~\bibnamefont {Martin}},\ }\href {https://pubs.aip.org/aip/jcp/article/158/8/084504/2868943} {\bibfield  {journal} {\bibinfo  {journal} {The Journal of Chemical Physics}\ }\textbf {\bibinfo {volume} {158}} (\bibinfo {year} {2023})}\BibitemShut {NoStop}%
\bibitem [{\citenamefont {Jain}\ and\ \citenamefont {McGaughey}(2015)}]{jain_effect_2015}%
  \BibitemOpen
  \bibfield  {author} {\bibinfo {author} {\bibfnamefont {A.}~\bibnamefont {Jain}}\ and\ \bibinfo {author} {\bibfnamefont {A.~J.}\ \bibnamefont {McGaughey}},\ }\href {https://doi.org/10.1016/j.commatsci.2015.08.014} {\bibfield  {journal} {\bibinfo  {journal} {Computational Materials Science}\ }\textbf {\bibinfo {volume} {110}},\ \bibinfo {pages} {115} (\bibinfo {year} {2015})}\BibitemShut {NoStop}%
\bibitem [{\citenamefont {Ko}\ \emph {et~al.}(2023)\citenamefont {Ko}, \citenamefont {Nassar}, \citenamefont {Miret}, \citenamefont {Liu},\ and\ \citenamefont {Ong}}]{ko_2023_8025189_matgl}%
  \BibitemOpen
  \bibfield  {author} {\bibinfo {author} {\bibfnamefont {T.~W.}\ \bibnamefont {Ko}}, \bibinfo {author} {\bibfnamefont {M.}~\bibnamefont {Nassar}}, \bibinfo {author} {\bibfnamefont {S.}~\bibnamefont {Miret}}, \bibinfo {author} {\bibfnamefont {E.}~\bibnamefont {Liu}},\ and\ \bibinfo {author} {\bibfnamefont {S.~P.}\ \bibnamefont {Ong}},\ }\href@noop {} {\bibfield  {journal} {\bibinfo  {journal} {Zenodo}\ } (\bibinfo {year} {2023})},\ \bibinfo {note} {\url{https://doi.org/10.5281/zenodo.8025189}}\BibitemShut {NoStop}%
\bibitem [{\citenamefont {Plimpton}(1995)}]{plimpton_lammps_1995}%
  \BibitemOpen
  \bibfield  {author} {\bibinfo {author} {\bibfnamefont {S.}~\bibnamefont {Plimpton}},\ }\href {https://doi.org/https://doi.org/10.1006/jcph.1995.1039} {\bibfield  {journal} {\bibinfo  {journal} {Journal of Computational Physics}\ }\textbf {\bibinfo {volume} {117}},\ \bibinfo {pages} {1} (\bibinfo {year} {1995})}\BibitemShut {NoStop}%
\bibitem [{\citenamefont {Hintze}\ and\ \citenamefont {Nelson}(1998)}]{hintze_violin_plots_1998}%
  \BibitemOpen
  \bibfield  {author} {\bibinfo {author} {\bibfnamefont {J.~L.}\ \bibnamefont {Hintze}}\ and\ \bibinfo {author} {\bibfnamefont {R.~D.}\ \bibnamefont {Nelson}},\ }\href {https://doi.org/10.1080/00031305.1998.10480559} {\bibfield  {journal} {\bibinfo  {journal} {The American Statistician}\ }\textbf {\bibinfo {volume} {52}},\ \bibinfo {pages} {181} (\bibinfo {year} {1998})}\BibitemShut {NoStop}%
\bibitem [{\citenamefont {Kresse}\ and\ \citenamefont {Hafner}(1993)}]{kresse_ab_1993}%
  \BibitemOpen
  \bibfield  {author} {\bibinfo {author} {\bibfnamefont {G.}~\bibnamefont {Kresse}}\ and\ \bibinfo {author} {\bibfnamefont {J.}~\bibnamefont {Hafner}},\ }\href {https://doi.org/10.1103/PhysRevB.47.558} {\bibfield  {journal} {\bibinfo  {journal} {Physical Review B}\ }\textbf {\bibinfo {volume} {47}},\ \bibinfo {pages} {558} (\bibinfo {year} {1993})}\BibitemShut {NoStop}%
\bibitem [{\citenamefont {Kresse}\ and\ \citenamefont {Furthmüller}(1996)}]{kresse_efficiency_1996}%
  \BibitemOpen
  \bibfield  {author} {\bibinfo {author} {\bibfnamefont {G.}~\bibnamefont {Kresse}}\ and\ \bibinfo {author} {\bibfnamefont {J.}~\bibnamefont {Furthmüller}},\ }\href {https://doi.org/10.1016/0927-0256(96)00008-0} {\bibfield  {journal} {\bibinfo  {journal} {Computational Materials Science}\ }\textbf {\bibinfo {volume} {6}},\ \bibinfo {pages} {15} (\bibinfo {year} {1996})}\BibitemShut {NoStop}%
\bibitem [{\citenamefont {Ghosez}\ \emph {et~al.}(1998)\citenamefont {Ghosez}, \citenamefont {Michenaud},\ and\ \citenamefont {Gonze}}]{ghosez_dynamical_1998}%
  \BibitemOpen
  \bibfield  {author} {\bibinfo {author} {\bibfnamefont {P.}~\bibnamefont {Ghosez}}, \bibinfo {author} {\bibfnamefont {J.-P.}\ \bibnamefont {Michenaud}},\ and\ \bibinfo {author} {\bibfnamefont {X.}~\bibnamefont {Gonze}},\ }\href {https://doi.org/10.1103/PhysRevB.58.6224} {\bibfield  {journal} {\bibinfo  {journal} {Physical Review B}\ }\textbf {\bibinfo {volume} {58}},\ \bibinfo {pages} {6224} (\bibinfo {year} {1998})},\ \bibinfo {note} {publisher: American Physical Society}\BibitemShut {NoStop}%
\bibitem [{\citenamefont {Marshall}\ and\ \citenamefont {Cleavelin}(1969)}]{marshall_elastic_1969}%
  \BibitemOpen
  \bibfield  {author} {\bibinfo {author} {\bibfnamefont {B.~J.}\ \bibnamefont {Marshall}}\ and\ \bibinfo {author} {\bibfnamefont {C.~R.}\ \bibnamefont {Cleavelin}},\ }\href {https://doi.org/10.1016/0022-3697(69)90164-4} {\bibfield  {journal} {\bibinfo  {journal} {Journal of Physics and Chemistry of Solids}\ }\textbf {\bibinfo {volume} {30}},\ \bibinfo {pages} {1905} (\bibinfo {year} {1969})}\BibitemShut {NoStop}%
\bibitem [{\citenamefont {Talirz}\ \emph {et~al.}(2020)\citenamefont {Talirz}, \citenamefont {Kumbhar}, \citenamefont {Passaro}, \citenamefont {Yakutovich}, \citenamefont {Granata}, \citenamefont {Gargiulo}, \citenamefont {Borelli}, \citenamefont {Uhrin}, \citenamefont {Huber}, \citenamefont {Zoupanos}, \citenamefont {Adorf}, \citenamefont {Andersen}, \citenamefont {Schütt}, \citenamefont {Pignedoli}, \citenamefont {Passerone}, \citenamefont {VandeVondele}, \citenamefont {Schulthess}, \citenamefont {Smit}, \citenamefont {Pizzi},\ and\ \citenamefont {Marzari}}]{talirz_materials_2020}%
  \BibitemOpen
  \bibfield  {author} {\bibinfo {author} {\bibfnamefont {L.}~\bibnamefont {Talirz}}, \bibinfo {author} {\bibfnamefont {S.}~\bibnamefont {Kumbhar}}, \bibinfo {author} {\bibfnamefont {E.}~\bibnamefont {Passaro}}, \bibinfo {author} {\bibfnamefont {A.~V.}\ \bibnamefont {Yakutovich}}, \bibinfo {author} {\bibfnamefont {V.}~\bibnamefont {Granata}}, \bibinfo {author} {\bibfnamefont {F.}~\bibnamefont {Gargiulo}}, \bibinfo {author} {\bibfnamefont {M.}~\bibnamefont {Borelli}}, \bibinfo {author} {\bibfnamefont {M.}~\bibnamefont {Uhrin}}, \bibinfo {author} {\bibfnamefont {S.~P.}\ \bibnamefont {Huber}}, \bibinfo {author} {\bibfnamefont {S.}~\bibnamefont {Zoupanos}}, \bibinfo {author} {\bibfnamefont {C.~S.}\ \bibnamefont {Adorf}}, \bibinfo {author} {\bibfnamefont {C.~W.}\ \bibnamefont {Andersen}}, \bibinfo {author} {\bibfnamefont {O.}~\bibnamefont {Schütt}}, \bibinfo {author} {\bibfnamefont {C.~A.}\ \bibnamefont {Pignedoli}}, \bibinfo {author} {\bibfnamefont {D.}~\bibnamefont {Passerone}}, \bibinfo {author} {\bibfnamefont {J.}~\bibnamefont {VandeVondele}}, \bibinfo {author} {\bibfnamefont {T.~C.}\ \bibnamefont {Schulthess}}, \bibinfo {author} {\bibfnamefont {B.}~\bibnamefont {Smit}}, \bibinfo {author} {\bibfnamefont {G.}~\bibnamefont {Pizzi}},\ and\ \bibinfo {author} {\bibfnamefont {N.}~\bibnamefont {Marzari}},\ }\href {https://doi.org/10.1038/s41597-020-00637-5} {\bibfield  {journal} {\bibinfo  {journal} {Scientific Data}\ }\textbf {\bibinfo {volume} {7}},\ \bibinfo {pages} {299} (\bibinfo {year} {2020})},\ \bibinfo {note} {publisher: Nature Publishing Group}\BibitemShut {NoStop}%
\end{thebibliography}
%

\end{document}